\documentclass[prd,eqsecnum,twocolumn,nofootinbib]{revtex4-2}

\usepackage{amsfonts}
\usepackage{amsmath}
\usepackage{amssymb}
\usepackage{amsthm}
\usepackage{bm}
\usepackage{dcolumn}
\usepackage{epsfig}
\usepackage{graphicx}
\usepackage{graphics}
\usepackage[latin1]{inputenc}
\usepackage{latexsym}
\usepackage{rotating}
\usepackage{xcolor}

\begin{document}

\title{Precisely computing bound orbits of spinning bodies around black holes\\I: General framework and results for nearly equatorial orbits}
\author{Lisa V. Drummond}
\affiliation{Department of Physics and MIT Kavli Institute, MIT, Cambridge, MA 02139 USA}
\author{Scott A. Hughes}
\affiliation{Department of Physics and MIT Kavli Institute, MIT, Cambridge, MA 02139 USA}
\begin{abstract}
Very large mass ratio binary black hole systems are of interest both as a clean limit of the two-body problem in general relativity, as well as for their importance as sources of low-frequency gravitational waves.  At lowest order, the smaller body moves along a geodesic of the larger black hole's spacetime.  Accurate models of such systems require post-geodesic corrections to this motion.  Post-geodesic effects that drive the small body away from the geodesic include the gravitational self force, which incorporates the backreaction of gravitational-wave emission, and the spin-curvature force, which arises from coupling of the small body's spin to the black hole's spacetime curvature. In this paper, we describe a method for precisely computing bound orbits of spinning bodies about black holes. Our analysis builds off of pioneering work by Witzany which demonstrated how to describe the motion of a spinning body to linear order in the small body's spin.  Exploiting the fact that in the large mass-ratio limit spinning-body orbits are close to geodesics (in a sense that can be made precise) and using closed-form results due to van de Meent describing precession of the small body's spin along black hole orbits, we develop a frequency-domain formulation of the motion which can be solved very precisely.  We examine a range of orbits with this formulation, focusing in this paper on orbits which are eccentric and nearly equatorial (i.e., the orbit's motion is $\mathcal{O}(S)$ out of the equatorial plane), but for which the small body's spin is arbitrarily oriented.  We discuss generic orbits with general small-body spin orientation in a companion paper.  We characterize the behavior of these orbits, contrasting them with geodesics, and show how the small body's spin shifts the frequencies $\Omega_r$ and $\Omega_\phi$ which affect orbital motion. These frequency shifts change accumulated phases which are direct gravitational-wave observables, illustrating the importance of precisely characterizing these quantities for gravitational-wave observations.
\end{abstract}
\maketitle

\section{Introduction and motivation}
\label{sec:intro}

\subsection{Extreme mass ratio inspirals of spinning bodies}
\label{sec:emri}

Extreme mass-ratio inspirals (EMRIs) are stellar-mass compact objects (of mass $\mu$) which orbit a massive black hole (mass $M$) and inspiral due to the backreaction of gravitational-wave (GW) emission.  They are predicted to be a key source of low-frequency gravitational waves, which will be targeted by the planned space-based Laser Interferometer Space Antenna (LISA) \cite{eLISA2013,Barausse2020}.  The mass ratios of EMRI systems are small; $\varepsilon \equiv \mu/M$ lies in the range $10^{-7}\text{--}10^{-4}$.  This means that the smaller object makes $\mathcal{O}(1/\varepsilon) \sim 10^4 \text{--} 10^7$ orbits during inspiral.  By matching phase with theoretical model waveforms (``templates'') over those many thousands or millions of orbits, it is expected that EMRI GWs will make possible very precise measurements.  Some of the science goals of EMRI measurements are to precisely determine the properties of the EMRI's black hole and its inspiraling companion \cite{Babak2017}, to probe that black hole's astrophysical environment \cite{Kocsis2011, Barausse2014, Derdzinski2019, Bonga2019}, and to robustly test the Kerr nature of the black hole spacetime \cite{Collins2004, Glampedakis2006,Barack2007, Vigeland2010, Gair2013}.

An EMRI's mass ratio means that these systems can be treated perturbatively.  This facilitates developing useful theoretical models, since models of the system can be developed using techniques from black hole perturbation theory --- we treat the binary as general relativity's exact Kerr solution \cite{Kerr1963}, and add a perturbation which describes the smaller body.  In addition to accurately describing systems with extreme mass ratios, applications of perturbation theory play a role in helping to understand intermediate mass ratio and even comparable mass binaries \cite{LeTiec2011, Nakano2011, LeTiec2013, vandeMeent2020, Rifat2020}.  Especially as the ground-based detectors uncover systems with very unequal mass components \cite{GW190814_2020, GW190412_2020}, there is great interest and potential in combining perturbation theory with numerical relativity \cite{Lousto2010} and analytic strong-field approaches \cite{Buonanno1999, Buonanno2000, Damour2008, Nagar2011, Balmelli2015_2, Khalil2020}.

At zeroth order in the mass ratio $\varepsilon$, the small body travels along a geodesic of the background spacetime of the massive black hole with four-momentum $p^{\alpha}$, obeying
\begin{equation}
\frac{Dp^{\alpha}}{d\tau} = 0\;,\label{eq:geodesic}
\end{equation}
where $D/d\tau$ is the covariant derivative computed along the orbit and $\tau$ is proper time.  When finite mass ratio and finite size effects are taken into account, the right-hand side of Eq.\ (\ref{eq:geodesic}) is replaced by a force $f^\alpha$.  An example of such a force is the gravitational self force, which describes the small body's interaction with its own spacetime curvature \cite{Pound2012, Isoyama2014, vandeMeent2015, Pound2015, Pound2017, Barack2019, Pound2020}.  The self force encodes the backreaction which drives GW-driven inspiral, as well as conservative effects that shift orbital properties relative to the geodesic.

In this paper, we examine the force that arises due to the coupling of the background curvature with the spin of the small body, the spin-curvature force $f_{S}^{\alpha}$.  The equation governing the small body's motion becomes
\begin{equation}
\frac{Dp^{\alpha}}{d\tau}  =f_{S}^{\alpha}\equiv-\frac{1}{2}{R^\alpha}_{\,\nu\lambda\sigma}u^{\nu}S^{\lambda\sigma}\;.
\label{eq:scf}
\end{equation}
This is one of the Mathisson-Papapetrou equations, and will be discussed in detail in Section \ref{sec:mpd}. Here ${R^\alpha}_{\,\nu\lambda\sigma}$ is the Riemann curvature tensor of the background spacetime, and $u^{\nu}$ is the 4-velocity associated with the smaller's orbital motion.  The tensor $S^{\lambda\sigma}$ describes the spin of the orbiting body.  If that body is a Kerr black hole, $S^{\lambda\sigma}\propto s\mu^2$ where $s$ is a dimensionless spin parameter with $s\leq1$.  The spin-curvature force thus affects the orbiting body's motion at next-to-leading-order in mass ratio, just like many important self force effects \cite{Pound2015,Barack2019,Pound2021}.

\subsection{Past work}
\label{sec:pastwork}

A great deal of work, both numerical and analytic, has gone into developing models for the dynamics of and gravitational waves produced by systems containing spinning members. Two limiting approaches have been used extensively for analytic modeling of such systems: the post-Newtonian PN approximation, formally good when members of the binary are widely separated and orbital speeds are small compared to light, and the extreme mass-ratio limit described in Sec.\ \ref{sec:emri}. The effective-one-body (EOB) framework synthesizes elements from post-Newtonian, extreme-mass-ratio, and numerical relativity results in order to construct a useful prescription for modeling inspirals across a wide parameter space.  The dynamics of comparable mass binaries with spinning components has been explored in many post-Newtonian studies \cite{Kesden2015, Gerosa2015, Gerosa2015_2, Cho2019, Mould2020, Tanay2021_1,Tanay2021_2}; complementary to this, binaries with spinning members have been investigated extensively in numerical relativity simulations \cite{Lousto2010_2, Hemberger2013, Boyle2014, Ossokine2015, Lousto2015, Lousto2016}.  Considerable work has also been undertaken to develop EOB models that include spin and quantify their reliability \cite{Damour2001, Damour2008, Nagar2011, Balmelli2013, Balmelli2015, Balmelli2015_2, Khalil2020}; a comparison of spinning effective one body Hamiltonians can be found in Ref.\ \cite{Rettegno2020}.  

In addition, studies of the relativistic three-body problem correspond to the spinning two-body problem in certain regimes.  For example, in hierarchical triple systems, there can be a correspondence between the orbital angular momentum of the so-called ``inner'' binary (a two-body system which itself orbits a massive black hole) and the spin of a test body.  This correspondence holds if the separation of the inner binary is much smaller than the curvature scale associated with the black hole about which the inner binary orbits \cite{Lim2020}.

A number of studies have examined the motion of spinning bodies orbiting black holes.  Many of these studies have focused either on numerical treatment of the Papapetrou equations (for example, Refs.\ \cite{Semerak1999, Plyatsko2011, Li2019}), or on constrained orbital geometries such as nearly circular or nearly equatorial orbits.  For example, Ref.\ \cite{Hinderer2013} finds analytic expressions for the radial, meridional, and spin precession frequencies, including terms quadratic in spin for the limit of nearly circular, nearly equatorial orbits (see in particular Sec.\ IV B of \cite{Hinderer2013}).

Treating the system to first order in the small body's spin has astrophysical relevance in the context of EMRIs.  A scheme of this type was outlined in Ref.\ \cite{Chicone2005} and elucidated further in Refs.\ \cite{Singh2008, Singh2008_2}. Spinning-body orbits have been computed to first order in spin using similar frameworks in Refs.\ \cite{Mashhoon2006, Bini2011_1, Bini2011_2}. A useful effective potential approach presented in Refs.\ \cite{1976Tod, Saijo1998, Hackmann2014} describes equatorial orbits when the spin of the small body is aligned with the orbit.  This method has been employed to compute corrections to orbital frequencies and explore resonance effects for equatorial orbits \cite{Abramowicz1979, Calvani1980, Mukherjee2019}. Corrections to the innermost stable circular orbit (ISCO) location of spinning-body motion have also been calculated \cite{Suzuki1998, Favata2011, Jefremov2015, Tsupko2016, Zhang2019, Zhang2019_2}. 

Another thread to this research is the use of a canonical Hamiltonian framework to describe the motion of a spinning body \cite{Tauber1988}.  An explicit Hamiltonian for the Newton-Wigner supplementary condition was presented to linear order in spin in Ref.\ \cite{Barausse2009}, and later extended to quadratic order by Vines et al.\ \cite{Vines2016}.  This canonical Hamiltonian picture provides the basis for certain spinning EOB models \cite{Barausse2010, Barausse2011}. Witzany et al.\ presented an overview of Hamiltonians for several commonly used spin supplementary conditions, including the Tulczyjew-Dixon condition, in Ref.\ \cite{Witzany2019}.  A Hamilton-Jacobi formulation of spinning-body motion, which exploits the separability of parallel transport in order to determine the turning points analytically, is also known and can be used to compute corrections to the orbital frequencies \cite{Witzany2019_2}.  A covariant Hamiltonian formalism has also been used to describe spinning-body motion \cite{Ambrosi2015, Ambrosi2016}.  This approach is used in Ref.\ \cite{Saravanan2021} to describe circular orbits of spinning bodies in Kerr without truncating higher order spin terms, as well as to study non-planar bound orbits in a Schwarzschild background.

Post-Newtonian analyses long ago indicated that spinning binaries exhibit chaotic dynamics \cite{Levin2000,Cornish2002,Levin2006}.  The integrability of eccentric, spinning back hole binaries up to second post-Newtonian order was demonstrated in Ref.\ \cite{Tanay2021_1}, with action angle variables presented explicitly in Ref.\ \cite{Tanay2021_2}.  In the extreme mass ratio limit, numerical studies in both Schwarzschild \cite{Suzuki1997} and Kerr \cite{Hartl2003, Hartl2003_2} backgrounds found evidence for chaotic motion.  However, the linear-in-spin Hamilton-Jacobi analysis of Witzany \cite{Witzany2019_2} found that the equations of motion ``almost'' separate --- the librational motion in the radial and polar directions is coupled only by the way in which the libration region varies over an orbit.  As such, Witzany shows that the equations of motion are amenable to computing important quantities such as frequencies associated with the orbits of spinning bodies.  This analysis indicates that terms beyond linear in spin are necessary in order for orbits to exhibit chaos.  Indeed, numerical studies have show that prolonged resonances leading to chaotic motion can be attributed to terms that are second order in spin \cite{Zelenka2020}.

Non-integrability and the possibility of chaotic dynamics in the orbits of spinning bodies has received particular attention due to the implications of this for gravitational wave detection \cite{Lukes2021}.  However, even if the motion remains perfectly predictable, it is crucial to understand and quantify the effect a small body's spin has on the dynamics of black hole orbits and the gravitational waves produced in spinning-body EMRI systems. The measurability of the secondary spin and its influence on EMRI parameter estimation has been assessed in previous studies \cite{Burko2015, Huerta2012, Piovano2020, Piovano2021}.  Quasi-circular equatorial orbits with the spin of the small body aligned with the orbit provide a useful limit that has been studied extensively, and is often used to verify new methods for calculating gravitational wave fluxes \cite{Han2010, Harms2016, Nagar2019, Piovano2020_2, Akcay2020_2}.  Gravitational-wave fluxes from equatorial orbits with aligned spin \cite{Saijo1998,Skoupy2021} and quasi-circular orbits with misaligned spin \cite{Tanaka1996} have also been well studied.  Warburton and collaborators investigated the gravitational wave emission of a spinning body with misaligned spin orbiting a non-rotating black hole in Ref.\ \cite{Warburton2017}. The impact of different spin supplementary conditions on gravitational wave fluxes has been explored for both Schwarzschild \cite{Harms2016_2} and Kerr \cite{Lukes2017} black holes.  Finally, as we were completing this analysis, Mathews et al.\ presented a detailed examination of the impact of a spinning secondary on the self force \cite{mathews2021selfforce}, focused on the simplest case (Schwarzschild background, spin parallel to orbit, circular configuration).

\subsection{This work: Synopsis of our formulation}
\label{subsubsec:freqdom}

In this work, we examine orbits under the influence of the spin-curvature force $f^\alpha_S$.  Because our focus is on extreme mass-ratio systems, we truncate all spin effects at leading order in the small body's spin.  Under the assumption that the small body is itself a Kerr black hole (an astrophysically plausible assumption for EMRI systems), the small body's spin has a magnitude that scales with its mass squared.  Terms beyond linear in spin thus scale very steeply with the system's mass ratio.  At this order, a closed-form description of the spin precession is known \cite{vandeMeent2020}, amounting to parallel transport of a vector along a Kerr geodesic.  With the precessional dynamics of the small body's spin in hand, we can straightforwardly compute the spin-curvature force.  From this, we find the spinning-body trajectory $[r(t), \theta(t), \phi(t)]$ consistent with the spin-curvature force by solving Eq.\ (\ref{eq:scf}).

Following Ref.\ \cite{vandeMeent2020}, we characterize the small body's spin using a set of quantities $\{S^1, S^2, S^3\}$ which represent the components of its spin vector projected onto three legs of a tetrad used in the closed-form analysis of its precession (see Sec.\ \ref{sec:ParallelTransport}).  (A fourth component $S^0$, corresponding to the remaining leg of the tetrad, is constrained to be zero by the spin supplementary condition discussed in Sec.\ \ref{sec:ssc}.)  We write its magnitude $S=\sqrt{S_\parallel^2+S_\perp^2}$, where $S_{\parallel}=S^3$ describes the component normal to the orbital plane, and $S_\perp=\sqrt{(S^1)^2+(S^2)^2}$ describes its magnitude within this plane.  If $S_\perp \ne 0$, then components of the spin vector oscillate in the orbital plane with a frequency $\Omega_s$, describing a precession of the spin vector along its orbit; this frequency is described in more detail in Sec.\ \ref{sec:ParallelTransport}, and computed in Ref.\ \cite{vandeMeent2020}.  At leading order in spin, the quantities $S_\perp$ and $S_{\parallel}$ (and thus $S$) are constants of motion along the spinning body's orbit.

Because we consider the small body's spin to be a small parameter, the spinning-body orbits we examine are ``close to'' geodesic orbits (in a sense made more precise later).  We begin our discussion of spinning-body orbits by examining how we parameterize bound Kerr geodesics.  The radial motion of bound geodesics is typically described using a semi-latus rectum $p$ and an eccentricity $e$, such that the orbit oscillates between apoastron at $r_1 = pM/(1-e)$ and periastron at $r_2 = pM/(1+e)$.  The polar angle $\theta$ of a bound orbit oscillates such that $-\sin{I} \le \cos\theta \le \sin{I}$.  Using these bounds, we write these motions
\begin{align}
    \hat r & =\frac{p M}{1 + e\cos\hat\chi_r}\;, \ \ \cos\hat\theta = \sin I\cos\hat\chi_\theta\;.
    \label{eq:geodparam1}
\end{align}
Here and throughout this paper, we use a ``hat'' accent (e.g.\ $\hat r$) to denote a quantity which is evaluated on a geodesic.  The definitions (\ref{eq:geodparam1}) introduce the angles $\hat\chi_r$ and $\hat\chi_\theta$, which are generalizations of ``true anomaly'' angles often used in discussions of orbits in Newtonian gravity.  The libration range of the geodesics does not change over an orbit, so that $p$, $e$ and $I$ are all constants of motion.  Geodesics can be equivalently characterized by another set of constants of motion: $\hat{E}$, $\hat{L}_z$ and $\hat{Q}$, which denote a geodesic's energy, axial angular momentum and Carter constant respectively.  These quantities are discussed in more detail in Sec.\ \ref{sec:kerrgeodesics}.

Spinning-body orbits cannot in general be parameterized in the same way as geodesics using Eq.\ (\ref{eq:geodparam1}).  For the ``nearly equatorial'' cases that we consider in this paper, we find the following parameterization robustly describes these orbits:
\begin{align}
    r & =\frac{p M}{1 + e\cos\chi_r}\;, \ \ \theta = \frac{\pi}{2}+\delta\vartheta_S\;.\label{eq:thetaparamfirst}
\end{align}
This radial motion has turning points at $r = pM/(1 \pm e)$, exactly as for geodesic orbits.  However, the anomaly angle $\chi_r$ is not the same as the anomaly angle $\hat\chi_r$ which describes geodesic motion.  We elaborate on the difference between these angles in Sec.\ \ref{sec:slightlyecc}.  The polar angle deviates from the equatorial plane by $\delta\vartheta_S$, a quantity with an amplitude $\mathcal{O}(S_\perp)$ which oscillates at harmonics of the frequency $\Omega_s$.  If $S_\perp = 0$, so that the small body's spin is aligned or anti-aligned with the orbital angular momentum, then $\delta\vartheta_S = 0$.  Aligned and anti-aligned orbits can be purely equatorial.

For generic orbits, we find that the libration regions in both $r$ and $\theta$ must be modified to include oscillations at precession frequency $\Omega_s$.  We defer the details of how this is handled to our companion analysis, Ref.\ \cite{Paper2}, which examines generic orbits of spinning bodies with generic spin-orbit configuration.

\subsection{Organization of this paper}

In the remainder of this paper, we present our method for precisely computing bound orbits of spinning bodies orbiting black holes.  We begin by outlining characteristics of geodesics around a Kerr black hole in Sec.\ \ref{sec:kerrgeodesics}.  We discuss the constants of motion, 4-velocities, and turning points associated with bound Kerr geodesics in \ref{subsec:kerrmetric} and \ref{subsec:fourvelocities_param}.  In \ref{subsec:geodesicsfreqdom}, we present a  frequency-domain description of motion in a Kerr spacetime that is particularly useful in our examination of spinning-body orbits.  In Sec.\ \ref{sec:mpd}, we move on to the equations of motion for a body when its spin couples to spacetime curvature.  We focus on the leading order in spin limit that has the most relevance to the astrophysical systems we are studying in Sec.\ \ref{sec:leadingorder}.  In this limit, the spin vector is parallel transported along the worldline.  Given this, we discuss parallel transport along Kerr geodesics in some detail in Sec.\ \ref{sec:ParallelTransport}.

We begin our detailed study of bound spinning-body motion by examining several simple cases.  In Sec.\ \ref{sec:simpleorbits}, we examine orbits which are circular and either equatorial or nearly equatorial, for which we can obtain closed form analytic solutions.  This simple case allows us to establish the general principles of the framework we use throughout the paper, as well as to compare with previously known results.  We present the circular, nearly equatorial case in detail and for general black hole spin.  In Sec.\ \ref{sec:slightlyecc}, we extend these circular cases by expanding in eccentricity in order to study slightly eccentric, nearly equatorial orbits.  For general Kerr, we develop closed-form solutions to first order in eccentricity.  We also present these solutions to second order in eccentricity for the Schwarzschild limit.

Finally, in Sec.\ \ref{sec:spinbodyfreqdom}, we use a frequency-domain treatment to compute orbits with arbitrary eccentricity and with the small body's spin arbitrarily oriented.  The frequency-domain expansion allows us to examine orbits with arbitrary eccentricity, provided we include enough harmonics in our expansion.  We calculate how the spin-curvature coupling shifts the orbital frequencies $\Omega_r$ and $\Omega_\phi$ from their geodesic expectations (using the fact that the parameterization for nearly equatorial spinning-body orbits is very similar to the parameterization of equatorial geodesic orbits), as well as how the coupling shifts the constants of motion $E^S$, $L_z^S$ and $Q^S$.

Section \ref{sec:summary} concludes with a summary of our results, and an outline of plans for future work that uses the orbits of spinning bodies.  We also briefly remark on results we present in our companion paper \cite{Paper2}, which describes how to extend this framework to model fully generic orbits (i.e., orbits of arbitrary eccentricity and inclination) with generic orientation of the small body's spin.

\section{Kerr Geodesics}
\label{sec:kerrgeodesics}

Because we describe orbits of spinning bodies as perturbations of the orbits of non-spinning bodies, we begin by briefly reviewing the properties of Kerr geodesics.  This content has been discussed at great length elsewhere  \cite{Schmidt2002, Kraniotis2004, DrascoHughes2004, Hackmann2008, Levin2008, Levin2009, FujitaHikida2009, Hackmann2010, Warburton2013, Rana2019}; here we provide a brief synopsis in order for the paper to be self-contained, and to introduce important notation and conventions.

\subsection{Kerr metric and constants of motion}
\label{subsec:kerrmetric}

The metric for a Kerr black hole with mass $M$ and spin parameter $a$ in Boyer-Lindquist coordinates $t$, $r$, $\theta$, $\phi$ \cite{Boyer1967} reads
\begin{align}
ds^2 & =-\left(1-\frac{2r}{\Sigma}\right)\,dt^2+\frac{\Sigma}{\Delta}\,dr^2-\frac{4Mar\sin^2\theta}{\Sigma}dt\,d\phi\nonumber\\
 &+\Sigma\,d\theta^2 +\frac{\left(r^2+a^2\right)^2-a^2\Delta\sin^2\theta}{\Sigma}\sin^2\theta\,d\phi^2,\label{eq:kerrmetric}
\end{align}
where 
\begin{equation}
\Delta =r^2-2Mr+a^2\;,\qquad \Sigma =r^2+a^2\cos^2\theta\;.
\end{equation}
(Here and throughout we use geometrized units, with $G = 1 = c$.)

Four constants of motion characterize Kerr geodesics.  The first is the rest mass $\mu$ of the orbiting body.  It is determined by requiring $\hat p^\alpha = \mu \hat u^\alpha$ (where $\hat p^\alpha$ is the geodesic's 4-momentum, and $\hat{u}^\alpha$ its 4-velocity; recall we use the hat accent to denote quantities defined along geodesics) and by requiring the norm of the 4-velocity to be $-1$.  The Kerr metric (\ref{eq:kerrmetric}) is independent of the coordinates $t$ and $\phi$, implying that the spacetime possesses two Killing vectors $\xi_{t}^{\alpha}$ and $\xi_{\phi}^{\alpha}$, corresponding to time translation and axial symmetries respectively. These Killing vectors yield two more constants of the motion, the energy per unit mass $\hat E$ and axial angular momentum per unit mass $\hat L_z$:
\begin{align}
\hat E & =-\xi_{t}^{\alpha}\hat u_{\mu}= -\hat u_{t}\;,\\
\hat L_z & =\xi_{\phi}^{\alpha}\hat u_{\mu} = \hat u_{\phi}\;.
\end{align}
Note that we have normalized these quantities by the mass $\mu$ of the orbiting body.

The Kerr metric also admits an anti-symmetric Killing-Yano tensor \cite{Penrose1973}, given by \cite{Tanaka1996}
\begin{equation}
\mathcal{F}_{\mu\nu}=a\cos\theta\left(\bar e_{\mu}^{1}\bar e_{\nu}^{0}-\bar e_{\mu}^{0}\bar e_{\nu}^{1}\right)+r\left(\bar e_{\mu}^2 \bar e_{\nu}^{3}-\bar e_{\mu}^{3}\bar e_{\nu}^2\right)\;,
\end{equation}
where
\begin{align}
\bar e_{\mu}^0 & =\left[\sqrt{\frac{\Delta}{\Sigma}},0,0,-a\sin^2\theta\sqrt{\frac{\Delta}{\Sigma}}\right],\\
\bar e_{\mu}^1 & =\left[0,\sqrt{\frac{\Sigma}{\Delta}},0,0\right],\\
\bar e_{\mu}^2 & =\left[0,0,\sqrt{\Sigma},0\right],\\
\bar e_{\mu}^3 & =\left[-\frac{a\sin\theta}{\sqrt{\Sigma}},0,0,\frac{\left(r^2+a^2\right)\sin\theta}{\sqrt{\Sigma}}\right].
\end{align}
This tensor has the defining property
\begin{equation}
\nabla_{\gamma}\mathcal{F}_{\alpha\beta}+\nabla_{\beta}\mathcal{F}_{\alpha\gamma}=0\;.
\label{eq:KillingYanoDerivs}
\end{equation}
Let us define the vector
\begin{equation}
 \hat{\mathcal{L}}^\nu=\mathcal{F}^{\mu\nu}\hat u_{\mu}\;.
    \label{eq:orbangmomdef}
\end{equation}
We will call this the orbital angular momentum 4-vector, since it has the dimensions of orbital angular momentum (per unit mass of the orbiting body), and reduces to the orbital angular momentum in the Schwarzschild limit.

Notice that in Refs.\ \cite{Witzany2019_2} and \cite{vandeMeent2019}, this vector is defined with the index contracted on the second index of $\mathcal{F}^{\mu\nu}$.  Because of the Killing-Yano tensor's antisymmetry, this results in an overall sign difference.  With the definition (\ref{eq:orbangmomdef}), equatorial orbits have $\hat{\mathcal{L}}^\theta \propto -\hat L_z$.  This is a sensible correspondence, since (by right-hand rule) one expects the angular momentum of a prograde equatorial orbit (for which $\hat L_z > 0$) to point opposite to the direction of increasing polar angle $\theta$.  We have found that this sign swap is needed to establish correspondence between our results and important examples of past literature.  In particular, past work which examined equatorial orbits of bodies with spin aligned with the large black hole's spin and with the orbital angular momentum typically designate the small body's spin as pointing along the ``$z$ direction.''  This correspondence requires the ``$z$ direction'' (i.e., parallel to the large black hole's spin) to point in the direction of decreasing $\theta$ at the equatorial plane.

From the antisymmetry of $\mathcal{F}^{\mu\nu}$ we see that
\begin{equation}
    \hat{\mathcal{L}}^\mu \hat u_\mu = 0\;.
    \label{eq:orbangmomconstraint}
\end{equation}
Further, using Eq.\ (\ref{eq:KillingYanoDerivs}), it is straightforward to show that $\hat{\mathcal{L}}^\mu$ is parallel-transported along geodesics:
\begin{equation}
\frac{D\hat{\mathcal{L}}^\beta}{d\tau} \equiv \hat u^{\alpha}\nabla_{\alpha}\hat{\mathcal{L}}^{\beta} = 0\;.
\end{equation}
It is also not hard to show that the square of this vector
\begin{equation}
\hat K = \hat{{\cal L}}^{\mu}\hat{{\cal L}}_{\mu}
\label{eq:carter1}
\end{equation}
is conserved, i.e.~that
\begin{equation}
    \frac{D\hat K}{d\tau} \equiv \hat u^\alpha\nabla_\alpha \hat K = 0\;.
\end{equation}
Carter \cite{Carter1968} first demonstrated the existence of a fourth conserved constant for Kerr geodesic motion.  This constant arises from a Killing tensor $K_{\mu\nu}$, which can be thought of as the ``square'' of $\mathcal{F_{\mu\nu}}$,
\begin{equation}
K_{\mu\nu}=\mathcal{F}_{\mu\alpha}{\mathcal{F}_\nu}^{\alpha}\;.
\end{equation}
The corresponding constant
\begin{equation}
\hat K = K_{\alpha\beta}\hat u^{\alpha}\hat u^{\beta}
\end{equation}
is identical to the $\hat K$ defined in (\ref{eq:carter1}), and is usually called the ``Carter constant.''  For many analyses, it is particularly convenient to combine $\hat K$, $\hat E$, and $\hat L_z$ into a related conserved quantity $\hat Q$ given by
\begin{align}
\hat Q &= \hat K - \left(\hat L_z-a\hat E\right)^2
\label{eq:Qdef}
\\
 &= \hat p_{\theta}^2 + a^2\cos^2{\hat\theta}\left(1 - \hat E^2\right)+\cot^2{\hat\theta}\,\hat L_z^2\;.
\end{align}
Confusingly, $\hat Q$ is also often called the Carter constant; we will use both $\hat K$ and $\hat Q$ from time to time in our analysis.  The constant $\hat Q$ is particularly useful for discussing geodesics, so we focus on this version of the Carter constant in the remainder of this section.

\subsection{4-velocities, turning points, and parameterization}
\label{subsec:fourvelocities_param}

Carter first showed that the existence of these conserved quantities permits the geodesic equations to be separated in Boyer-Lindquist coordinates \cite{Carter1968}.  These separated equations are given by
\begin{align}
\Sigma^2\left(\frac{d{\hat r}}{d\tau}\right)^2 &= [\hat E({\hat r}^2+a^2)-a\hat L_z]^2\nonumber\\
 & \qquad-\Delta[{\hat r}^2+(\hat L_z-a\hat E)^2+\hat Q]\nonumber\\
 & \equiv R({\hat r})\;,\label{eq:geodr}\\
\Sigma^2\left(\frac{d{\hat\theta}}{d\tau}\right)^2 &= \hat Q-\cot^2{\hat\theta} \hat L_z^2-a^2\cos^2{\hat\theta}(1-\hat E^2)\nonumber\\
 & \equiv\Theta({\hat\theta})\;,\label{eq:geodtheta}\\
\Sigma\frac{d  {\hat\phi}}{d\tau} &= \csc^2{\hat\theta} \hat L_z + a\hat E\left(\frac{{\hat r}^2 + a^2}{\Delta} - 1\right) - \frac{a^2\hat L_z}{\Delta}\nonumber\\
 & \equiv\Phi({\hat r},{\hat\theta})\;,\label{eq:geodphi}\\
\Sigma\frac{d  {\hat t}}{d\tau} &= \hat E\left(\frac{({\hat r}^2 + a^2)^2}{\Delta} - a^2\sin^2{\hat\theta}\right)\nonumber\\
&\qquad + a \hat L_z\left(1 - \frac{{\hat r}^2 + a^2}{\Delta}\right)\nonumber\\
 & \equiv T({\hat r},{\hat\theta})\;.\label{eq:geodt}
\end{align}
Because these are evaluated strictly along geodesic orbits, we parameterize them using the coordinates $(\hat r, \hat\theta, \hat\phi, \hat t)$ of such an orbit.  Equations (\ref{eq:geodr}) -- (\ref{eq:geodt}) are parameterized using proper time $\tau$ along the orbit.  As written, these equations are not completely separated: the factor $\Sigma = {\hat r}^2 + a^2\cos^2{\hat\theta}$ couples the radial and polar motions.  By introducing a new time parameter $\lambda$, commonly called ``Mino time'' and defined by $d\lambda=d\tau/\Sigma$ \cite{Mino2003}, the radial and polar equations of motion decouple, yielding
\begin{align}
\left(\frac{d{\hat r}}{d\lambda}\right)^2 &= R({\hat r})\;,\qquad \left(\frac{d{\hat\theta}}{d\lambda}\right)^2=\Theta({\hat\theta})\;,
\nonumber\\
\frac{d  {\hat\phi}}{d\lambda} &= \Phi({\hat r},{{\hat\theta}})\;,\qquad \frac{d  {\hat t}}{d\lambda}=T({\hat r},{\hat\theta})\;.
\label{eq:geods_mino}
\end{align}
Mino-time $\lambda$ is a very convenient parameterization for describing the strong-field dynamics of Kerr black hole orbits.  By using $d\hat t/d\lambda$, it is not difficult to convert from $\lambda$ to Boyer-Lindquist time $t$, which naturally describes quantities as measured by a distant observer.


To understand the turning points of bound geodesics and the parameterization that we use, begin by carefully examining the functions $R({\hat r})$ and $\Theta({\hat\theta})$.  For bound orbits, $R({\hat r})$ can be written
\begin{equation}
R({\hat r})=(1-\hat E^2)(r_{1}-{\hat r})({\hat r}-r_{2})({\hat r}-r_{3})({\hat r}-r_{4})\;,
\end{equation}
where the roots are ordered such that $r_4 \le r_3 \le r_2 \le {\hat r} \le r_1$.  The roots $r_1$ and $r_2$ are turning points of the motion.  Likewise, $\Theta({\hat\theta})$ can be written
\begin{equation}
\Theta({\hat\theta})=\frac{a^2}{\sin^2{\hat\theta}}\left(1 - \hat E^2\right)\left(z_{+} - \cos^2{\hat\theta}\right)\left(z_{-} - \cos^2{\hat\theta}\right)\;,
\end{equation}
where we have introduced ${\hat z} \equiv \cos^2{\hat\theta}$.  These roots are ordered such that $0 \le z_- \le 1 \le z_+$; turning points of the motion occur where ${\hat z} = z_-$.  This occurs when ${\hat\theta} = \theta_-$ and ${\hat\theta} = \pi - \theta_-$, defined by $\cos^2\theta_- = z_-$.

Bound geodesics are thus confined to a torus, bounded in radius by $r_2 \le {\hat r} \le r_1$ and in polar angle by $\theta_- \le {\hat\theta} \le (\pi - \theta_-)$.  We can build these bounds into the orbiting body's motion by defining
\begin{align}
    {\hat r} & =\frac{p M}{1 + e\cos\hat\chi_r}\;,
    \label{eq:rdef}\\
    \cos{\hat\theta} &= \sin I\cos\hat\chi_\theta\;.
    \label{eq:thdef}
\end{align}
The angles $\hat\chi_r$ and $\hat\chi_\theta$ are relativistic generalizations of the ``true anomaly'' angles often used in Newtonian orbital dynamics; these angles increase monotonically over an orbit.  The parameters $p$ and $e$ are the orbit's semi-latus rectum and eccentricity, respectively; in the Newtonian limit, they correspond to the equivalent parameters which define a Keplerian ellipse.  By inspection, one can see that
\begin{equation}
    r_1 = \frac{pM}{1 - e}\;,\qquad r_2 = \frac{pM}{1 + e}\;.
\end{equation}
The angle $I$ defines the inclination of the orbit; it is related to the angle $\theta_-$ according to
\begin{equation}
    I = \pi/2 - \mbox{sgn}(\hat L_z)\theta_-\;.
\end{equation}
This angle automatically encodes a notion of prograde ($\hat L_z > 0$, $I < 90^\circ$) and retrograde ($\hat L_z < 0$, $I > 90^\circ$) orbits.  Equatorial orbits ($\theta_- = 90^\circ$) have $I = 0^\circ$ (prograde) or $I = 180^\circ$ (retrograde).

Up to initial conditions, an orbit can be specified by either the set of constants of the motion ($\hat E$, $\hat L_z$, $\hat Q$) or the quantities ($p$, $e$, $I$) which determine the orbit's geometry (being careful to choose values which do not go inside the ``last stable orbit,'' the locus of parameter space inside which bound orbits are unstable and rapidly plunge into the black hole; see \cite{Stein2020} for discussion).  In this analysis, we use ($p$, $e$, $I$), and then use expressions given in Refs.\ \cite{FujitaHikida2009, vandeMeent2019} (see also App.\ A of Ref.\ \cite{Hughesetal2021}) to determine $\hat E$, $\hat L_z$, and $\hat Q$.  Once these parameters are known, we can use closed-form expressions for the solutions to the geodesic equations (\ref{eq:geodr}--\ref{eq:geodt}), formulated in terms of elliptic functions \cite{FujitaHikida2009}.  We also use solutions for bound geodesic trajectories as functions of Mino-time, ${\hat r(\lambda)}$ and ${\hat z(\lambda)}$, using the simplified form given by van de Meent \cite{vandeMeent2019}.  Formulae for computing geodesic trajectories are implemented in the {\tt KerrGeodesics} {\it Mathematica} package of the Black Hole Perturbation Toolkit (hereafter ``the Toolkit'') \cite{Kerrgeodesics}.

\subsection{Frequency-domain description of geodesic motion}
\label{subsec:geodesicsfreqdom}

Bound Kerr geodesics are triperiodic, with three frequencies describing their radial, polar, and azimuthal motions. Denote by $\hat\Lambda_{r}$, $\hat\Lambda_{\theta}$, and $\hat\Lambda_{\phi}$ the radial, polar, and axial Mino-time periods (i.e., the interval of Mino time it takes for the orbit to move from $r_1$ to $r_2$ back to $r_1$; the interval to move from $\theta_-$ to $\pi - \theta_-$ back to $\theta_-$; and the interval to move through $2\pi$ radians of axial angle). Denote by $\hat\Upsilon_{r}$, $\hat\Upsilon_{\theta}$, and $\hat\Upsilon_{\phi}$ the corresponding frequencies, with $\hat\Upsilon_x = 2\pi/\hat\Lambda_x$. First derived in this form in Ref.\ \cite{DrascoHughes2004}, we used closed-form expressions for these quantities given in Ref.\ \cite{FujitaHikida2009}, and coded into the {\tt KerrGeodesics} package of the Toolkit \cite{Kerrgeodesics}.


From these Mino-time expressions, we can find their Boyer-Lindquist coordinate-time analogues using a factor $\hat\Gamma$ which is the orbit-averaged factor relating an interval of Mino-time $\lambda$ to an element of coordinate time $t$.  Let $\hat T_x$ be the coordinate time orbital period for motion in coordinate $x$, and let $\hat\Omega_x = 2\pi/\hat T_x$ be the corresponding frequency.  Then,
\begin{equation}
\hat\Omega_{r,\theta,\phi}=\frac{\hat\Upsilon_{r,\theta,\phi}}{\hat\Gamma}\;,\qquad
\hat T_{r,\theta,\phi} = \hat\Gamma\,\hat\Lambda_{r,\theta,\phi}\;.
\end{equation}
Expressions for $\hat\Gamma$ (and thus for $\hat\Omega_{r,\theta,\phi}$) are also provided in Ref.\ \cite{FujitaHikida2009} and encoded in the {\tt KerrGeodesics} package of the Toolkit \cite{Kerrgeodesics}


The Mino-time frequencies are particularly useful for our purposes because they make possible Fourier expansions of functions evaluated along Kerr orbits.  Let $f(\lambda)=f\left[{\hat r(\lambda)},{\hat\theta(\lambda)}\right]$ be a function of ${\hat r(\lambda)}$ and ${\hat \theta(\lambda)}$.  As shown in Ref.\ \cite{DrascoHughes2004}, we can write
\begin{equation}
f = \sum_{k=-\infty}^\infty\sum_{n = -\infty}^\infty f_{kn}e^{-i\left(k\hat\Upsilon_{\theta} + n\hat\Upsilon_{r}\right)\lambda}\;,
\end{equation}
where the Fourier coefficient $f_{kn}$ is given by
\begin{widetext}
\begin{equation}
f_{kn} = \frac{1}{\hat\Lambda_{r}\hat\Lambda_{\theta}}\int_{0}^{\hat\Lambda_{r}}\int_{0}^{\hat \Lambda_{\theta}} f\left[{\hat r(\lambda_{r})},{\hat\theta(\lambda_{\theta})}\right] e^{ik\hat\Upsilon_{\theta}\lambda_\theta} e^{in\hat\Upsilon_{r}\lambda_r}d\lambda_{\theta}d\lambda_{r}\;.
\end{equation}
\end{widetext}
The component $f_{00}$ represents the orbit-average of the function $f[{\hat r(\lambda)}, {\hat \theta(\lambda)}]$.  It's worth noting that the quantities $\hat\Upsilon_\phi$ and $\hat\Gamma$ are orbit averages of the functions $\Phi( {\hat r}, {\hat \theta})$ and $T( {\hat r},  {\hat \theta})$ defined in Eq.\ (\ref{eq:geods_mino}):
\begin{align}
    \hat\Upsilon_\phi &= \frac{1}{\hat\Lambda_r\hat\Lambda_\theta}\int_0^{\hat\Lambda_r}\int_0^{\hat\Lambda_\theta} \Phi[{\hat r(\lambda_r)}, {\hat \theta(\lambda_\theta)}]d\lambda_r\,d\lambda_\theta\;,
    \label{eq:Upsphi_def}\\
    \hat\Gamma &= \frac{1}{\hat\Lambda_r\hat\Lambda_\theta}\int_0^{\hat\Lambda_r}\int_0^{\hat\Lambda_\theta} T[{\hat r(\lambda_r)}, {\hat \theta(\lambda_\theta)}]d\lambda_r\,d\lambda_\theta\;.
    \label{eq:Gamma_def}
\end{align}
We will use a variant of these definitions to compute $\Upsilon_\phi$ and $\Gamma$ along orbits of spinning bodies. 

\section{The motion of a spinning body}
\label{sec:mpd}
Strictly speaking, geodesics describe only the motion of zero-mass point particles. Any mass deforms the spacetime, pushing its trajectory away from the geodesic; any structure beyond a point can couple to spacetime curvature, also pushing its trajectory away from the geodesic. The leading example of such structure is the body's spin.  We now consider the orbital motion of a pointlike body endowed with spin angular momentum.

\subsection{Spin-curvature coupling}
\label{sec:scc}

A small spinning body moving in a curved spacetime precesses as it moves along its trajectory, and couples to the curvature of the background spacetime.  The equations governing this precession and motion are known as the Mathisson-Papapetrou equations \cite{Papapetrou1951, Mathisson2010,Mathisson2010G_2,Dixon1970}, and are given by
\begin{align}
\frac{Dp^{\alpha}}{d\tau} & =-\frac{1}{2}{R^\alpha}_{\,\nu\lambda\sigma}u^{\nu}S^{\lambda\sigma}\;,\label{eq:mp1}\\
\frac{DS^{\alpha\beta}}{d\tau} & =p^{\alpha}u^{\beta}-p^{\beta}u^{\alpha}\;.\label{eq:mp2}
\end{align}
In these equations, the operator $D/d\tau$ denotes a covariant derivative along the small body's worldline, ${R^\alpha}_{\,\nu\lambda\sigma}$ is the Riemann curvature of the spacetime in which the small body orbits, $S^{\lambda\sigma}$ is the small body's spin tensor (about which we say more below), $p^\alpha$ is the small body's 4-momentum, and $u^\nu = dx^\nu/d\tau$ is its 4-velocity.  In general, a spinning body's 4-momentum and 4-velocity are not parallel to each other, but are related by
\begin{equation}
p^{\alpha}=\mu u^{\alpha}-u_{\gamma}\frac{DS^{\alpha\gamma}}{d\tau}\;.\label{eq:momvel}
\end{equation}
Including additional structure on the small body leads to more complicated equations of motion.  For example, the small body's quadrupole moment couples to the gradient of curvature \cite{Bini2008,Bini2014,Steinhoff2010} and introduces additional torque terms \cite{Rudiger1981}.  The Mathisson-Papapetrou equations represent the ``pole-dipole'' approximation, in which the small body is treated as a monopolar point mass supplemented with a dipolar spin.



For each spacetime Killing vector $\xi^{\alpha}$ there is constant of motion along the spinning body's worldline given by
\begin{equation}
\mathcal{C}=p_{\alpha}\xi^{\alpha}-\frac{1}{2}S^{\alpha\beta}\nabla_{\beta}\xi_{\alpha}\;.
\end{equation}
Using this, one finds that the conserved energy and axial angular momentum per unit mass for a spinning body moving in a Kerr spacetime are given by
\begin{align}
E^S & = -u_t+\frac{1}{2\mu}\partial_{\beta}g_{t\alpha}S^{\alpha\beta} \label{eq:Espin},\\ 
L_z^S & = u_{\phi}-\frac{1}{2\mu}\partial_{\beta}g_{\phi\alpha}S^{\alpha\beta}.\label{eq:Lspin}
\end{align}
There is no Carter constant for a spinning body, though (as we discuss below) there is a generalization of the Carter constant which is conserved to linear order in the small body's spin.




\subsection{Spin supplementary conditions}
\label{sec:ssc}

Equations (\ref{eq:mp1}) and (\ref{eq:mp2}) do not completely specify the evolution of all degrees of freedom in the orbit of a spinning body; we must impose an additional constraint in order to close the system of equations.  This constraint is called the Spin Supplementary Condition (SSC), and can be regarded as fixing internal degrees of freedom associated with the extended structure of the small body.  In the non-relativistic limit, the center of mass can be identified as the natural place for the worldline to pass through the extended body.  However, the center of mass is observer dependent in relativistic dynamics.  The role of the SSC is thus to select one of the infinite choices of worldlines passing through the small body.  Since there is in general no natural choice for the worldline, the SSC is intrinsically arbitrary.  Excellent discussion of the physical meaning of the SSC can be found in Ref.\ \cite{Costa2014}; comparisons of different SSCs and investigation of their equivalence can be found in Refs.\ \cite{Lukes2014, Kyrian2007, Mikoczi2017, Lukes2017_2, Timogiannis2021}. 

An SSC commonly used in studies of gravitational wave sources is due to Tulczyjew \cite{Tulczyjew1959}, and is given by
\begin{equation}
p_{\alpha}S^{\alpha\beta}=0\;.\label{eq:TD}
\end{equation}
Using (\ref{eq:TD}), we find the relationship between the four-velocity and the four-momentum (\ref{eq:momvel}) is now given by
\begin{equation}
u^{\mu}=\frac{\mathcal{M}}{\mu^2}\left(p^{\mu}+\frac{2S^{\mu\nu}R_{\nu\rho\sigma\tau}p^{\rho}S^{\sigma\tau}}{4\mu^2+R_{\alpha\beta\gamma\delta}S^{\alpha\beta}S^{\gamma\delta}}\right)\;,
\end{equation}
where 
\begin{align}
\mu & \equiv\sqrt{-p_{\alpha}p^{\alpha}}\;,\\
\mathcal{M} & \equiv-p_{\alpha}u^{\alpha}\;. \label{eq:mathcalM}
\end{align}
These relationships tell us that $p^\alpha = \mu u^\alpha + \mathcal{O}(S^2)$, and $\mu = \mathcal{M} + \mathcal{O}(S^2)$, a result we will exploit shortly.

The spin tensor is antisymmetric, which facilitates defining the spin vector \cite{Kyrian2007}
\begin{equation}
S^{\mu}=-\frac{1}{2\mu}{\epsilon^{\mu\nu}}_{\alpha\beta}p_{\nu}S^{\alpha\beta},
\label{eq:spinvec}
\end{equation}
where
\begin{equation}
\epsilon_{\alpha\beta\gamma\delta}=\sqrt{-g}[\alpha\beta\gamma\delta]
\end{equation}
and where $\sqrt{-g}$ is the metric determinant, reducing to $\Sigma\sin\theta$ for Kerr, and $[\alpha\beta\gamma\delta]$ is the totally antisymmetric symbol.  By combining these results, one can show that the magnitude of the spin is another constant of the motion, given by
\begin{equation}
S^2=S^{\alpha}S_{\alpha}=\frac{1}{2}S_{\alpha\beta}S^{\alpha\beta}\;.\label{eq:smag}
\end{equation}

\subsection{Leading order in small body's spin}
\label{sec:leadingorder}

The magnitude $S$ of the small body's spin can be defined using a dimensionless spin parameter $s$:
\begin{equation}   
S=s\mu^2\;.
\end{equation}
If the small body is itself a Kerr black hole, then $0 \le s \le 1$, which tells us that $S \le \mu^2$.  Linear-in-spin effects are thus effectively quadratic in the system's mass ratio, affecting a system's dynamics at the same formal order as important self force effects \cite{Pound2015,Barack2019,Pound2021}.  The next order in spin scales with the fourth power of the system's mass ratio, practically negligible at extreme mass ratios.  A linear-in-spin analysis is thus formally interesting as well as of astrophysical relevance.  As such, we focus on the linear-in-spin limit, neglecting terms in all of our equations that are $\mathcal{O}(S^2)$ or higher.

In this limit, the Matthisson-Papapetrou equations (\ref{eq:mp1}) -- (\ref{eq:mp2}) and the Tulczyjew SSC (\ref{eq:TD}) take a particularly useful form.  Revisiting various relations in Secs.\ \ref{sec:scc} and \ref{sec:ssc} but dropping all terms beyond linear in $S$, the Tulczyjew SSC (\ref{eq:TD}) becomes
\begin{equation}
p^{\alpha}=\mu u^{\alpha}\;.
\label{eq:momvelfirstorder}
\end{equation}
The orbit's 4-velocity and 4-momentum are parallel at this order.  With this, the Mathisson-Papapetrou equations can be written
\begin{align}
\frac{Du^{\alpha}}{d\tau} & =-\frac{1}{2\mu}{R^\alpha}_{\,\nu\lambda\sigma}u^{\nu}S^{\lambda\sigma}\;,
\label{eq:mp1linear}\\
\frac{DS^{\alpha\beta}}{d\tau} &= 0\;.\label{eq:mp2linear}
\end{align}
The second of these equations tells us that the spin tensor is parallel transported along the worldline at this order.

Linearizing in $S$, Eq.\ (\ref{eq:spinvec}) becomes
\begin{equation}
S^{\mu}=-\frac{1}{2}{\epsilon^{\mu\nu}}_{\alpha\beta}\hat{u}_{\nu}S^{\alpha\beta}\;,\label{eq:spinveclinear}
\end{equation}
or equivalently,
\begin{equation}
S^{\alpha\beta}=\epsilon^{\alpha\beta\mu\nu}\hat{u}_{\mu}S_{\nu}\;.\label{eq:spintenslinear}
\end{equation}
Using these linear-in-spin forms, the SSC (\ref{eq:TD}) becomes
\begin{equation}
\hat{u}_{\alpha}S^{\alpha\beta}=0\;,\label{eq:TDlinear}
\end{equation}
or
\begin{equation}
\hat{u}_{\alpha}S^{\alpha}=0\;.\label{eq:TDlinear2}
\end{equation}
Equation (\ref{eq:TDlinear2}) helps us understand the meaning of the SSC, at least in a linear-in-spin analysis: it tells us that in a freely-falling frame that moves with the geodesic whose 4-velocity is $\hat u^\alpha$, the small body's spin is purely spatial.  Combining Eqs.\ (\ref{eq:mp2linear}) and (\ref{eq:spintenslinear}), we find
\begin{equation}
\frac{DS^{\mu}}{d\tau}=0\;,
\label{eq:mp2linear2}
\end{equation}
so the spin vector is also parallel transported along the worldline at this order.

\subsection{Parallel transport in Kerr}
\label{sec:ParallelTransport}

Since the small body's spin vector is parallel transported along its orbit, as described by Eq.\ (\ref{eq:mp2linear2}), let us examine such parallel transport in detail.  Past work \cite{Ruangsri2016} showed how to build a solution describing this transport using a frequency-domain expansion, demonstrating that an additional frequency emerges which characterizes the timescale associated with the spin's precession.  Van de Meent \cite{vandeMeent2019} has since then produced an elegant closed-form tetrad-based solution for describing the parallel transport of vectors along Kerr geodesics, following methods first developed Marck \cite{Marck1983, Marck1983_2, Kamran1986}; see also work by Bini and collaborators, which explores and clarifies the geometrical properties of Marck's procedure \cite{Bini2008,Bini2019,Bini2017}, as well as Mashoon and collaborators \cite{Mashhoon2006,Chicone2006}.  Following Ref.\ \cite{vandeMeent2019}, we summarize the procedure for constructing this tetrad and describe how to use it to describe a spinning body moving along its orbit.

We write the tetrad $\{e_{0\alpha}(\lambda), \tilde{e}_{1\alpha}(\lambda), \tilde{e}_{2\alpha}(\lambda), e_{3\alpha}(\lambda)\}$.  Take its first leg, $e_{0\alpha}(\lambda)$, to be the geodesic's 4-velocity; take its last leg, $e_{3\alpha}(\lambda)$, to be the (normalized) orbital angular momentum 4-vector defined in Eq.\ (\ref{eq:orbangmomdef}).  Our tetrad so far consists of the vectors
\begin{equation}
    e_{0\alpha}(\lambda) = \hat u_\alpha(\lambda)\;,\qquad e_{3\alpha}(\lambda) = \frac{1}{\sqrt{\hat K}}\hat{\mathcal{L}}_\alpha(\lambda)\; \label{eq:tetradleg03},
\end{equation}
where $\hat{\mathcal{L}}_\alpha(\lambda)$ is the orbital angular momentum 4-vector along the geodesic with 4-velocity $\hat u_\alpha(\lambda)$.  By the properties of $\hat u^\alpha(\lambda)$, $\hat{\mathcal{L}}^\alpha(\lambda)$, and $\hat K$, these tetrad legs are orthogonal to each other and parallel transported along $\hat u^\alpha(\lambda)$.  We then construct $\tilde{e}_{1\alpha}(\lambda)$ and $\tilde{e}_{2\alpha}(\lambda)$ by choosing two vectors which lie in the plane orthogonal to $e_{0\alpha}(\lambda)$ and $e_{3\alpha}(\lambda)$; see Ref.\ \cite{vandeMeent2019}, Eqs.\ (50) and (51), for explicit formulas.

The resulting tetrad is in general not parallel transported.  However, by defining
\begin{align}
    e_{1\alpha}(\lambda) &= \cos\psi_p(\lambda)\,\tilde{e}_{1\alpha}(\lambda) + \sin\psi_p(\lambda)\,\tilde{e}_{2\alpha}(\lambda)
    \label{eq:tetradleg1}\\
    e_{2\alpha}(\lambda) &= -\sin\psi_p(\lambda)\,\tilde{e}_{1\alpha}(\lambda) + \cos\psi_p(\lambda)\,\tilde{e}_{2\alpha}(\lambda)
    \label{eq:tetradleg2}
\end{align}
and requiring that the precession phase $\psi_p(\lambda)$ satisfies
\begin{equation}
\frac{d\psi_p}{d\lambda}=\sqrt{\hat{K}}\left(\frac{(r^2+a^2)\hat{E}-a\hat{L}_z}{\hat{K}+r^2}+a\frac{\hat{L}_z-a(1-z^2)\hat{E}}{\hat{K}-a^2z^2}\right)
\label{eq:precphaseeqn}
\end{equation}
we obtain a tetrad $\{e_{0\alpha}(\lambda), e_{1\alpha}(\lambda), e_{2\alpha}(\lambda), e_{3\alpha}(\lambda)\}$ that is parallel transported along the geodesic \cite{Marck1983, Marck1983_2, vandeMeent2019}.  Van de Meent further finds a closed form solution to Eq.\ (\ref{eq:precphaseeqn}) of the form
\begin{equation}
    \psi_p(\lambda) = \Upsilon_s\lambda + \psi_r(\hat\Upsilon_r\lambda) + \psi_\theta(\hat\Upsilon_\theta\lambda)\;,
    \label{eq:precphasesol}
\end{equation}
where $\Upsilon_s$ (denoted $\Upsilon_\psi$ in Ref.\ \cite{vandeMeent2019}) is the frequency (conjugate to Mino-time) describing the precession of this tetrad along the orbit; the functions $\psi_r(\hat\Upsilon_r\lambda)$ and $\psi_\theta(\hat\Upsilon_\theta\lambda)$ are phases associated with the orbit's radial and polar motions. We define the Mino-time precession period as $\Lambda_s = 2\pi/\Upsilon_s$.

This solution makes setting the spin of the small body easy: We write the small body's spin vector
\begin{equation}
    S_\alpha = S^0 e_{0\alpha}(\lambda) + S^1 e_{1\alpha}(\lambda) + S^2 e_{2\alpha}(\lambda) + S^3 e_{3\alpha}(\lambda)\;,
    \label{eq:spinvectetrad}
\end{equation}
where $\{S^0, S^1, S^2, S^3\}$ are all constants with the dimension of angular momentum.  The requirement that $\hat u^\alpha S_\alpha = 0$ means that $S^0 = 0$ for all configurations.  A component $S^3 \equiv S_{\parallel}$ denotes a component of the small body's spin parallel or antiparallel to the orbital angular momentum, normal to the orbital plane; $S^1$ and $S^2$ define components perpendicular to the orbital angular momentum, in the orbital plane.  A spin vector with $S^1 = S^2 = 0$ does not precess, and so its motion has no frequency components at harmonics of the spin-precession frequency $\Upsilon_s$.  By contrast, when $S^1$ or $S^2$ are non-zero, the small body's spin precesses over an orbit, and harmonics of the frequency $\Upsilon_s$ appear in a frequency-domain description of the small body's orbit.

Code for computing these tetrad legs is implemented as part of the {\tt KerrGeodesics} package in the Toolkit \cite{Kerrgeodesics}.

\subsection{Spin deviation from geodesic trajectory}
\label{sec:SpinDev}

As argued in Sec.\ \ref{sec:leadingorder}, our focus is on computing orbits to linear order in the small body's spin.  For the configurations that we study, the spin is a small parameter, and these trajectories can be regarded as perturbative deviations from bound Kerr geodesics.  We discuss the nature of an orbit's ``spin shift'' in detail later as we analyze specific orbit and spin configurations.  In general, the small body's trajectory can be written in the form
\begin{equation}
    x^\alpha(\lambda) = \hat x^\alpha(\lambda) + \delta x_S^\alpha(\lambda)\;,
    \label{eq:trajectoryshift}
\end{equation}
where $\hat x^\alpha(\lambda)$ is the coordinate-space trajectory of an appropriately chosen geodesic, and $\delta x_S^\alpha(\lambda)$ is the $\mathcal{O}(S)$ shift due to the spin.  Similarly, we write the small body's 4-velocity
\begin{equation}
u^{\alpha}=\hat{u}^{\alpha}+u_{S}^{\alpha}\;,
\label{eq:4vellinear}
\end{equation}
where $\hat u^\alpha$ solves the geodesic equation, and $u_S^\alpha=\mathcal{O}(S)$.

One important point to note is that $\hat{x}^\alpha(\lambda)$ will in general have different periods than $x^\alpha(\lambda)$: the periods $\Lambda_{r,\theta,\phi}$ which characterize bound orbits of spinning bodies differ from the geodesic periods $\hat\Lambda_{r,\theta,\phi}$ by $\mathcal{O}(S)$.  As such, a naive definition of $\delta x^\alpha_S$ necessarily contain unbounded, secularly growing terms.  Such terms ruin the perturbative expansion that we use.

As such, we do not use the explicit form Eq.\ (\ref{eq:trajectoryshift}) directly when we compute spinning-body orbits in Secs. \ref{sec:slightlyecc} and \ref{sec:spinbodyfreqdom}.  We instead characterize these orbits using amplitude-phase variables.  Doing so, the frequency shift is incorporated into the parameterization; see Eq.\ (\ref{eq:rparam}) or (\ref{eq:rparam2}) and nearby text.  Once we have solved for the frequency shift and phase variables, we can then compute $\delta x^\alpha_S$.  These quantities are particularly useful for finding the concomitant ``spin shifts'' to constants of motion, which we describe below.  In Appendix \ref{sec:secularterms}, we provide the explicit form of $\delta x^\alpha_S$ in terms of variables that we use in this work, as well as further discussion of the secular terms.

As the orbit evolves, we must preserve the norm of its 4-velocity.  Using Eq.\ (\ref{eq:4vellinear}), demanding that $\hat u^\alpha\hat u_\alpha = -1$, and enforcing $u^{\alpha}u_{\alpha}=-1$ yields the constraint 
\begin{equation}
\hat{u}^\alpha u^S_\alpha+\hat{u}_\alpha u_S^\alpha=0\;.\label{eq:udotu}
\end{equation}
Writing $u_\alpha=g_{\alpha \beta}u^\beta$, and noting that $g_{\alpha \beta}$ is evaluated along the spinning-body orbits for which $r=\hat{r}+\delta r_S$ and $\theta=\hat{\theta}+\delta\vartheta_S$, the spin-corrected covariant 4-velocity has the form
\begin{equation}
    u^S_\alpha=g_{\alpha\beta} u^\beta_S+\delta r_S \partial_r g_{\alpha\beta}\hat u^\beta +\delta \theta_S \partial_\theta g_{\alpha\beta}\hat u^\beta\;.
\end{equation}
This allows us to write constraint (\ref{eq:udotu}) entirely in terms of the contravariant spin-correction to the 4-velocity, viz.,
\begin{equation}
2g_{\alpha\beta}\hat u^\alpha u^\beta_S + \delta r_S \partial_r g_{\alpha\beta}\hat u^\alpha\hat u^\beta +\delta \theta_S \partial_\theta g_{\alpha\beta}\hat u^\alpha\hat u^\beta=0\;.\label{eq:udotucontra}
\end{equation}
We use this constraint throughout our analysis. We also define the leading order in spin corrections to the energy $\delta E^S$ and axial angular momentum $\delta L_z^S$ due to the spin using (\ref{eq:Espin}) and (\ref{eq:Lspin}):
\begin{align}
E^{S}&=\hat E + \delta E^S\;, \label{eq:deltaEspin}\\
L_{z}^{S}&=\hat{L}_z + \delta L_z^S\;.\label{eq:deltaLspin}
\end{align}

As mentioned in Sec.\ \ref{sec:scc}, an analogue to the Carter constant is preserved at linear order in spin.  Normalizing by the orbiting body's rest mass squared, it is given by \cite{Rudiger1981}
\begin{equation}
K^S=K_{\alpha\beta}u^\alpha u^\beta+\delta\mathcal{C}^S,
\label{eq:Kspin}
\end{equation}
where 
\begin{equation}
\delta\mathcal{C}^S= -\frac{2}{\mu}\hat u^{\mu}S^{\rho\sigma}\left( {\mathcal{F}^\nu}_{\sigma}\nabla_{\nu}\mathcal{F}_{\mu \rho } - {\mathcal{F}_\mu}^\nu\nabla_{\nu}\mathcal{F}_{\rho\sigma}\right)\;.
\label{eq:Cspin}
\end{equation}
We define the first order in spin correction to $K$ by
\begin{equation}
K^{S}=\hat{K}+\delta K^S\;, \label{eq:deltaKspin}\\
\end{equation}
where $\hat K$ is the Carter constant along the geodesic whose 4-velocity is $\hat u^\alpha$, and $\delta K^S$ is $\mathcal{O}(S)$.  Combining Eqs.\ (\ref{eq:trajectoryshift}), (\ref{eq:4vellinear}) and (\ref{eq:Kspin}) with the definition (\ref{eq:deltaKspin}) and truncating at linear order in $S$, we find
\begin{align}
    \delta K^S &= 2K_{\alpha\beta}\hat u^\alpha u^\beta_S + \delta r_S \partial_r K_{\alpha\beta}\hat u^\alpha\hat u^\beta + \delta \theta_S \partial_\theta K_{\alpha\beta}\hat u^\alpha\hat u^\beta
    \nonumber\\
    &  + \delta\mathcal{C}^S\;.
    \label{eq:deltaKspin2}
\end{align}
The first line of Eq.\ (\ref{eq:deltaKspin2}) includes two terms which are due to the shift of the small body's orbit that we find when examining spinning-body orbits. Applying Eq.\ (\ref{eq:Qdef}), we then find the first-order shift in $Q$:
\begin{equation}
    \delta Q^S = \delta K^S - 2(\hat L_z - a\hat E)(\delta L^S_z - a\delta E^S)\;.
    \label{eq:deltaQspin}
\end{equation}

For nearly equatorial orbits with polar motion defined by $\theta=\pi/2+\delta\vartheta_S$ in Eq.\ (\ref{eq:thetaparamfirst}), $\delta\vartheta_S$ and $\delta\theta_S$ may be used interchangeably (which we do throughout this paper). However, in general, $\delta\vartheta_S$ corresponds only to the corrections to the \textit{libration region} of the polar motion, while $\delta\theta_S$ denotes the entire spin-perturbation associated with $\theta$, as defined in Eq.\ (\ref{eq:trajectoryshift}). This distinction becomes important in our companion study \cite{Paper2},

\subsection{General characteristics of spinning-body orbits}
\label{sec:spinningbodyorbitoverview}

In the remainder of this paper, we examine several examples of the orbits of spinning bodies about Kerr black holes.  Before exploring these specific cases in detail, we briefly lay out and summarize general characteristics of the orbits that we find.

Consider first an orbit that would be equatorial if the orbiting body were non-spinning.  If this body's spin is normal to the equatorial plane (i.e., parallel or antiparallel to both the orbital angular momentum and the large black hole's spin), then its orbit is quite simple.  Just as in the geodesic case, we can use the parameterization $r = pM/(1 + e\cos\chi_r)$.  The radial turning points are fixed for the duration of the orbit at $pM/(1 \pm e)$, and the orbit's dynamics maps onto a true anomaly angle $\chi_r$.  This true anomaly differs from the true anomaly that describes geodesics, $\hat\chi_r$; details of this difference are presented in Sec.\ \ref{sec:slightlyecc}.  The orbit's radial frequency is shifted compared to the geodesic by an amount $\mathcal{O}(S)$; we write the radial frequency $\Upsilon_r = \hat\Upsilon_r + \Upsilon^S_r$.  This case is discussed in quantitative detail in Secs.\ \ref{sec:secondorderine} and \ref{sec:eqplanealign}, with the special case of circular equatorial orbits presented in Sec.\ \ref{subsec:circeqalign}.

Consider next such an orbit but with the spin misaligned with respect to the orbital plane.  This misalignment introduces $\mathcal{O}(S)$ oscillations centered about the equatorial plane: The polar motion acquires a correction $\delta \vartheta_S$ whose Fourier expansion is at harmonics of the spin frequency $\Upsilon_s$ and the radial frequency $\Upsilon_r = \hat{\Upsilon}_r + \Upsilon_r^S$.  The radial motion, however, remains exactly as it was in the spin-aligned case.  We discuss this case in detail in Secs.\ \ref{sec:leadingorderine} and \ref{sec:ecceqprecess}; an explicit analytic solution for circular, nearly equatorial motion is calculated in Sec.\ \ref{subsec:circeqmisalign}.

We focus on these equatorial and nearly equatorial cases in this paper. For orbits that are not ``nearly equatorial'', the parameterization becomes rather more complicated.  In particular, the ``geodesic-like'' parameterization of the nearly equatorial case must be modified, adding a spin-induced contribution to the orbit's libration region in both the radial and polar motions.  This holds even if the spin-vector is aligned with the orbital angular momentum.  We discuss these more complicated cases in a companion paper \cite{Paper2}.

\section{Spinning-body orbits I:\\ Circular, nearly equatorial orbits}
\label{sec:simpleorbits}

We begin our study of spinning-body orbits by examining several simple cases for which we can find closed-form, fully analytic solutions.  These cases allow us to introduce the main principles we use to describe and parameterize our solutions, and provide limiting examples which can be compared against other results in the literature.  We begin with the simplest possible orbit: a circular orbit of radius $r$, confined to the equatorial plane ($I = 0^\circ$ or $I = 180^\circ$).

Many of the results we find are derived in Ref.\ \cite{Tanaka1996}, which focuses on circular orbits of spinning bodies, as well as elsewhere in the literature.  The results we present in Sec.\ \ref{subsec:circeqalign} can also be obtained using the effective potential derived in Ref.\ \cite{Saijo1998} (see also Refs.\ \cite{1976Tod} and \cite{Hackmann2014}).  To facilitate the comparison to this literature, we discuss the method of Ref.\ \cite{Saijo1998} in detail in Appendix \ref{sec:Saijocomparison}.

\subsection{Aligned spin}
\label{subsec:circeqalign}

Start with the small body spin parallel or antiparallel to the orbit: we set the spin components $S^1 = S^2 = 0$, and set $S^3 =s_\parallel \mu^2$, with $-1 \le s_\parallel \le 1$.  The small body's spin is parallel to the orbit if $s_\parallel > 0$, and antiparallel if $s_\parallel < 0$. The geodesic integrals of motion are
\begin{align}
    \hat E &= \frac{1 - 2v^2 \pm qv^3}{\sqrt{1 - 3v^2 \pm 2qv^3}}\;,
    \label{eq:Ecirceq}\\
    \hat L_z &= \pm \sqrt{rM}\frac{1 \mp 2qv^3 + q^2v^4}{\sqrt{1 - 3v^2 \pm 2qv^3}}\;,
    \label{eq:Lzcirceq}\\
    \hat Q &= 0\;.
\end{align}
We have introduced $v = \sqrt{M/r}$ (equivalently $r = M/v^2$) and $q = a/M$.  Where there is a choice, the upper sign is for prograde orbits ($I = 0^\circ$) and the lower is for retrograde ($I = 180^\circ$).  The small body's background 4-velocity is given by $\hat u_\alpha = (-\hat E, 0, 0, \hat L_z)$.

The small body's spin 4-vector is given by
\begin{equation}
    S_\alpha =s_\parallel\mu^2 e_{3\alpha} = (0, 0, \mp r s_\parallel \mu^2, 0)\;.
\end{equation}
This result comes from the fact that for an equatorial circular orbit \cite{vandeMeent2019},
\begin{align}
    e_{3\alpha} &= \left(0,0,-r\frac{(\hat L_z - a\hat E)}{|\hat L_z - a\hat E|},0\right)
    \nonumber\\
    &= \left(0,0,\mp r,0\right)\;.
\end{align}
If the orbit is prograde and $s_\parallel > 0$, or the orbit is retrograde and $s_\parallel < 0$, then the small body's spin points in the direction of decreasing $\theta$; vice versa if {$s_\parallel$} and the orbit have the opposite signs and orientations.

Let us examine (\ref{eq:mp1linear}) for this case.  Using Eq.\ (\ref{eq:4vellinear}), we start by expanding the covariant derivative:
\begin{align}
    \frac{Du^\alpha}{d\tau} &= (\hat u^\beta + u^\beta_S)\nabla_\beta\left(\hat u^\alpha + u^\alpha_S\right)
    \nonumber\\
    &= \frac{d\hat u^\alpha}{d\tau} + \frac{du^\alpha_S}{d\tau} + {\Gamma^\alpha}_{\beta\gamma}\hat u^\beta\hat u^\gamma + 2{\Gamma^\alpha}_{\beta\gamma}\hat u^\beta u^\gamma_S + \mathcal{O}(S^2)
    \nonumber\\
    &= \frac{du^\alpha_S}{d\tau} + 2{\Gamma^\alpha}_{\beta\gamma}\hat u^\beta u^\gamma_S\;.
    \label{eq:expand4velderiv}
\end{align}
Here, ${\Gamma^\alpha}_{\beta\gamma}$ is the Christoffel connection for the Kerr geometry evaluated along the orbit.  In going from the second line to the third line, we used the fact that $\hat u^\alpha$ solves the geodesic equation, and we linearized in $S$.  We also used the fact that, for this orbit, the spinning body remains confined to the equatorial plane $\theta = \pi/2$ at radius $r$.  For the misaligned case we consider next, the orbit oscillates in the polar direction, and there is a correction term that involves $\partial_\theta{\Gamma^\alpha}_{\beta\gamma}$.

Requiring the spinning body's orbit to be circular and equatorial means that
\begin{equation}
    u^r_S = u^\theta_S = 0\;.
\end{equation}
Further, the requirement that $u^\alpha_S\hat u_\alpha = 0$ tells us that
\begin{equation}
    u^t_S = \frac{\hat L_z}{\hat E}u^\phi_S\;.
    \label{eq:utScirceq}
\end{equation}
The only unique component we must determine is thus $u^\phi_S$.  Note that we must have $du^\phi_S/d\tau = 0$.  If we observe the system in a frame that co-rotates with the orbit, it appears static; the symmetries of the spin-curvature coupling in this case do not introduce any dynamics.

Combining Eqs.\ (\ref{eq:mp2linear2}) and (\ref{eq:expand4velderiv}) with $u^r_S = u^\theta_S = 0 = du^\phi_S/d\tau$, we find the equation which governs the spin correction to the small body's orbital velocity is given by
\begin{equation}
    2{\Gamma^r}_{\beta\gamma}\hat u^\beta u^\gamma_S = -\frac{1}{2\mu}{R^r}_{\nu\lambda\sigma}\hat u^\nu S^{\lambda\sigma}\;;
    \label{eq:mp2circeqparallel}
\end{equation}
all other components of this equation vanish.  Expanding the right-hand and left-hand sides of (\ref{eq:mp2circeqparallel}), we find
\begin{widetext}
\begin{align}
    2{\Gamma^r}_{\beta\gamma}\hat u^\beta u^\gamma_S &= \mp\frac{2v\sqrt{1 - 3v^2 \pm 2qv^3}(1 - 2v^2 + q^2v^4)u^\phi_S}{1 - 2v^2 \pm qv^3}\;,
    \\
    -\frac{1}{2\mu}{R^r}_{\nu\lambda\sigma}\hat u^\nu S^{\lambda\sigma} &= \frac{3s_\parallel\mu}{M^2}\frac{v^7(1 \mp q v)(1 - 2v^2 + q^2v^4)}{1 - 3v^2 \pm 2qv^3}\;.
\end{align}
\end{widetext}
Using this to evaluate Eq.\ (\ref{eq:mp2circeqparallel}) yields
\begin{equation}
u^\phi_S = \mp\frac{3s_\parallel\mu}{2M^2}\frac{v^6(1 \mp q v)(1 - 2v^2 \pm q v^3)}{(1 - 3v^2 \pm 2qv^3)^{3/2}}\;.
\label{eq:uphiScirceqalign}
\end{equation}
Using Eq.\ (\ref{eq:utScirceq}), this in turn yields a simple result for $u^t_S$.

An observationally important aspect of this solution is its influence on the system's orbital frequency.  Using
\begin{equation}
\Omega_\phi = \frac{u^\phi}{u^t} = \frac{\hat u^\phi + u^\phi_S}{\hat u^t + u^t_S}\;,\label{eq:Omegaphicirc}
\end{equation}
expanding in $S$, using $\hat\Omega_\phi = \hat u^\phi/\hat u^t$, and finally defining $\Omega_\phi = \hat\Omega_\phi + \delta\Omega_\phi$, we find the correction to the frequency due to the spin-curvature force:
\begin{equation}
\delta\Omega_\phi = \hat\Omega_\phi\left(\frac{u^\phi_S}{\hat u^\phi} - \frac{u^t_S}{\hat u^t_S}\right)\;.
\label{eq:deltaOmPhicirceq}
\end{equation}
For circular and equatorial orbits,
\begin{equation}
    \hat\Omega_\phi = \pm\frac{v^3}{M(1 \pm q v^3)}\;.
    \label{eq:OmPhicirceq}
\end{equation}
Combining these various results, we find the shift to the axial frequency:
\begin{equation}
    \delta\Omega_\phi = \mp\frac{3s_\parallel}{2M}\frac{\mu}{M}\frac{(1 \mp qv)}{(1 \pm q v^3)^2}v^6\;.
    \label{eq:deltaOmPhicirceqalign}
\end{equation}
This agrees exactly with Eq.\ (4.26) in Ref.\ \cite{Tanaka1996}.

The orbiting body's energy, axial angular momentum, and Carter constant are also shifted.  Combining Eqs.\ (\ref{eq:Espin}), (\ref{eq:Lspin}), (\ref{eq:deltaEspin}), and (\ref{eq:deltaLspin}) with the results in this section and using Eqs.\ (\ref{eq:deltaKspin2}) and (\ref{eq:deltaQspin}), we find
\begin{widetext}
\begin{align}
    \delta E^S &= -\frac{s_\parallel}{2}\frac{\mu}{M}\frac{(1 \mp qv)(1 \mp 4q^3 + 3q^2v^4)}{(1 - 3v^2 \pm 2qv^3)^{3/2}}v^5\;,
    \label{eq:deltaEScirceqaligned}\\
    \delta L^S_z &= \pm\frac{s_\parallel\mu}{2}\frac{(2 - 13v^2 + 18v^4) \pm 3q(3 - 7v^2)v^3 + 2q^2(1 + 2v^2)v^6 \pm q^3(3 - 7v^2)v^7 + 3q^4v^{10}}{(1 - 3v^2 \pm 2qv^3)^{3/2}}\;,
    \label{eq:deltaLzScirceqaligned}\\
    \delta K^S &= s_\parallel\mu \frac{(2 - 13v^2 + 18v^4) \mp 2qv(2 - 17v^2 + 28v^4) - q^2v^4(17 - 45v^2) \mp 6q^3v^7 - 3q^4v^8}{v(1 - 3v^2 \pm 2qv^3)^2}\;,
    \label{eq:deltaKScirceqaligned}\\
    \delta Q^S &= \mp 2 s_\parallel\mu a\;.
    \label{eq:deltaQScirceqaligned}
\end{align}
\end{widetext}
These expressions for the conserved quantities $\delta E^S$ and $\delta L_z^S$ match exactly with Eqs.\ (\ref{eq:EscircSchw}) and (\ref{eq:LSscircSchw}) derived using the alternative approach outlined in Appendix \ref{sec:Saijocomparison}.  It is interesting that there is a non-zero $\delta Q^S$ even though there is no change to the polar motion of the small body in this case.  We note that Witzany has provided a modified definition of $\delta Q^S$ (see text near Eq.\ (48) of Ref.\ \cite{Witzany2019_2}) such that it is zero for cases in which there is no polar motion; we are likely to adopt this definition in future work.
In any case, our result agrees with that reported in Ref.\ \cite{Tanaka1996}, after translating the somewhat different notation.


\subsection{Misaligned spin}
\label{subsec:circeqmisalign}

Now consider the small body's spin misaligned from the orbit.  The background 4-velocity and integrals of motion are identical to those used in Sec.\ \ref{subsec:circeqalign}, but the small body's spin becomes
\begin{equation}
    S_\alpha = \mu^2\bigl(s_\perp\cos\phi_s\,e_{1\alpha} + s_\perp\sin\phi_s\,e_{2\alpha} + s_\parallel\,e_{3\alpha}\bigr)\;.
    \label{eq:Smisalign1}
\end{equation}
We have broken the spin into a component parallel to the orbit (out of the orbital plane) with magnitude $s_\parallel$, and into components normal to the orbit (in the orbital plane) with magnitude $s_\perp$. The angle $\phi_s$ describes the orientation of the spin components normal to the orbit.  Setting $s = \sqrt{s_\perp^2 + s_\parallel^2}$, we require $0 \le s \le 1$.

Using (\ref{eq:tetradleg1}) and (\ref{eq:tetradleg2}), Eq.\ (\ref{eq:Smisalign1}) can be rewritten
\begin{align}
    S_\alpha &= \mu^2\biggl[s_\perp\Bigl(\cos(\phi_s + \psi_p)\tilde{e}_{1\alpha} + \sin(\phi_s + \psi_p)\tilde{e}_{2\alpha}\Bigr)
    \nonumber\\
    &\qquad + s_\parallel e_{3\alpha}\biggr]\;,
    \label{eq:Smisalign2}
\end{align}
where $\psi_p$ is the precession phase, which grows with time.  The tetrad leg $e_{3\alpha}$ is the same as in Sec.\ \ref{subsec:circeqalign}.  Continuing to use the parameterization $q \equiv a/M$, $v = \sqrt{M/r}$, the tetrad legs $\tilde{e}_{1\alpha}$ and $\tilde{e}_{2\alpha}$ are given by
\begin{widetext}
\begin{align}
    \tilde{e}_{1\alpha} &= \left(0, \frac{1}{\sqrt{1 - 2v^2 + a^2v^4}}, 0, 0\right)\;,
    \label{eq:tildee1_circeq}    \\
    \tilde{e}_{2\alpha} &= \left(v\sqrt{\frac{1 - 2v^2 + q^2v^4}{1 - 3v^2 \pm 2qv^3}}, 0, 0,\mp r(1 \pm qv^3)\sqrt{\frac{1 - 2v^2 + q^2v^4}{1 - 3v^2 \pm 2qv^3}}\right)\;.
    \label{eq:tildee2_circeq}
\end{align}
\end{widetext}
For circular and equatorial orbits, the precession phase $\psi_p$ can be written as functions of Mino-time $\lambda$, proper time $\tau$, or Boyer-Lindquist time $t$:
\begin{equation}
    \psi_p = \Upsilon_s\lambda  = \omega_s\tau = \Omega_s t\;,
\end{equation}
with
\begin{align}
    \Upsilon_s = \sqrt{rM} &= M/v \;,\qquad\omega_s = \sqrt{M/r^3} = v^3/M\;,
    \nonumber\\
    \Omega_s &= \omega_s\frac{\sqrt{1 - 3v^2 \pm 2 qv^3}}{(1 \pm q v^3)}\;.
\end{align}
This limiting form for $\Upsilon_s$ was found in Ref.\ \cite{Ruangsri2016}, and is confirmed by the general expression derived in Ref.\ \cite{vandeMeent2019}.  The factor $\Sigma$ which converts from Mino-time frequencies to proper-time frequencies takes the constant value $r^2$ for circular and equatorial orbits; likewise, the factor
\begin{equation}
    \Gamma = r^2\frac{1 \pm q v^3}{\sqrt{1 - 3v^2 \pm 2 qv^3}}
\end{equation}
which converts between Mino-time frequencies and coordinate-time quantities is constant for circular and equatorial orbits.

To proceed, we again examine Eq.\ (\ref{eq:mp1linear}) and use Eq.\ (\ref{eq:thetaparamfirst}), i.e., $\theta=\pi/2+\delta\vartheta_S$, with $\delta\vartheta_S=\mathcal{O}(S)$.  Expanding the covariant derivative yields a slightly different result as compared to what we found in the aligned case:
\begin{align}
    \frac{Du^\alpha}{d\tau} &= (\hat u^\beta + u^\beta_S)\nabla_\beta\left(\hat u^\alpha + u^\alpha_S\right)
    \nonumber\\
    &= \frac{d\hat u^\alpha}{d\tau} + \frac{du^\alpha_S}{d\tau} + {\Gamma^\alpha}_{\beta\gamma}\hat u^\beta\hat u^\gamma + \delta\vartheta_S\partial_\theta{\Gamma^\alpha}_{\beta\gamma}\hat u^\beta \hat u^\gamma
    \nonumber\\
    &\qquad + 2{\Gamma^\alpha}_{\beta\gamma}\hat u^\beta u^\gamma_S + \mathcal{O}(S^2)
    \nonumber\\
    &= \frac{du^\alpha_S}{d\tau} + \delta\vartheta_S\partial_\theta{\Gamma^\alpha}_{\beta\gamma}\hat u^\beta \hat u^\gamma + 2{\Gamma^\alpha}_{\beta\gamma}\hat u^\beta u^\gamma_S\;.
    \label{eq:expand4velderiv_thetaoscillate}
\end{align}
The misaligned spin causes the small body to oscillate about the equatorial plane by $\delta\vartheta_S$.  This shifts the connection term at $\mathcal{O}(S)$, leading to the term in $\partial_\theta{\Gamma^\alpha}_{\beta\gamma}$.

Expanding the covariant derivatives and Riemann components of Eq.\ (\ref{eq:mp1linear}) for this case, making use of Eq.\ (\ref{eq:expand4velderiv_thetaoscillate}) we find
\begin{widetext}
\begin{align}
    \frac{du^r_S}{d\tau} &\pm \frac{2v(1 - 2v^2 + q^2v^4)\sqrt{1 - 3v^2 \pm 2qv^3}}{1 - 2v^2 \pm qv^3}u^\phi_S = \frac{3s_\parallel\mu}{M^2}\frac{v^7(1\mp qv)(1 - 2v^2  + q^2v^4)}{1 - 3v^2 \pm 2qv^3}\;,
    \label{eq:circeqprec_rcomp}\\
    \frac{du^\phi_S}{d\tau} &+ \frac{2v^5(1 - 2v^2 \pm qv^3)}{M^2(1 - 2v^2 + q^2v^4)\sqrt{1 - 3v^2 \pm 2qv^3}}u^r_S = 0\;,
    \label{eq:circeqprec_phicomp}
\end{align}
\end{widetext}
for the equations governing $u^r_S$ and $u^\phi_S$.  Notice that these equations do not couple to the precessing orbit's polar motion.  Notice also that since $\hat u^r = \hat u^\theta = 0$, Eq.\ (\ref{eq:utScirceq}) holds for the misaligned case, and we do not need a separate equation governing $u^t_S$.

We require the orbit to remain circular, so we put $u^r_S = 0 = du^r_S/d\tau$.  This allows us to immediately solve Eq.\ (\ref{eq:circeqprec_rcomp}):
\begin{equation}
    u^\phi_S = \mp\frac{3s_\parallel\mu}{2M^2}\frac{v^6(1 \mp q v)(1 - 2v^2 \pm q v^3)}{(1 - 3v^2 \pm 2qv^3)^{3/2}}\;.
    \label{eq:uphiScirceqmisalign}
\end{equation}
Since this does not vary with time, Eq.\ (\ref{eq:circeqprec_phicomp}) is also satisfied.  Equation (\ref{eq:uphiScirceqmisalign}) is identical to the result we found in the spin-aligned case, Eq.\ (\ref{eq:uphiScirceqalign}).  Our solution for $u^t_S$ is likewise identical to its aligned counterpart.  From this it follows that Eq.\ (\ref{eq:deltaOmPhicirceqalign}) describes the change to the orbital frequency in this case as well.

The polar motion for this misaligned case requires more attention.  As stated above, we put $\theta = \pi/2 + \delta\vartheta_S$, where $\delta\vartheta_S$ denotes the spin-induced polar motion about the equatorial plane. Because $\hat u^\theta = 0$, we put $u^\theta = u^\theta_S = d\delta\vartheta_S/d\tau$.  The polar component of Eq.\ (\ref{eq:mp1linear}) thus becomes
\begin{widetext}
\begin{equation}
    \frac{d^2\delta\vartheta_S}{d\tau^2} + \frac{v^6}{M^2}\frac{(1 \mp 4qv^3 + 3 q^2v^4)}{(1 - 3v^2 \pm 2qv^3)}\delta\vartheta_S = -\frac{3s_\perp\mu}{M^3}\frac{v^9(1 \mp qv)\sqrt{1 - 2v^2 + q^2v^4}}{1 - 3v^2 \pm 2qv^3}\cos(\phi_s + \psi_p)\;.
    \label{eq:circeqprec_thetacomp}
\end{equation}
The coefficient of $\delta\vartheta_S$ on the left-hand side of Eq.\ (\ref{eq:circeqprec_thetacomp}) is the square of the polar proper-time frequency for circular equatorial geodesic orbits, which we denote $\omega_\theta$. The solution to Eq.\ (\ref{eq:circeqprec_thetacomp}) has the form 
\begin{align}
\delta\vartheta_S=A(\tau)\sin(\omega_\theta \tau)+B(\tau)\cos(\omega_\theta \tau)\;,
\end{align}
where
\begin{align}
\omega_\theta = \frac{v^3}{M}\sqrt{\frac{1 \mp 4qv^3 + 3q^2v^4}{1-3v^2\pm2qv^3}}\;,
\label{eq:omegatheta}
\end{align}
and where $A(\tau)$ and $B(\tau)$ are given by
\begin{align}
A(\tau) = c_1 - \frac{3s_\perp\mu}{2M^2}\frac{v^{6}(1 \mp qv)}{\sqrt{1 - 3v^2 \pm 2qv^3}}\sqrt{\frac{1 - 2v^2 + q^2v^4}{1 \mp 4qv^3 + 3q^2v^4}}\left[\frac{\sin(\phi_s + (\omega_s - \omega_\theta)\tau)}{\omega_s - \omega_\theta} + \frac{\sin(\phi_s + (\omega_s + \omega_\theta)\tau)}{{\omega_s + \omega_\theta}}\right]\;,
\label{eq:Atau_circ}\\
B(\tau) = c_2 + \frac{3s_\perp\mu}{2M^2}\frac{v^{6}(1 \mp qv)}{\sqrt{1 - 3v^2 \pm 2qv^3}}\sqrt{\frac{1 - 2v^2 + q^2v^4}{1 \mp 4qv^3 + 3q^2v^4}}\left[\frac{\cos(\phi_s + (\omega_s - \omega_\theta)\tau)}{\omega_s - \omega_\theta} - \frac{\cos(\phi_s + (\omega_s + \omega_\theta)\tau)}{{\omega_s + \omega_\theta}}\right]\;.
\label{eq:Btau_circ}
\end{align}
\end{widetext}
The constants $c_1$ and $c_2$ must be determined by matching to the initial conditions $\delta\vartheta_S|_{\tau=0}$ and $u^\theta_S|_{\tau=0}$. The precession of the small body's spin as it orbits the black hole causes the orbital plane to likewise precess.  Note that the frequency combination $\omega_s - \omega_\theta$ never passes through zero anywhere over the domain of allowed orbits.  As such, the functions $A(\tau)$ and $B(\tau)$ defined in Eqs.\ (\ref{eq:Atau_circ}) and (\ref{eq:Btau_circ}) are well behaved everywhere.


The changes to the integrals of motion we find are identical to those in the aligned case, Eqs.\ (\ref{eq:deltaEScirceqaligned}) -- (\ref{eq:deltaQScirceqaligned}).  The fact that the changes $\delta E^S$ and $\delta L^S_z$ are identical is consistent with other patterns that this analysis uncovered.  However, the fact that $\delta Q^S$ is identical --- in particular, that $\delta Q^S$ is insensitive to $s_\perp$ --- is somewhat surprising, since the small body does in fact move in the polar direction when the spin and orbit are misaligned.  The precession of the smaller body's spin nonetheless keeps the orbit equatorial on average, which appears to be sufficient for $Q$ to take its equatorial value.  This again is consistent with results found in Ref.\ \cite{Tanaka1996}.

\section{Spinning-body orbits II:\\ Slightly eccentric, nearly equatorial orbits}
\label{sec:slightlyecc}

Slightly eccentric equatorial orbits are simple enough that, by expanding in both eccentricity $e$ and spin $s$, we can develop and present mostly closed-form results for this case.  In our discussion below, we show leading-order results, $\mathcal{O}(e,s)$, for orbits of bodies with general spin orientation in the Kerr spacetime.  We go to higher order, $\mathcal{O}(e^2,s)$ for Schwarzschild only, confining ourselves to the case of small body spin aligned with the orbit.  Though no issue of principle prevents us from developing a more generic analysis at higher order, the formulas describing Kerr orbits become cumbersome as we go to higher order in $e$.  As we will see below, our leading-order analysis is sufficient for us to understand the impact of misaligned spin on spinning-body orbital dynamics.

The results in the aligned spin section, Sec.\ \ref{sec:secondorderine}, can be obtained using an alternative method we describe in Appendix \ref{sec:Saijocomparison}. This method is discussed in Refs.\ \cite{1976Tod, Saijo1998, Hackmann2014}, and involves using conserved quantities $E^S$, $L_z^S$, $\mu^2$ and $S^2$ to develop an effective potential for the radial motion.

\subsection{General principles}
\label{sec:genprinciples}

In this section and in what follows, we switch from using proper time $\tau$ to Mino time $\lambda$ for our parameterization of these orbits.  This switch is not necessary for equatorial or nearly equatorial orbits, but will be necessary for the generic cases that we study in a companion paper.  Using this parameterization now allows us to set up the calculation in this framework, and to examine the form of the solutions which emerge in this relatively simple limit.

The governing equation for the orbits is Eq.\ (\ref{eq:mp1linear}), which we repeat here and use to define the spin-curvature force $f^\alpha_S$:
\begin{equation}
    \frac{Du^\alpha}{d\tau} = -\frac{1}{2\mu}{R^\alpha}_{\nu\lambda\sigma}u^\nu S^{\lambda\sigma} \equiv f^\alpha_S/\mu\;.
\end{equation}
Expanding the covariant derivative, this becomes
\begin{equation}
    \frac{du^\alpha}{d\tau} + {\Gamma^\alpha}_{\beta\gamma} u^\beta u^\gamma = f^\alpha_S/\mu\;,
    \label{eq:forcedgeodesic1}
\end{equation}
where ${\Gamma^\alpha}_{\beta\gamma}$ is the Christoffel connection for the Kerr spacetime, evaluated along the orbit.

Let us define
\begin{equation}
    U^\alpha \equiv \frac{dx^\alpha}{d\lambda} = \Sigma u^\alpha\;;
    \label{eq:Udef}
\end{equation}
this follows from $u^\alpha = dx^\alpha/d\tau$, as well as the definition of Mino-time: $d/d\lambda = \Sigma d/d\tau$.  From (\ref{eq:Udef}), it follows that
\begin{equation}
    \frac{du^\alpha}{d\lambda} = \frac{1}{\Sigma}\frac{dU^\alpha}{d\lambda} - \frac{U^\alpha}{\Sigma^2}\frac{d\Sigma}{d\lambda}\;.
    \label{eq:dudlambda}
\end{equation}

Next multiply (\ref{eq:forcedgeodesic1}) by $\Sigma^2$.  Doing so and using Eq.\ (\ref{eq:dudlambda}), we put the equation which governs spinning-body orbits into the form
\begin{equation}
    \frac{dU^\alpha}{d\lambda} - \frac{U^\alpha}{\Sigma}\frac{d\Sigma}{d\lambda} + {\Gamma^\alpha}_{\beta\gamma}U^\beta U^\gamma = \Sigma^2 f^\alpha/\mu\;.
    \label{eq:forcedgeodesic2}
\end{equation}

Note that in general,
\begin{equation}
    \frac{d\Sigma}{d\lambda} = 2r\,U^r - 2a^2\cos\theta\sin\theta\,U^\theta\;.
    \label{eq:dSigmadlambda}
\end{equation}
For the equatorial and nearly equatorial orbits which are our focus in this section, the second term in (\ref{eq:dSigmadlambda}) is $\mathcal{O}(S^2)$, which we neglect.  The factor $(1/\Sigma)d\Sigma/d\lambda$ in Eq.\ (\ref{eq:forcedgeodesic2}) becomes $2U^r/r$.

For misaligned orbits, the orbiting body oscillates about the equatorial plane, just as we discussed for the circular misaligned case in Sec.\ \ref{subsec:circeqmisalign}.  Setting the polar angle to $\theta = \pi/2 + \delta\vartheta_S$, with $\delta\vartheta_S = \mathcal{O}(S)$, the connection term in Eq.\ (\ref{eq:forcedgeodesic2}) becomes
\begin{align}
    {\Gamma^\alpha}_{\beta\gamma}U^\beta U^\gamma &= \left({\Gamma^\alpha}_{\beta\gamma}\right)_{\theta = \pi/2}U^\beta U^\gamma \nonumber\\
    &+ \delta\vartheta_S\left(\partial_\theta{\Gamma^\alpha}_{\beta\gamma}\right)_{\theta = \pi/2}\hat U^\beta \hat U^\gamma\;.
\end{align}
Notice that it is the geodesic 4-velocity $\hat U^\beta$ that appears in the term with the derivative of the connection.  Because $\delta\vartheta_S$ is itself $\mathcal{O}(S)$, contributions from the non-geodesic parts of $U^\beta$ enter this term at $\mathcal{O}(S^2)$ or higher.

Let us write the small body's spin in the form
\begin{align}
    S_\alpha &= \mu^2\biggl[s_\perp\Bigl(\cos(\phi_s + \psi_p)\tilde{e}_{1\alpha} + \sin(\phi_s + \psi_p)\tilde{e}_{2\alpha}\Bigr)
    \nonumber\\
    &\qquad + s_\parallel e_{3\alpha}\biggr]\;,
    \label{eq:Smisalign_form1}\\
    &= \left(s_\perp \mu^2 \sigma_t, s_\perp \mu^2 \sigma_r, \mp s_\parallel\mu^2r, s_\perp\mu^2 \sigma_\phi\right)\;.
    \label{eq:Smisalign_form2}
\end{align}
Both the precession phase $\psi_p$ and the tetrad elements $\tilde{e}_{1\alpha}$ and $\tilde{e}_{2\alpha}$ are more complicated than they were in the circular limit; we defer discussion of their detailed forms until they are needed later in our analysis.  The form (\ref{eq:Smisalign_form2}) is a useful rewriting of (\ref{eq:Smisalign_form1}); the components $\sigma_{t,r,\phi}$ can be read out of $\tilde{e}_{1\alpha}$ and $\tilde{e}_{2\alpha}$.

With everything in place, it is now not difficult to evaluate all the terms appearing in Eq.\ (\ref{eq:forcedgeodesic2}) and write out the equations governing the small body's 4-velocity $U^\alpha$. First, we write out the equations for $U^r$, $U^t$ and $U^\phi$.
\begin{widetext}
\begin{align}
    \frac{dU^t}{d\lambda} &- \frac{2U^r\left[\left(r^3 - 3Mr^2 +a^2(r-M)\right)U^r + aM(3r^2 + a^2)U^\phi\right]}{r^2\Delta} = \frac{3s_\parallel\mu(\hat L_z - a\hat E)M(r^2 + a^2)\hat U^r}{r^2\Delta}\;,
    \label{eq:forcet_eqeccgen}\\
    \frac{dU^r}{d\lambda} &+ \frac{\Delta\left[M(U^t - a U^\phi)^2 - r^3(U^\phi)^2\right]}{r^4} - \frac{(2r^2 - 3Mr - a^2)(U^r)^2}{r\Delta} = \frac{3s_\parallel\mu(\hat L_z - a\hat E)M\left[\hat E(r^2 + a^2) - a\hat L_z\right]}{r^2}\;,
    \label{eq:forcer_eqeccgen}\\
    \frac{dU^\phi}{d\lambda} &+ \frac{2U^r\left[aMU^r + (r^3 - 2Mr^2 - a^2M)U^\phi\right]}{r^2\Delta} = \frac{3as_\parallel\mu(\hat L_z - a\hat E)M\hat U^r}{r^2\Delta}\;.
    \label{eq:forcephi_eqeccgen}
\end{align}
No term involving $\delta\vartheta_S$ enters these equations at $\mathcal{O}(S)$.  Indeed, note that the equations for $U^t$, $U^r$, and $U^\phi$ are completely independent of $U^\theta$ at this order.  We can therefore solve $U^{t,r,\phi}$ independently from our solution for $U^\theta$.

It is worth remarking that Eqs.\ (\ref{eq:forcet_eqeccgen}) and (\ref{eq:forcephi_eqeccgen}) turn out to simplify further by converting them to equations for $u_t$ and $u_\phi$.  Doing so using by converting from $U^{t,\phi}$ to $u^{t,\phi}$, lowering an index, and then using $u_t = -\hat{E} + u_t^S$, $u_\phi = \hat{L}_z + u_\phi^S$, where $u^S_{t,\phi} = \mathcal{O}(S)$, we find
\begin{align}
\frac{du^S_t}{d\lambda} &= -\frac{3s_{\parallel}\mu (\hat L_z - a\hat{E})M\hat{U}^r}{r^4}\;,
\label{eq:forcet_eqeccgen_v2}
\\
\frac{du^S_\phi}{d\lambda} &= \frac{3as_{\parallel}\mu (\hat L_z - a\hat{E})M\hat{U}^r}{r^4}\;.
\label{eq:forcephi_eqeccgen_v2}
\end{align}
Solving Eqs.\ (\ref{eq:forcet_eqeccgen_v2}) and (\ref{eq:forcephi_eqeccgen_v2}) is equivalent to solving (\ref{eq:forcet_eqeccgen}) and (\ref{eq:forcephi_eqeccgen}), respectively.

Finally, the equation we find for $U^\theta$ is
\begin{align}
    &\frac{dU^\theta}{d\lambda}+\frac{2a^4 r \hat E^2 - 4 a^3r \hat E \hat L_z + (r - 2M) r^3\hat L_z^2 + a^2(2r^3 \hat E^2 + 2r \hat L_z^2 - (\hat U^r)^2)}{r^2\Delta}\delta\vartheta_S \nonumber\\&= -\frac{3s_\perp\mu(\hat L_z - a\hat E)M}{r^3\Delta}\left(\sigma_t(r^2+a^2)\hat U^r + \sigma_r\left[\hat E(r^2 + a^2) - a\hat L_z\right]\Delta + \sigma_\phi a \hat U^r\right)\;,
    \label{eq:forcetheta_eqeccgen}
\end{align}
\end{widetext}
Notice that $dU^\theta/d\lambda$ only couples to $s_\perp$, and $dU^{t,r,\phi}/d\lambda$ only couple to $s_\parallel$.  Notice further that we have not yet introduced an expansion in eccentricity.  This means that for {\it all} nearly equatorial orbits, the small body's motion in the equatorial plane is totally decoupled from its out-of-plane dynamics.

For equatorial and nearly equatorial orbits, we take the small body to move on a trajectory whose radial motion is given by
\begin{equation}
    r = \frac{pM}{1 + e\cos\chi_r}\;.
    \label{eq:rparam}
\end{equation}
We introduce here the orbit's the semi-latus rectum $p$ and eccentricity $e$, as well as the radial true anomaly $\chi_r$.  This anomaly can be written
\begin{equation}
    \chi_r = w_r + \delta\chi_r\;,\label{eq:chir}
\end{equation}
where $w_r$ is the radial {\it mean anomaly}.  The difference between the radial mean and true anomalies, $\delta\chi_r$, is an oscillatory function whose mean value is zero.  In the Mino-time parameterization, $w_r = \Upsilon_r\lambda$.

As discussed in Sec.\ \ref{sec:kerrgeodesics}, the parameterization (\ref{eq:rparam}) is used extensively in studies of geodesic motion.  As we will show, it works perfectly for nearly equatorial orbits of spinning bodies as well.  This form does not work so well for generic orbits of spinning bodies; for general orbit inclination, we need to allow the radial libration region to oscillate as the orbit precesses. This case is discussed in the companion analysis, Ref.\ \cite{Paper2}.

We now solve for the orbit by introducing simultaneous expansions in the small body's spin and the orbit's eccentricity $e$.  By requiring that Eqs.\ (\ref{eq:forcet_eqeccgen}) -- (\ref{eq:forcephi_eqeccgen}) hold order by order, we construct a full solution for the orbit of the small body's motion to that order in our expansion.

\subsection{Leading order in eccentricity}
\label{sec:leadingorderine}

We begin by considering Kerr orbits at $\mathcal{O}(e,s)$.  In this limit, it suffices to put $\chi_r = w_r = \Upsilon_r\lambda = (\hat\Upsilon_r + \Upsilon^S_r)\lambda$.  [Although there is a linear-in-eccentricity correction to $\chi_r$, its impact on the small body's motion enters at $\mathcal{O}(e^2)$.]

To first order in $e$, the radial motion of the small body is thus given by
\begin{equation}
    r = pM\left(1 - e\cos w_r\right) = pM\left[1 - \frac{e}{2}\left(e^{iw_r} + e^{-iw_r}\right)\right]\;.
\end{equation}
The second form proves to be particularly useful for our purposes.

Our goal is to compute how the spin-curvature interaction affects all of the important parameters of our system.  Just as in our study of circular and equatorial orbits, we assume that the constants of the motion take the form $\mathcal{X}^S = \hat{\mathcal{X}} + \delta\mathcal{X}^S$ (with $\mathcal{X} \in [E, L_z, K, Q]$), and that
\begin{align}
    \Upsilon_r &= \hat\Upsilon_r + \Upsilon^S_r\;,
    \\
    \Upsilon_\phi &= \hat\Upsilon_\phi + \Upsilon^S_\phi\;,
    \\
    \Gamma &= \hat\Gamma + \Gamma^S\;.
\end{align}
First consider just the leading-order geodesic motion.  The integrals of motion are
\begin{align}
    \hat E &= \frac{1 - 2v^2 \pm qv^3}{\sqrt{1 - 3v^2 \pm 2qv^3}} + \mathcal{O}(e^2)\;,
    \label{eq:Ecirceq2}\\
    \hat L_z &= \pm \frac{M}{v}\sqrt{\frac{1 \mp 2qv^3 + q^2v^4}{1 - 3v^2 \pm 2qv^3}}+ \mathcal{O}(e^2)\;,
    \label{eq:Lzcirceq2}\\
    \hat Q &= 0\;.
\end{align}
As before, $q \equiv a/M$, but now we have $v = \sqrt{1/p}$. We also have
\begin{align}
    \hat\Upsilon_r &= \frac{M}{v}\sqrt{\frac{1 - 6v^2 \pm 8qv^3 - 3q^2v^4}{1 - 3v^2 \pm 2 qv^3}} + \mathcal{O}(e^2)\;,
    \\
    \hat\Upsilon_\phi &= \pm \frac{M}{v}\sqrt{\frac{1}{1 - 3v^2 \pm 2qv^3}} + \mathcal{O}(e^2)\;,
    \\
    \hat\Gamma &= \frac{M^2(1 \pm qv^3)}{v^4\sqrt{1 - 3v^2 \pm 2qv^3}} + \mathcal{O}(e^2)\;.
\end{align}
Let us first consider the components which describe the in-plane orbital motion, $U^{t,r,\phi}$.  We write these components
\begin{align}
    U^t &= U^t_0 + s_{\parallel}e\left(U^t_{-1}e^{iw_r} + U^t_{+1}e^{-iw_r}\right)\;,
    \label{eq:kerrecceqtime1}\\
    U^\phi &= U^\phi_0 + s_{\parallel}e\left(U^\phi_{-1}e^{iw_r} + U^\phi_{+1}e^{-iw_r}\right)\;,  \label{eq:kerrecceqaxial1}\\
    U^r &= \frac{dr}{d\lambda} = -\frac{iepM}{2}\left(\hat\Upsilon_r + \Upsilon_r^S\right)\left(e^{iw_r} - e^{-iw_r}\right)\;.
    \label{eq:kerreccradial1}
\end{align}
In our assumed form of $U^r$, we used the fact that for small eccentricity equatorial orbits, $dw_r/d\lambda = \hat\Upsilon_r + \Upsilon_r^S$.

We next insert Eqs.\ (\ref{eq:kerrecceqtime1}), (\ref{eq:kerrecceqaxial1}), and (\ref{eq:kerreccradial1}) into Eqs.\ (\ref{eq:forcet_eqeccgen}), (\ref{eq:forcer_eqeccgen}), and (\ref{eq:forcephi_eqeccgen}), also enforcing the constraint (\ref{eq:udotu}) in order to solve to each order in $s$ and $e$.  This exercise yields
\begin{widetext}
\begin{align}
    U^t_0 &= \frac{M^2(1 \pm qv^3)}{v^4\sqrt{1 - 3v^2 \pm 2qv^3}} \mp \left(\frac{3s_\parallel\mu}{2}\right)\frac{Mv(1 \mp qv)(1 \mp 2qv^3 + q^2v^4)}{(1 - 3v^2 \pm 2qv^3)^{3/2}}\;,
    \label{eq:kerrecc_timelikesol_0freq}\\
    U^t_{-1} &= U^t_{+1} = \mp\left(\frac{3s_\parallel\mu}{2}\right)\frac{qMv^4(1 \mp qv)^2(1 \mp 2qv^3 + q^2 v^4)}{(1 - 2v^2 + q^2v^4)(1 - 3v^2 \pm 2qv^3)^{3/2}}\;,
    \label{eq:kerrecc_timelikesol_1freq}\\
    U^\phi_0 &=  \pm \frac{M}{v}\sqrt{\frac{1 }{1 - 3v^2 \pm 2qv^3}} - \left(\frac{3s_\parallel\mu}{2}\right)\frac{v^2(1 \mp qv)(1 - 2v^2 \pm qv^3)}{(1 - 3v^2 \pm 2qv^3)^{3/2}}\;,
    \label{eq:kerrecc_axialsol_zerofreq}\\
    U^\phi_{-1} &= U^\phi_{+1} = -\left(\frac{3s_\parallel\mu}{2}\right)\frac{qv^5(1 - 2v^2 \pm qv^3)}{(1 - 2v^2 + q^2v^4)(1 - 3v^2 \pm 2qv^3)^{3/2}}\;,
    \label{eq:kerrecc_axialsol_1freq}\\
    \Upsilon^S_r &= \left(\frac{3s_\parallel\mu}{2}\right)\frac{v^2(1 \mp qv)\left(1 - 2v^2 \mp qv^3(5 - 14v^2) + 5v^4q^2(1 - 4v^2) \pm 7q^3v^7\right)}{(1 - 3v^2 \pm 2qv^3)^{3/2}\sqrt{1 - 6v^2 \pm 8qv^3 - 3q^2v^4}}\;.
    \label{eq:kerrlinecc_UpsilonS_r}
\end{align}
Eq.\ (\ref{eq:kerrlinecc_UpsilonS_r}) matches with the expression Eq.\ (\ref{eq:UpsilonrSKerrexactine}) derived using the exact-in-$e$ approach discussed in Appendix \ref{sec:Saijocomparison}.  The integrals of the motion for these orbits are identical to those what we found in the circular case, Eqs.\ (\ref{eq:deltaEScirceqaligned}) -- (\ref{eq:deltaQScirceqaligned}), but with $v = \sqrt{1/p}$.

\begin{figure*}
\centerline{\includegraphics[scale=0.58]{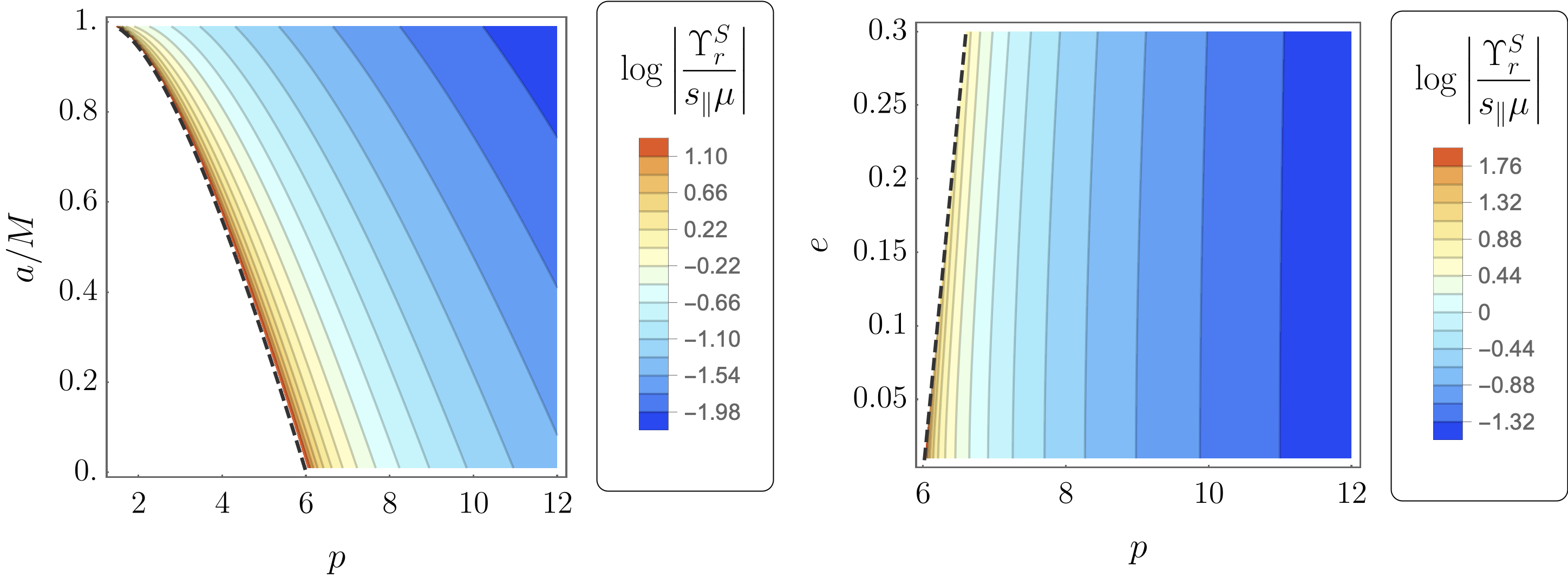}}
\caption{Example of the spin contribution $\Upsilon_r^S$ to the radial Mino-time frequency $\Upsilon_r$.  Left panel shows $\Upsilon_r^S$ to leading order in $e$ as a function of semi-latus rectum $p$ and spin parameter $a$ for prograde orbits ($I = 0^\circ$); see Eq.\ (\ref{eq:kerrlinecc_UpsilonS_r}).  Right panel shows $\Upsilon_r^S$ to second-order in $e$ for Schwarzschild black hole orbits ($a = 0$) as a function of $p$ and $e$.  In both cases, the last stable orbit is indicated by the black dashed line.
\label{fig:contourmap}}
\end{figure*}

Turn now to the out-of-plane motion.  To make progress here, we first must more completely describe the tetrad elements.  They take the form
\begin{align}
    \tilde{e}_{1\alpha} &= \tilde{e}_{1\alpha}^0 + e\,\tilde{e}_{1\alpha}^1\;,
    \\
    \tilde{e}_{2\alpha} &= \tilde{e}_{2\alpha}^0 + e\,\tilde{e}_{2\alpha}^1\;.
\end{align}
The terms $\tilde{e}_{1\alpha}^0$ and $\tilde{e}_{2\alpha}^0$ are exactly as defined in Eqs.\ (\ref{eq:tildee1_circeq}) and (\ref{eq:tildee2_circeq}), but with $v = \sqrt{1/p}$ rather than $v = \sqrt{M/r}$. The eccentricity corrections are given by

\begin{align}
    \tilde{e}_{1\alpha}^1 &= \left(-\frac{v^2}{M}\sqrt{\frac{1 - 3v^2 \pm 2qv^3}{1 - 2v^2 + q^2v^4}}\hat\Upsilon_r\sin w_r, \frac{v^2(1 - q^2v^2)}{(1 - 2v^2 + q^2v^4)^{3/2}}\cos w_r, 0, qv^2\sqrt{\frac{1 - 3v^2 \pm 2qv^3}{1 - 2v^2 + q^2v^4}}\hat\Upsilon_r\sin w_r\right)\;,
    \label{eq:etilde1correction}
    \\
    \tilde{e}_{2\alpha}^1 &= \left(v\sqrt{\frac{1 - 3v^2 \pm 2qv^3}{1 - 2v^2 + q^2v^4}}\cos w_r, -\frac{v^3(1 \mp qv)}{(1 - 2v^2 - q^2v^4)^{3/2}}\hat\Upsilon_r\sin w_r, 0, pM(1 \mp qv^3)\sqrt{\frac{1 - 3v^2 \pm 2qv^3}{1 - 2v^2 + q^2v^4}}\cos w_r\right)\;.
    \label{eq:etilde2correction}
\end{align}
We used $dw_r/d\lambda = \hat\Upsilon_r$ rather than $dw_r/d\lambda = \Upsilon_r = \hat \Upsilon_r + \Upsilon^S_r$ because these tetrad elements are used to build the spin vector $S_\alpha$; any contribution from $\Upsilon^S_r$ is at $\mathcal{O}(S^2)$.

\end{widetext}

To complete our description of the out-of-plane motion, we first note that because $\hat U^\theta = 0$
\begin{equation}
    U^\theta = \Sigma u^\theta = \Sigma \frac{d\delta\vartheta_S}{d\tau} = \frac{d\delta\vartheta_S}{d\lambda}\;,
\end{equation}
and so
\begin{equation}
    \frac{dU^\theta}{d\lambda} = \frac{d^2\delta\vartheta_S}{d\lambda^2}\;.
\end{equation}
Using this in Eq.\ (\ref{eq:forcetheta_eqeccgen}), along with Eqs.\ (\ref{eq:Ecirceq2}), (\ref{eq:Lzcirceq2}), and  (\ref{eq:kerreccradial1}) for $\hat E$, $\hat L_z$, and $\hat U^r$, and finally using Eqs.\ (\ref{eq:etilde1correction}) and (\ref{eq:etilde2correction}) to work out the components $\sigma_t$, $\sigma_r$, and $\sigma_\phi$ yields
\begin{equation}
    \frac{d^2\delta\vartheta_S}{d\lambda^2} + \Upsilon_\theta^2 \delta\vartheta_S = F_S^\theta(\lambda)\;,\label{eq:deltathetade}
\end{equation}
where
\begin{equation}
    \Upsilon_\theta =\frac{M}{v}\sqrt{\frac{1\mp4qv^3+3q^2v^4}{1-3v^2\pm2qv^3}},
\end{equation}
is the Mino-time polar frequency for nearly equatorial circular orbits, and where the forcing term is given by
\begin{widetext}
\begin{align}
    F^\theta_S(\lambda) &= 3s_\perp\mu M\biggl[\mp\frac{v(1\mp qv)\left[1 - 2v^2 + q^2v^4 + e\left(1 - v^2 \mp 2qv^3 + 2q^2v^4\right)\cos w_r\right]}{(1 - 3v^2 \pm 2qv^3)\sqrt{1 - 2v^2 + q^2v^4}}\cos(\phi_s + \psi_p)
    \nonumber\\
    & \equiv 3s_\perp\mu M(\alpha_1 + e\alpha_2\cos w_r)\cos(\phi_s + \psi_p)\;.
    \label{eq:dUthetadlambda1}
\end{align}
\end{widetext}
For notational convenience, we have introduced
\begin{align}
    \alpha_1 &= \mp\frac{v(1 \mp qv)\sqrt{1 - 2v^2 + q^2v^4}}{(1 - 3v^2 \pm 2qv^3)}\;,
    \\
    \alpha_2 &= \mp\frac{v(1 \mp qv)(1 - v^2 \mp 2qv^3 + 2q^2v^4)}{(1 - 3v^2 \pm 2qv^3)\sqrt{1 - 2v^2 + q^2v^4}}\;.
\end{align}
For eccentric equatorial orbits, the precession phase takes the form
\begin{equation}
    \psi_p = \Upsilon_s\lambda + \psi_r\;,
\end{equation}
where
\begin{equation}
    \Upsilon_s = M\sqrt{p} + \mathcal{O}(e^2) = \frac{M}{v} + \mathcal{O}(e^2)\;,
\end{equation}
and where $\psi_r$ is a contribution to the precession phase that varies along the orbit's radial motion. Van de Meent \cite{vandeMeent2019} provides a general expression for $\psi_r$; for small eccentricity, this expression reduces to
\begin{align}
    \psi_r &= -\frac{2ev^2(1 \mp qv)^2}{(1 - 2v^2 + q^2v^4)}\sqrt{\frac{1 - 3v^2 \pm 2qv^3}{1 - 6v^2 \pm 8qv^3 - 3q^2v^4}}\sin w_r
    \nonumber\\
    &\equiv e\varpi(q, v)\sin w_r\;.
\end{align}
Note that $\psi_r \propto ev^2$, and so by definition $\psi_r$ is a small quantity in the small eccentricity limit.  This allows us to usefully expand $\cos(\phi_s + \psi_p)$:
\begin{widetext}
\begin{align}
    \cos(\phi_s + \psi_p) &= \cos(\phi_s + \Upsilon_s\lambda + e\varpi\sin w_r)
    \nonumber\\
    &= \cos(\phi_s + \Upsilon_s\lambda)\cos(e\varpi\sin w_r) - \sin(\phi_s + \Upsilon_s\lambda)\sin(e\varpi\sin w_r)
    \nonumber\\
    & \simeq \cos(\phi_s+\Upsilon_s\lambda) - e\varpi\sin w_r\sin(\phi_s+\Upsilon_s\lambda)\;.
    \label{eq:cospsi_p_linecc}
\end{align}
Combining Eqs.\ (\ref{eq:dUthetadlambda1}) and (\ref{eq:cospsi_p_linecc}), and then linearizing in $e$ yields
\begin{equation}
   F^\theta_S(\lambda) = 3s_\perp\mu M\Bigl\{\alpha_1\cos(\phi_s + \Upsilon_s\lambda) + e\Bigl[\alpha_2\cos w_r\cos(\phi_s + \Upsilon_s\lambda) - \alpha_1\varpi\sin w_r\sin(\phi_s + \Upsilon_s\lambda)\Bigr]\Bigr\}\;.
\end{equation}
As in Sec.\ \ref{subsec:circeqmisalign}, we use variation of constants to solve Eq.\ (\ref{eq:deltathetade}), yielding
\begin{align}
\delta\vartheta_S=A(\lambda) \cos(\Upsilon_\theta\lambda)+B(\lambda)\sin(\Upsilon_\theta\lambda)\;, \label{eq:Utheta_linecc}
\end{align}
where
\begin{align}
A(\lambda)&=c_1-\frac{3\mu M  s_{\perp}}{8 \Upsilon_\theta } \biggl[\frac{4 \alpha_1 \cos (\lambda  (\Upsilon_s-\Upsilon_\theta ))}{\Upsilon_s-\Upsilon_\theta }-\frac{4 \alpha_1 \cos (\lambda  (\Upsilon_\theta +\Upsilon_s))}{\Upsilon_\theta +\Upsilon_s}+\frac{2 e (\alpha_2 - \alpha_1\varpi) \cos (\lambda  (-\Upsilon_\theta +\Upsilon_r-\Upsilon_s))}{-\Upsilon_\theta +\hat\Upsilon_r-\Upsilon_s}\nonumber\\&+\frac{2e (\alpha_2 + \alpha_1\varpi) \cos (\lambda  (-\Upsilon_\theta +\Upsilon_r+\Upsilon_s))}{-\Upsilon_\theta +\hat\Upsilon_r+\Upsilon_s}-\frac{2e(\alpha_2 -\alpha_1\varpi) \cos (\lambda  (\Upsilon_\theta +\Upsilon_r-\Upsilon_s))}{\Upsilon_\theta +\hat\Upsilon_r-\Upsilon_s}\nonumber\\&-\frac{2e(\alpha_2 + \alpha_1 \varpi) \cos (\lambda  (\Upsilon_\theta +\Upsilon_r+\Upsilon_s))}{\Upsilon_\theta +\hat\Upsilon_r+\Upsilon_s}\biggr]\;,
\label{eq:Alambda}\\
B(\lambda)&=c_2+\frac{3\mu M s_\perp}{8 \Upsilon_\theta } \biggl[\frac{4 \alpha_1 \sin (\lambda  (\Upsilon_s-\Upsilon_\theta ))}{\Upsilon_s-\Upsilon_\theta }+\frac{4 \alpha_1 \sin (\lambda  (\Upsilon_\theta +\Upsilon_s))}{\Upsilon_\theta +\Upsilon_s}+\frac{2e(\alpha_2 -\alpha_1\varpi) \sin (\lambda  (-\Upsilon_\theta +\Upsilon_r-\Upsilon_s))}{-\Upsilon_\theta +\hat\Upsilon_r-\Upsilon_s}
\nonumber\\
&+\frac{2e(\alpha_2 + \alpha_1\varpi) \sin (\lambda  (-\Upsilon_\theta +\Upsilon_r+\Upsilon_s))}{-\Upsilon_\theta +\hat\Upsilon_r+\Upsilon_s}+\frac{2e(\alpha_2 - \alpha_1\varpi) \sin (\lambda  (\Upsilon_\theta +\Upsilon_r-\Upsilon_s))}{\Upsilon_\theta +\hat\Upsilon_r-\Upsilon_s}
\nonumber\\
&+\frac{2e(\alpha_2 + \alpha_1\varpi) \sin (\lambda  (\Upsilon_\theta +\Upsilon_r+\Upsilon_s))}{\Upsilon_\theta +\hat\Upsilon_r+\Upsilon_s}\biggr]\;.
\label{eq:Blambda}
\end{align}
\end{widetext}
We have put $\phi_s = 0$ here for simplicity.  Notice that the total radial frequency $\Upsilon_r$ appears inside the sine and cosine functions, but the {\it geodesic} radial frequency $\hat\Upsilon_r$ appears outside these functions in these solutions.  This is because $A(\lambda)$ and $B(\lambda)$ are used to build the $\mathcal{O}(S)$ out of plane precessional motion of the small body, and $\Upsilon_r = \hat\Upsilon_r + \mathcal{O}(S)$.  Using $\Upsilon_r$ instead of $\hat\Upsilon_r$ outside of the sines and cosines would affect the solution at $\mathcal{O}(S^2)$, and we neglect terms at this order.

It is important to note that the combination $\hat\Upsilon_r + \Upsilon_s - \Upsilon_\theta$ can pass through zero.  For example, when $a = 0$, this occurs for orbits that have $v = \sqrt{(2\sqrt{3} - 3)/3}$, for which $p \simeq 6.464$; for $a = M$, this occurs for orbits that have $v = (1/2)(\pm 1/\sqrt{3} + \sqrt{1/3 + 2/\sqrt{3}}$, for which $p \simeq 1.238$ (prograde) and $p \simeq 9.690$ (retrograde).  The general case smoothly connects these limiting forms as a function of $a$.  At least naively, Eq.\ (\ref{eq:Alambda}) appears to be poorly behaved at such ``resonant'' orbits, with certain terms diverging as this combination of frequencies passes through zero.  It is not difficult to show, however, that the combination $\alpha_2 + \alpha_1\varpi$ passes through zero at exactly the same orbits for which $\hat\Upsilon_r + \Upsilon_s = \Upsilon_\theta$.  Such resonances thus have no dynamical impact on the system.  This is consistent with recent work \cite{Witzany2019_2,Zelenka2020} which shows that spinning body orbits are integrable at leading order in the smaller body's spin.

Equations (\ref{eq:Utheta_linecc}), (\ref{eq:Alambda}) and (\ref{eq:Blambda}) show that the out-of-plane motion of the small body depends on $s_\perp$, is uncoupled from the in-plane motion, and is periodic, with structure at harmonics of the precession frequency $\Upsilon_s$, the radial frequency $\Upsilon_r$, and the polar frequency $\Upsilon_\theta$. As we consider more general configurations, we expect qualitatively similar behavior.  We thus design our algorithm for describing the small body's orbital motion in the general case in order to capture such behavior.

\subsection{Next order in eccentricity}
\label{sec:secondorderine}

As our final ``simple'' case, we examine equatorial and eccentric orbits to second order in eccentricity.  To keep the expressions relatively simple, we do this only for orbits of Schwarzschild black holes, and only examine the spin-aligned case.  As we saw for the equatorial and nearly equatorial orbits discussed in the previous section, non-aligned small body spin decouples from all components of the body's orbit except the out-of-plane motion component $U^\theta$, which is itself decoupled from the aligned spin and from all other components of the orbital motion.  Focusing on the Schwarzschild limit of aligned spin orbits will be sufficient for us to develop a strategy for solving for this motion to high precision for more generic cases.

The two most important changes versus our previous analyses are that it will turn out we need to know many quantities describing geodesics to {\it fourth} order in $e$ in order to compute corrections to the orbits of spinning bodies; and, we need a more complete accounting for the difference between the true anomaly $\chi_r$ and the mean anomaly $w_r \equiv \Upsilon_r\lambda$.  The need to go to fourth order in $e$ may be somewhat surprising.  The reason is that the radial velocity introduces a factor $e$; certain terms in the analysis which scale with $\hat U^r \hat U_r$ or $\hat U^r U^S_r$ have their order in eccentricity ``boosted'' by a factor of $e^2$.

To describe the true anomaly, we generalize a functional form that is well known from studies of Keplerian orbits, writing
\begin{align}
    \chi_r &= w_r + \left[e\left(\beta_{11} + \beta^S_{11}\right) + e^3\left(\beta_{31} + \beta^S_{31}\right)\right]\sin w_r\nonumber\\
    &+ e^2\left(\beta_{22} + \beta^S_{22}\right)\sin2w_r + e^3\left(\beta_{33} + \beta^S_{33}\right)\sin3w_r
    \nonumber\\
    &\equiv w_r + \delta\hat{\chi}_r + \delta\chi^S_r\;.
    \label{eq:anomaly_2ndorderecc}
\end{align}
The quantity $\delta\hat{\chi}_r$ stands for all the oscillatory geodesic terms (i.e., the terms with $\beta_{ab}$) that take us from the mean anomaly to the true anomaly.  The quantity $\delta\chi^S_r$ stands for the equivalent terms which arise from spin-curvature coupling (the terms with $\beta^S_{ab}$). 

Other quantities we need are the integrals of the motion and the radial frequency:
\begin{align}
    \hat E &= \frac{1 - 2v^2}{\sqrt{1 - 3v^2}} + \frac{e^2v^2}{2}\frac{(1 - 4v^2)^2}{(1 - 2v^2)(1 - 3v^2)^{3/2}}
    \nonumber\\
    &+ \frac{e^4v^4}{8}\frac{(1-4v^2)^2(3-8v^2)}{(1-2v^2)^3(1-3v^2)^{5/2}}\;,\label{eq:Ehat4thorder}
    \\
    \hat L_z &= M\sqrt{p}\biggl(\frac{1}{\sqrt{1 - 3v^2}} + \frac{e^2v^2}{2}\frac{1}{(1 - 3v^2)^{3/2}}\nonumber\\
    &+ \frac{3e^4v^4}{8}\frac{1}{(1 - 3v^2)^{5/2}}\biggr)\;,\label{eq:Lzhat4thorder}
    \\
    \hat\Upsilon_r &= M\sqrt{p}\biggl(\sqrt{\frac{1 - 6v^2}{1 - 3v^2}} + \frac{e^2v^2}{4}\frac{(1 - 9v^2)(2 - 9v^2)}{(1 - 3v^2)^{3/2}(1 - 6v^2)^{3/2}}\nonumber\\
    &+ \frac{3e^4v^4}{64}\frac{[8 - 25v^2(1 - 3v^2)(8 - 49v^2 + 147v^4)]}{(1 - 3v^2)^{5/2}(1 - 6v^2)^{7/2}}\biggr)\;.
\end{align}

The true anomaly (\ref{eq:anomaly_2ndorderecc}), coupled with the form $r = p/(1 + e\cos\chi_r)$, suffices to fully describe the radial motion.  Turn next to the small body's motion in $t$ and $\phi$.  We parameterize this motion using the 4-velocity components
\begin{align}
    u_t = -\hat E + u^S_t\;,
    \nonumber\\
    u_\phi = \hat L_z + u^S_\phi\;\label{eq:uphiut}.
\end{align}
Raising the index and multiplying by $\Sigma = r^2$, these components can be easily converted to the forms $U^{t,\phi}$.  We assume that the spin corrections to these 4-velocity components take the form
\begin{equation}
    u^S_t = \sum_{n = -3}^3 u^s_{t,n} e^{-inw_r}\;,\quad
    u^S_\phi = \sum_{n = -3}^3 u^s_{\phi, n} e^{-inw_r}\;.
\end{equation}
We generically find that $u^s_{(t,\phi),n} \propto e^{|n|}$.  We find that we don't have enough information to pin down these components for $|n| > 3$; presumably we need to describe the geodesic motion to higher order in order to do this.

We solve for the various unknown quantities we have introduced by enforcing Eqs.\ (\ref{eq:forcet_eqeccgen}) -- (\ref{eq:forcephi_eqeccgen}) and the constraint (\ref{eq:udotu}), and then gathering terms in spin and eccentricity.  Terms at order $(s_\parallel)^0$ are geodesic, and can be used to find the coefficients which make $\delta\hat{\chi}_r$, defined in Eq.\ (\ref{eq:anomaly_2ndorderecc}):
\begin{align}
    \beta_{11} &= -\frac{v^2}{1 - 6v^2}\;,
    \\
    \beta_{22} &= \frac{v^4}{8(1 - 6v^2)^2}\;,
    \\
    \beta_{31} &= -\frac{19v^6}{16(1 - 6v^2)^3}\;,
    \\
    \beta_{33} &= -\frac{v^6}{48(1 - 6v^2)^3\;.}
\end{align}

Turn now to various aspects of the solution at order $s_\parallel$.  First, we find the following coefficients which define $\delta\chi^S_r$:
\begin{align}
    \beta^S_{11} &= \frac{s_\parallel\mu}{M} v^3\frac{(1 - 2v^2)}{(1 - 6v^2)^2}\;,
    \\
    \beta^S_{22} &= -\frac{s_\parallel\mu}{4M}\frac{v^5(1 - 2v^2)}{(1 - 6v^2)^3}\;,
    \\
    \beta^S_{31} &= \frac{s_\parallel\mu}{16M}\frac{v^7(25 + 156v^2 - 924v^4)}{(1 - 2v^2)(1 - 6v^2)^4}\;,
    \\
    \beta^S_{33} &= \frac{s_\parallel\mu}{16M}\frac{v^7(1 - 2v^2)}{(1 - 6v^2)^4}\;.
\end{align}
We next find the terms which define $u^S_\phi$ and $u^S_t$:
\begin{widetext}
\begin{align}
    u^S_\phi &= -s_\parallel\mu\left(\frac{3v^2}{2}\frac{(1 - 2v^2)}{(1 - 3v^2)^{3/2}} + e^2\frac{v^2}{4}\frac{(2 - 5v^2 - 16v^4 + 48v^6)}{(1 - 2v^2)(1 - 3v^2)^{5/2}}\right)\;,
    \\
    u^S_t &= \frac{s_\parallel\mu}{M}\biggl(\frac{3v^5}{2}\frac{(1 - 2v^2)}{(1 - 3v^2)^{3/2}} + \frac{3ev^5\cos w_r}{(1 - 3v^2){1/2}} + \frac{e^2v^5}{4}\frac{\left[2 - 25v^2 + 126v^4 + 234v^6 + 6(1 - 3v^2)^2(1 - 7v^2)\cos2w_r\right]}{(1 - 6v^2)(1 - 3v^2)^{5/2}}
    \nonumber\\
    &+ \frac{e^3v^5}{8}\frac{\cos w_r\left[4 - 24v^2 - 81v^4 + 459v^6 + (4 - 84v^2 + 513v^4 - 891v^6)\cos2w_r\right]}{(1 - 6v^2)^2(1 - 3v^2)^{3/2}}\biggr)\;.
\end{align}
Finally, we compute the shift to the radial frequency due to spin-curvature coupling:
\begin{equation}
    \Upsilon_r^S = \frac{3s_\parallel\mu}{2}\left(\frac{v^2(1 - 2v^2)}{(1 - 3v^2)^{3/2}\sqrt{1 - 6v^2}} - \frac{e^2v^2}{12}\frac{(4 - 106v^2 + 985v^4 - 4275v^6 + 8928v^8 - 7452v^{10})}{(1 - 2v^2)(1 - 3v^2)^{5/2}(1 - 6v^2)^{5/2}}\right)\;.\label{eq:UpsilonrS2ndine}
\end{equation}
Neglecting the terms in $e^2$, this is consistent with the result we found previously, Eq.\ (\ref{eq:kerrlinecc_UpsilonS_r}) in the limit $q \to 0$.  In addition, Eq.\ (\ref{eq:kerrlinecc_UpsilonS_r}) agrees exactly with the $\Upsilon^r_S$ in Eq.\ (\ref{eq:UpsilonrsexactineSchw}) obtained using the approach presented in Ref.\ \cite{Saijo1998}; see Appendix \ref{sec:Saijocomparison} for details of this comparison.

Several other important quantities can be derived from what we computed here.  Two that are particularly important are the axial frequency $\Upsilon_\phi$, and the quantity $\Gamma$ which converts from Mino-time frequencies and periods to coordinate-time frequencies and periods.  As discussed in Sec.\ \ref{subsec:geodesicsfreqdom}, the axial frequency $\Upsilon_\phi$ is the orbit average of $U^\phi$:
\begin{equation}
    \Upsilon_\phi = \frac{1}{2\pi}\int_0^{2\pi}U^\phi(w_r)dw_r\;.
\end{equation}
Using $U^\phi = \Sigma g^{\phi\phi}u_\phi$, we find
\begin{equation}
    \Upsilon_\phi = \frac{M}{v\sqrt{1 - 3v^2}}\left(1 + \frac{e^2v^2}{2(1 - 3v^2)} - \frac{3s_\parallel v^2}{2}\frac{1 - 2v^2}{1 - 3v^2} - \frac{s_\parallel e^2v^3}{4}\frac{(2 - 5v^2 - 16v^4 + 48v^6)}{(1 - 2v^2)(1 - 3v^2)^2}\right)\;.
\end{equation}
Likewise, $\Gamma$ is found by orbit averaging $U^t = \Sigma g^{tt}u_t$:
\begin{align}
    \Gamma &= \frac{M^2}{v^4\sqrt{1 - 3v^2}}\biggl(1 + \frac{e^2}{2}\frac{(3 - 38v^2 + 148v^4 - 186v^6)}{(1 - 11v^2 + 36v^4 - 36v^6)}
    \nonumber\\
    & - \frac{3s_\parallel v^5}{2}\frac{1}{(1 - 3v^2)} - \frac{s_\parallel e^2 v^3}{4}\frac{(4 - 43v^2 + 160v^4 - 186 v^6 - 144v^8 + 216v^{10})}{(1 - 2v^2)(1 - 3v^2)^2(1 - 6v^2)^2}\biggr)\;.
\end{align}
With these quantities in hand, it is straightforward to compute $\Omega_{r,\phi} = \Upsilon_{r,\phi}/\Gamma$. Finally, the shifts to the conserved integrals due to the spin-curvature interaction become.
\begin{align}
    \delta E^S &= -\frac{s_\parallel \mu v^5}{2M(1 - 3v^2)^{3/2}}\left(1 - e^2\frac{(4 - 15v^2)}{2(1 - 3v^2)}\right)\;,\label{eq:deltaEs2ndine}
    \\
    \delta L^S_z &= \frac{s_\parallel\mu(2 - 13v^2 + 18v^4)}{2(1 - 3v^2)^{3/2}}\left(1 - \frac{e^2v^4}{2}\frac{(17 - 96v^2 + 144v^4)}{(1 - 2v^2)^2(1 - 3v^2)(2 - 9v^2)}\right)\;.\label{eq:deltaLzs2ndine}
\end{align}
All of these quantities agree with Eqs.\ (\ref{eq:ESexactineSchw}) and (\ref{eq:LzSexactineSchw}) which were obtained using the exact-in-eccentricity approach outlined in Appendix \ref{sec:Saijocomparison}.

\section{Spinning-body orbits III: Frequency-domain treatment }
\label{sec:spinbodyfreqdom}

We now consider nearly equatorial orbits with \textit{arbitrary} eccentricity, using a frequency-domain treatment of the spinning body's motion. As described in Sec.\ \ref{sec:ParallelTransport}, the spin of the small body introduces the precession frequency $\Upsilon_{s}$ into the analysis.  The small body also shifts the orbital frequencies by an amount $\mathcal{O}(S)$ which we denote $\Upsilon^S_r$ and  $\Upsilon^S_{\theta}$.  Functions evaluated on a spinning body's orbit can thus be written as a Mino-time Fourier expansion in terms of frequencies $\Upsilon_r = \hat{\Upsilon}_{r}+ \Upsilon^S_r$, $\Upsilon_\theta = \hat{\Upsilon}_{\theta}+\Upsilon^S_{\theta}$ and $\Upsilon_{s}$:
\begin{equation}
f(\lambda)=\sum_{j=-1}^{1}\sum_{n,k=-\infty}^{\infty}f_{jnk}e^{-ij\Upsilon_{s}\lambda}e^{-in(\hat\Upsilon_{r}+\Upsilon^S_{r})\lambda}e^{-ik(\hat\Upsilon_{\theta}+\Upsilon^S_{\theta})\lambda}\;. \label{eq:flambdaFourier}
\end{equation}
The Fourier coefficient $f_{jnk}$ is given by
\begin{equation}
f_{jnk} = \frac{1}{\Lambda_{r}\Lambda_{\theta}\Lambda_{s}}\int_{0}^{\Lambda_{r}}\int_{0}^{ \Lambda_{\theta}}\int_{0}^{\Lambda_{s}} f\left( \lambda_r,\lambda_\theta,\lambda_s\right)e^{ij\Upsilon_{s}\lambda_s}e^{in(\hat\Upsilon_{r}+\Upsilon^S_{r})\lambda_r}e^{ik(\hat\Upsilon_{\theta}+\Upsilon^S_{\theta})\lambda_\theta}d\lambda_{\theta}d\lambda_{r}d\lambda_{s}\;,
\end{equation}
where $\Lambda_{r,\theta,s} = 2\pi/\Upsilon_{r,\theta,s}$.  By writing all relevant quantities as expansions of this form, we can compute the properties of spinning-body orbits to arbitrary precision, and develop a natural way of computing the frequency shifts $\Upsilon^S_r$ and  $\Upsilon^S_{\theta}$.  As written, Eq.\ (\ref{eq:flambdaFourier}) is appropriate for generic spinning-body orbits.  In this analysis, we examine orbits of arbitrary eccentricity that are equatorial or nearly equatorial; the generic case is developed and presented in a companion analysis \cite{Paper2}.

\end{widetext}

\subsection{Aligned spin}
\label{sec:eqplanealign}
\begin{figure}
\centerline{\includegraphics[scale=0.53]{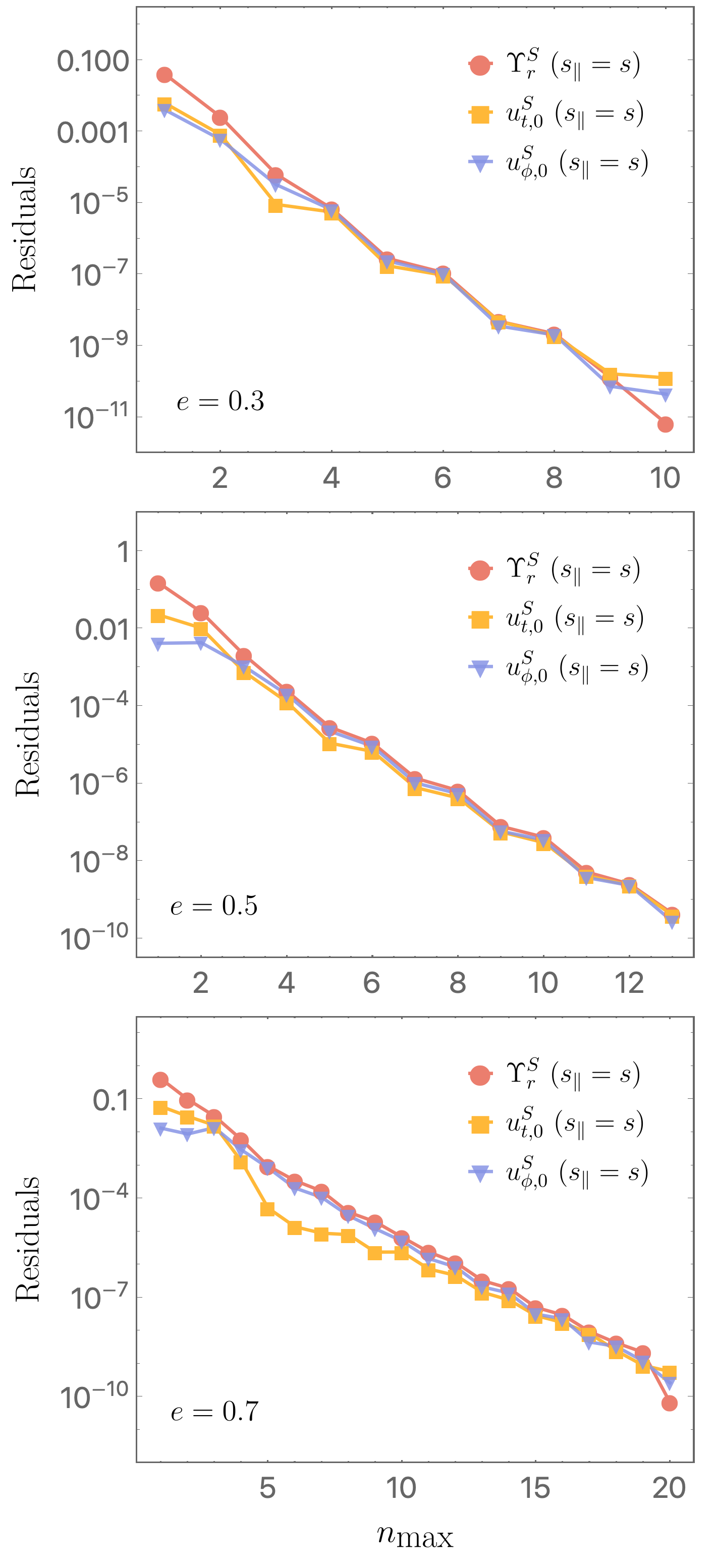}}
\caption{Plot of residuals versus $n_{\text{max}}$ with $s_{\parallel}=s$ for $u^S_{t,0}$ (orange), $u^S_{\phi,0}$ (blue), $\Upsilon_r^S$ (red). These residuals are computed by comparing our frequency-domain expansion to results found using an approach which, for the spin-aligned case, is exact in eccentricity; see Ref.\ \cite{Saijo1998} and Appendix \ref{sec:Saijocomparison} for detailed discussion.  Top panel shows $e=0.3$; middle is $e=0.5$; and bottom is $e=0.7$.  In all cases, the large black hole has spin parameter $a = 0.9M$, and the orbit has $p=10$ and $I=0^{\circ}$.
\label{fig:residualplote}}
\end{figure}

We first consider eccentric orbits with the spin of the small body aligned with the orbit.  The orbit's geometry in this case is exactly as in Sec.\ \ref{sec:secondorderine}, but we now allow for arbitrary eccentricity.  In this case, only radial oscillations are present in the motion, so all orbits can be described using expansions of the form 
\begin{align}
f(\lambda) & =\sum_{n=-\infty}^{\infty}f_n e^{-in (\hat{\Upsilon}_r+\Upsilon_r^S)\lambda} \label{eq:radialexp}\;.
\end{align}
To evaluate these expressions, we truncate the Fourier expansion at a finite value $n_{\text{max}}$.  In Fig.\ \ref{fig:residualplote}, we examine the convergence of important properties of the orbit as we increase $n_{\text{max}}$.  These residuals are computed by comparing our frequency-domain expansion for these quantities with an alternate method which is exact in eccentricity, but only applies to the spin-aligned case.  This method, which is based on that described by Saijo et al.\ (Ref.\ \cite{Saijo1998}) is described in detail in Appendix \ref{sec:Saijocomparison}.  Our results indicate that we can accurately handle large eccentricities (up to at least $e \sim 0.8$) by increasing $n_{\text{max}}$, though larger $e$ requires larger values of $n_{\text{max}}$ in order in order to meet a prescribed level of truncation error.

As described in Sec.\ \ref{sec:genprinciples}, we parameterize the radial motion as
\begin{equation}
    r = \frac{pM}{1 + e\cos\chi_r}\;.\label{eq:rparam2}
\end{equation}
This form guarantees that the motion is constrained to the interval $p/(1+e)\leq r \leq p/(1-e)$. As in Eq.\ (\ref{eq:chir}), we write the true anomaly $\chi_r$ in Eq.\ (\ref{eq:rparam2}) as 
\begin{equation}
    \chi_r=w_r+\delta\chi_r\;,
\end{equation}
where $w_r$ is the mean anomaly and $\delta\chi_{r}$ is an oscillating contribution to $\chi_r$. The oscillating contribution in turn has a piece associated with geodesic motion, $\delta \hat{\chi}_r$, and another piece that arises from spin-curvature coupling $\delta \chi^S_r=\mathcal{O}(S)$, 
\begin{align}
\delta\chi_{r} & =\delta \hat{\chi}_r + \delta \chi^S_r\;.\label{eq:deltachir}
\end{align}
The mean anomaly also has geodesic and spin-curvature contributions:
\begin{equation}
w_{r} = \left(\hat{\Upsilon}_r + \Upsilon^S_r\right)\lambda\;,\label{eq:meananom}
\end{equation}
where $\Upsilon^S_r$ is the $\mathcal{O}(S)$-correction to the radial Mino-time frequency. It is useful to write the true anomaly angles $\delta \hat{\chi}_r$ and $\delta \chi^S_r$ as Fourier expansions\footnote{Note that if the function we are Fourier expanding already has a subscript, we use a comma to denote the specific Fourier mode. For example, $\delta \hat\chi_{r,1}$ is the $n=1$ harmonic of function $\delta \hat \chi_r$.},
\begin{align}
\delta \hat{\chi}_{r} & =\sum_{n=-\infty}^{\infty}\delta \hat{\chi}_{r,n} e^{-in w_r\lambda}\;, \label{eq:deltahatchi} \\
\delta \chi^S_{r} & =\sum_{n=-\infty}^{\infty}\delta \chi^S_{r,n}e^{-in w_r\lambda}\;. \label{eq:deltachiS}
\end{align}
We set $\chi^S_{r,0}=0$; this amounts to a choice of initial true anomaly.  Note that the geodesic Fourier coefficients $\delta \hat{\chi}_{r,n} $ are known, as described in Sec.\ \ref{sec:kerrgeodesics}.  Observe, however, that $w_r$ includes the frequency correction $\Upsilon_r^S$, meaning that $w_r + \delta \hat{\chi}_{r}$, with $\delta \hat{\chi}_{r}$ given by Eq.\ (\ref{eq:deltahatchi}), is \textit{not} the same as the true anomaly for the corresponding geodesic orbit with the same radial turning points.  We treat the non-oscillating part of the spinning body's true anomaly as almost identical to the non-oscillating part of the true anomaly belonging to the geodesic with the same turning points, differing only by an appropriate shift to the orbit's frequency.  This cures a pathology associated with the fact that the rate at which the mean anomaly accumulates for geodesic orbits differs at $\mathcal{O}(S)$ from the rate at which it accumulates for spinning-body orbits.  This issue is described in more detail in Appendix \ref{sec:secularterms}.

\begin{figure*}
\centerline{\includegraphics[scale=0.58]{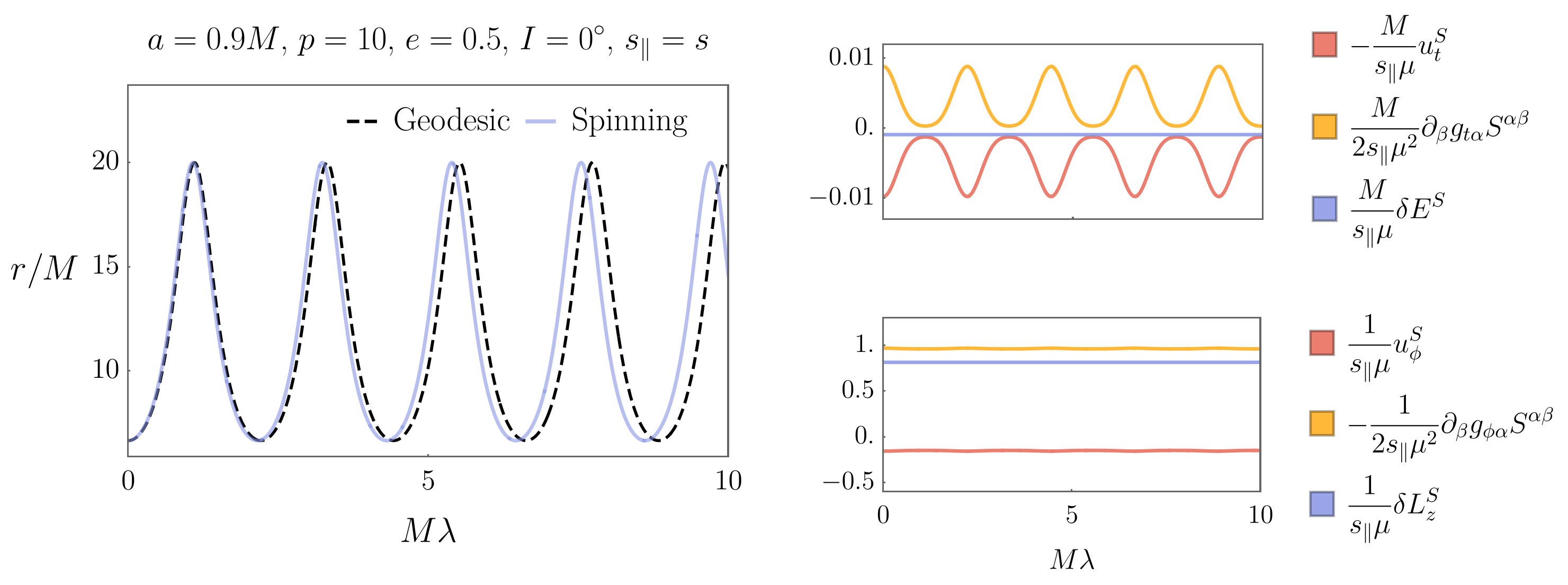}}

\caption{Example of radial motion for an aligned, spinning body in an equatorial orbit of a Kerr black hole ($a = 0.9M$).  Left panel shows $r$ versus $\lambda$ for a geodesic (black dashed) and for a spinning-body orbit (blue solid).  These orbits share radial turning points, corresponding to semi-latus rectum $p = 10M$, eccentricity $e = 0.5$.  Top right panel shows the spinning body's $-u_t^S$ (red), $\partial_{\beta}g_{t\alpha} S^{\alpha\beta}/(2\mu)$ (orange), and $\delta E^S$ (blue) versus $\lambda$.  Bottom right panel shows the spinning body's $u_{\phi}^S$ (red), $-\partial_{\beta} g_{\phi\alpha} S^{\alpha\beta}/(2\mu)$ (orange), $\delta L_z^S$ (blue) versus $\lambda$.  Notice that the shifts in the integrals of motion $E$ and $L_z$ are constants, even though the terms which contribute to them oscillate.  (The oscillations in the terms which contribute to $\delta L_z^S$ are so small they can barely be seen on this plot.)  In all cases, the Fourier expansions have been taken to $n_{\rm max} = 8$; for the left panel, we have used $\mu s/M=0.5$.
\label{fig:exampleorbitseqparallel}}
\end{figure*}

As in Eq.\ (\ref{eq:uphiut}), we define the $\mathcal{O}(S)$-corrections to the temporal and axial components of the 4-velocity by
\begin{align}
u_t=-\hat{E}+ u_t^S\;, \ \ \ \
u_{\phi}=\hat{L}_z+ u_{\phi}^S\;,\label{eq:utuphi}
\end{align}
where $u_t^S$ and $u_{\phi}^S$ can also be written as Fourier expansions,
\begin{align}
u_t^S & =\sum_{n=-\infty}^{\infty}u_{t,n}^Se^{-in w_r\lambda}\;,\\
u_{\phi}^S  &=\sum_{n=-\infty}^{\infty}u_{\phi,n}^Se^{-in w_r\lambda}\label{eq:uphis}\;.
\end{align}
We divide both $u_t^S$ and $u_{\phi}^S$ into a piece that is constant, and a piece that oscillates:
\begin{align}
u_t^S & =u_{t,0}^S+\delta u_{t}^S(\lambda)\;,\ \ \ u_{\phi}^S =u_{\phi,0}^S+\delta u_{\phi}^S(\lambda)\;.
\label{eq:deltautphis}
\end{align}
We can solve for the oscillating pieces using the $t$- and $\phi$- components of Eq.\ (\ref{eq:mp1linear}). Combining the axial and temporal components yields two equations of the form
\begin{align}
\frac{d u^S_{\phi}}{d\lambda}= \mathcal{R}_{\phi}\;, \ \ \ \frac{d u^S_{t}}{d\lambda}= \mathcal{R}_{t}\; \label{eq:Reqs},
\end{align}
where $\mathcal{R}_{\phi}$ and $\mathcal{R}_t$ are functions of known geodesic quantities.  For the equatorial and nearly equatorial cases, Eqs.\ (\ref{eq:Reqs}) are equivalent to Eqs.\ (\ref{eq:forcet_eqeccgen_v2}) -- (\ref{eq:forcephi_eqeccgen_v2}), and we can read out the functions $\mathcal{R}_\phi$ and $\mathcal{R}_t$ from there.  The equations in (\ref{eq:Reqs}) allow us to immediately solve for $\delta u^S_t$ and $\delta u^S_{\phi}$.  The constants $u^S_{t,0}$ and $u^S_{\phi,0}$ are determined by the system's initial conditions; as described below, we solve for these quantities along with with the other unknowns, $\delta \chi_r^S$ and $\Upsilon_r^S$.

\begin{widetext}

To make further progress, we insert Eqs.\ (\ref{eq:rparam2}) and (\ref{eq:utuphi}) into Eq.\ (\ref{eq:mp1linear}) and linearize in spin.  By gathering in terms of unknown quantities, the radial component of Eq.\ (\ref{eq:mp1linear}) has the form 
\begin{align}
\mathcal{F}_r \frac{d^2 \delta\chi_r^S}{d \lambda^2}+\mathcal{G}_r\frac{d \delta\chi_r^S}{d \lambda}+\mathcal{H}_r \delta\chi_r^S+\mathcal{I}_{1r} \Upsilon_r^S +\mathcal{I}_2 u^S_{t,0} +\mathcal{I}_3 u^S_{\phi, 0}+\mathcal{J}=0\;.\label{eq:radiallinMP}
\end{align}
In this equation, we have gathered all the terms and functional behavior which are known (i.e., they depend on the behavior of the geodesic with $p$ and $e$) into the functions $\mathcal{F}_r$, $\mathcal{G}_r$, $\mathcal{H}_r$, $\mathcal{I}_{1r}$, $\mathcal{I}_2$, $\mathcal{I}_3$ and $\mathcal{J}$.  The explicit expressions for these functions in the Schwarzschild spacetime can be found in Appendix \ref{sec:coefficientfunctions}.  For Kerr, the expressions become rather unwieldy.  We include a \textit{Mathematica} notebook in the supplementary material which computes the expressions for $a\neq0$.  Note that we solved for $\delta u^S_t$ and $\delta u^S_\phi$ when we solve (\ref{eq:Reqs}); these functions are incorporated into $\mathcal{J}$. 

We also use $u^{\alpha}u_{\alpha}=-1$ linearized in spin [i.e., Eq. (\ref{eq:udotu})], as an additional constraint.  This yields an equation of the form
\begin{align}
\mathcal{K}_r \frac{d \delta\chi_r^S}{d \lambda}+\mathcal{M}_r \delta\chi_r^S+\mathcal{N}_{1r}\Upsilon_r^S +\mathcal{N}_2 u^S_{t,0} +\mathcal{N}_3u^S_{\phi, 0}+\mathcal{P}=0\;,\label{eq:udotucoeff}
\end{align}
where $\mathcal{K}_r$, $\mathcal{M}_r$, $\mathcal{N}_{1r}$, $\mathcal{N}_2$, $\mathcal{N}_3$ and $\mathcal{P}$ are again all functions\footnote{The functions $\mathcal{F}_r$, $\mathcal{G}_r$, etc.\ follow a mostly alphabetic sequence; however, we skip the letter $\mathcal{L}$ in our scheme to avoid confusion with the angular momentum 4-vector defined in Eq.\ (\ref{eq:orbangmomdef}).}
of known quantities, and are listed in Appendix \ref{sec:coefficientfunctions} for Schwarzschild (with the Kerr versions included in supplemental material).  The solutions for $\delta u_t^S$ and  $\delta u_{\phi}^S$ are here incorporated into the function $\mathcal{P}$. 

\end{widetext}

To solve for the unknown aspects of the spinning body's orbit, we write $\mathcal{F}_r$, $\mathcal{G}_r$, $\mathcal{H}_r$, $\mathcal{I}_{1r}$, $\mathcal{I}_{2}$, $\mathcal{I}_{3}$, $\mathcal{J}$, $\mathcal{K}_r$, $\mathcal{M}_r$, $\mathcal{N}_{1r}$, $\mathcal{N}_2$, $\mathcal{N}_{3}$ and $\mathcal{P}$ as Fourier expansions of the form shown in Eq.\ (\ref{eq:radialexp}).  We insert these expansions, along with Eq.\ (\ref{eq:deltachiS}), into Eqs.\ (\ref{eq:radiallinMP}) and (\ref{eq:udotucoeff}).  Evaluating Eqs.\ (\ref{eq:radiallinMP}) and  (\ref{eq:udotucoeff}) in the frequency domain, we turn this differential equation into a system of linear equations which can be expressed in the form
\begin{equation}
    \mathbf{M}\cdot\mathbf{v}+\mathbf{c}=0\;,
\end{equation}
where $\mathbf{M}$ is a matrix whose entries are related to the Fourier expansions of several of the functions appearing in Eqs.\ (\ref{eq:radiallinMP}) and (\ref{eq:udotucoeff}), and where $\mathbf{c}$ is a column vector whose entries are related to the Fourier expansion of the functions $\mathcal{K}$ and $\mathcal{P}$.  The entries of the column vector $\mathbf{v}$ are the problem's various unknown quantities, such as the spin-induced shift in the radial frequency $\Upsilon^S_r$.  As an illustration of this equation's form, we have written out the explicit form of $\mathbf{M}$, $\mathbf{v}$, and $\mathbf{c}$ in Appendix \ref{sec:matrixsystem} for $n_{\text{max}}=1$.  Note that this value of $n_{\rm max}$ is far too small to achieve numerical convergence, and is used only for illustrative purposes.  The matrix equation is ungainly when written out for realistic values of $n_{\rm max}$, though it poses no difficulties for numerical analysis.  We then solve this system of linear equations for the unknown variables $\delta\chi_r^S$, $\Upsilon_r^S$, $u_{\phi,0}^S$ and $u_{t,0}^S$.  This yields a complete solution for the motion of the spinning body to first order in spin. 

When the small body's spin is aligned with the orbit, an alternative method based on Ref.\ \cite{Saijo1998} allows us to calculate $\Upsilon^S_r$ exactly as a function of eccentricity; this method is described in detail in Appendix \ref{sec:Saijocomparison}.  Figure \ref{fig:residualplote} shows how $\Upsilon_r^S$, $u_{\phi, 0}^S$ and $u_{t,0}^S$ converge to the exact result as we increase the value of $n_{\text{max}}$. For higher eccentricities, we need to include more harmonics (use a larger value of $n_{\text{max}}$) in order for the solution to converge to the same level of accuracy as the lower eccentricity orbit. For example, for an eccentricity of $e=0.7$ (bottom panel of  Fig.\ \ref{fig:residualplote}) we need $n_{\text{max}}=20$ to obtain the same discrepancy between the exact and frequency-domain result as for $e=0.3$ (top panel of Fig.\ \ref{fig:residualplote}) with $n_{\text{max}}=9$.

An example of an aligned spinning body's equatorial orbit is shown in the left panel of Fig.\ \ref{fig:exampleorbitseqparallel}. The geodesic orbit with the same radial turning points is overplotted for comparison.  Notice the two ways in which the spinning body's radial motion differs from that of the geodesic.  First, the radial frequency is shifted by $\Upsilon_r^S$.  This effect can be very clearly seen in Fig.\ \ref{fig:exampleorbitseqparallel}.  Second, the shape of the orbit is modified due to the impact of the oscillatory term in the true anomaly $\delta\chi_r^S$.  This effect is quite a bit smaller, and is not obvious in the figure for this choice of parameters.

In the right panel of Fig.\ \ref{fig:exampleorbitseqparallel}, we show $u_t^S$ and $u_{\phi}^S$, as well as corrections to the spinning body's energy $\delta E^S$ and axial angular momentum $\delta L_z^S$ [using Eqs.\ (\ref{eq:deltaEspin}) and (\ref{eq:deltaLspin})].  As expected, the oscillations  $\partial_{\beta}g_{t\alpha} S^{\alpha\beta}/(2\mu)$ and $\partial_{\beta}g_{t\alpha} S^{\alpha\beta}/(2\mu)$ precisely cancel oscillations in $\delta u_t^S$ and $\delta u_{\phi}^S$; upon summing, $\delta E^S$ and $\delta L_z^S$ are indeed constant.  The values for the spinning body's energy and axial angular momentum match those obtained using the alternative approach described in Appendix \ref{sec:Saijocomparison}; see App.\ \ref{sec:kerrSaijoecc} in particular.

\subsection{Misaligned spin}
\label{sec:ecceqprecess}

\begin{figure}
\centerline{\includegraphics[scale=0.53]{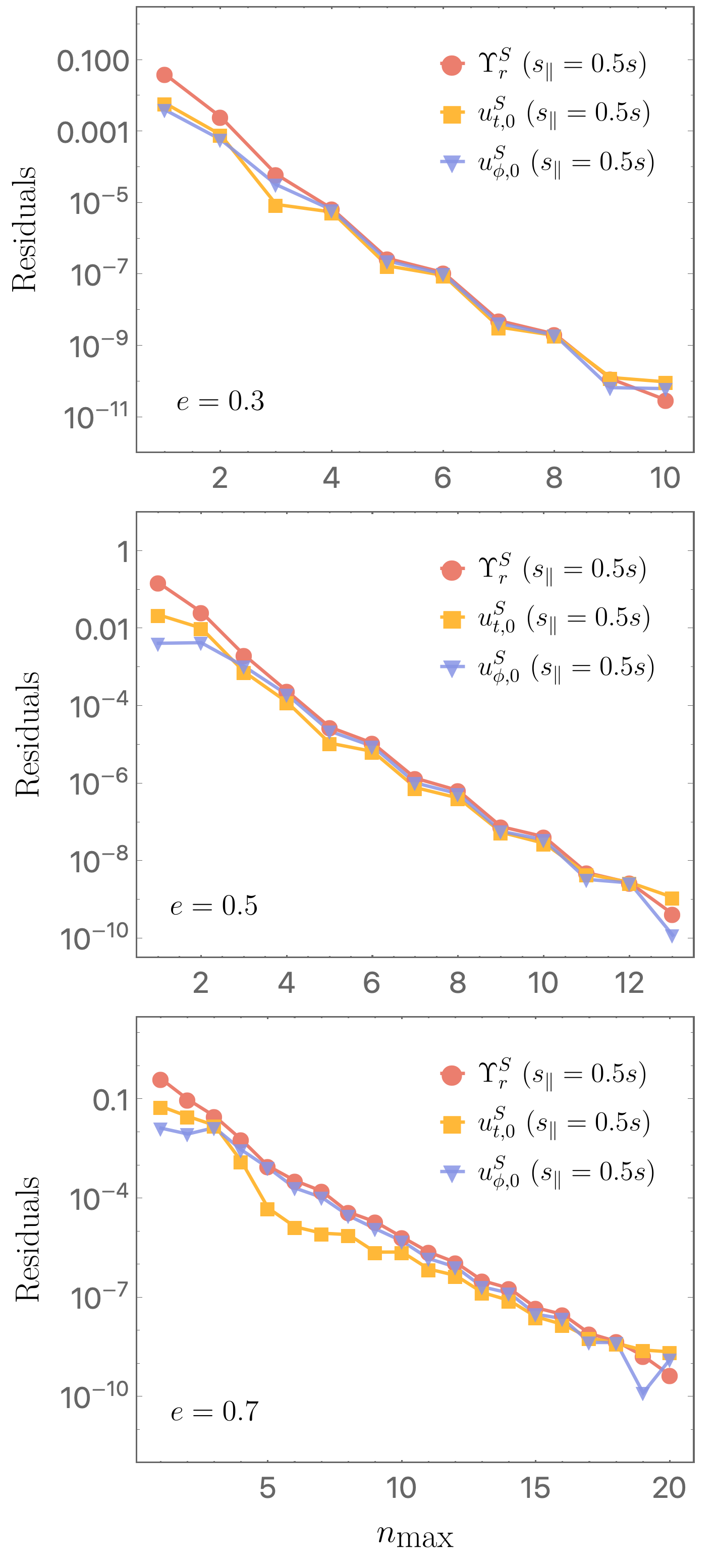}}
\caption{Plot of residuals versus $n_{\text{max}}$ for a nearly equatorial orbit of a misaligned spinning body.  The body's spin in this case has $s_{\parallel}=0.5s$, $s_\perp = \sqrt{3}s/2$.  We show residuals for $u^S_{t,0}$ (orange), $u^S_{\phi,0}$ (blue), $\Upsilon_r^S$ (red).  To compute these residuals, we use the fact that the equations for this case are identical to the equations for the spin-aligned case, but substituting $s_\parallel$ for the small body's spin $s$.  Because of this, the exact-in-eccentricity solution (described in Ref.\ \cite{Saijo1998} and Appendix \ref{sec:Saijocomparison}) that describes aligned orbits can be used to compute the quantities which describe the radial part of misaligned spinning body's orbit, provided we use only the parallel component $s_\parallel$ all of the relevant expressions.  As in Fig.\ \ref{fig:residualplote}, top panel shows $e=0.3$, middle shows $e=0.5$, and bottom is $e=0.7$.  In all cases, the large black hole has spin parameter $a = 0.9M$, and the orbit has $p=10$ and $I=0^{\circ}$.
\label{fig:residualplotemisaligned}}
\end{figure}

\begin{figure*}
\centerline{\includegraphics[scale=0.62]{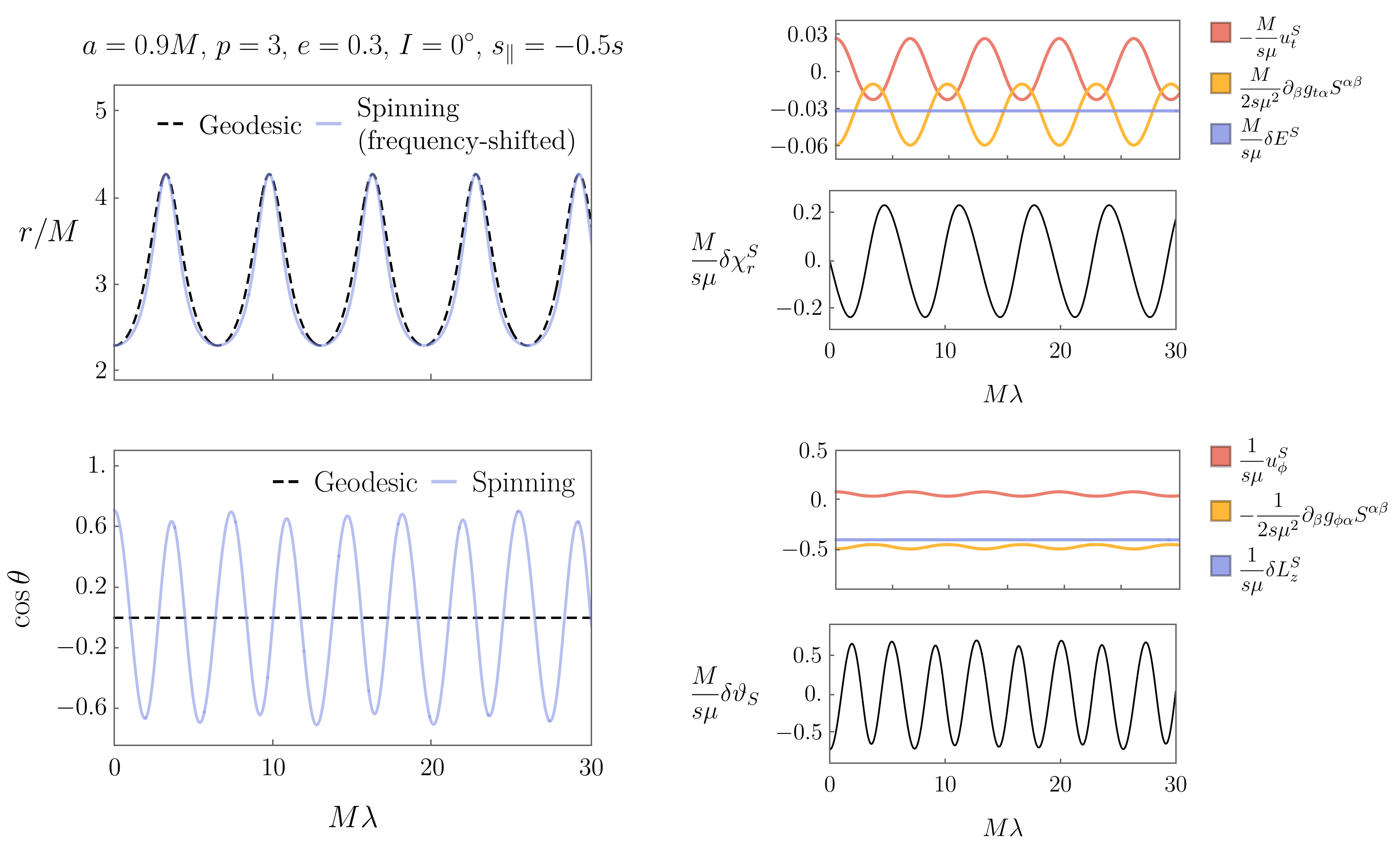}}
\caption{Example of the motion of a nearly equatorial prograde ($I = 0^\circ$) orbit for a non-aligned spinning test body around a Kerr black hole with $a = 0.9M$.  Top left panel shows $r$ versus $\lambda$ for a geodesic (black dashed) and a spinning test body (blue solid) orbit.  These orbits share radial turning points, corresponding to $p = 3M$, $e = 0.3$.  Note that, in the left two panels, we have used an unphysically high spin $\mu s /M=1$ in order make the spin-curvature effects clearly visible.  Also note that for making this plot, the spinning-body orbit has been shifted slightly: its radial frequency $\Upsilon_r= \hat\Upsilon_r + \Upsilon_r^S$ has been replaced with $\hat\Upsilon_r $. This is done so that in the plot the geodesic and the spinning-body orbit pass through their radial turning points at the same times, which helps to illustrate differences in their motion between each turning point. Bottom left panel shows $\cos\theta$ versus $\lambda$ for a geodesic (black dashed) and the spinning-body (blue solid) orbit.  Top right shows $-u_t^S$ (red), $\partial_{\beta} g_{t\alpha} S^{\alpha\beta}/(2\mu)$ (orange), and $\delta E^S$ (blue), as well as $\delta\chi_r^S$ (black), all versus $\lambda$.  Finally, the bottom right panel shows $u_{\phi}^S$ (red), $-\partial_{\beta} g_{\phi\alpha} S^{\alpha\beta}/(2\mu)$ (orange), and $\delta L_z^S$ (blue), as well as $\delta\vartheta_S$ (black), all versus $\lambda$.  Notice that the spin-induced shifts to the integrals of motion $E$ and $L_z$ are constants, although each such term has contributions that oscillate.  In making these plots, we have used $s_{\parallel}=-0.5s$, $\phi_s=0$ and $n_{\text{max}}=5$.  In the two left panels, we have used $\mu s /M=1$.
\label{fig:exampleorbitseqprec}}
\end{figure*}

We now consider eccentric, nearly equatorial orbits, allowing the spin of the small body to have arbitrary orientation. As we saw in Secs.\ \ref{subsec:circeqalign}, \ref{sec:secondorderine} and \ref{sec:eqplanealign}, if the spin of the small body is aligned with the orbit, the motion remains in the equatorial plane. However, if the spin of the test body is misaligned, the spin vector precesses, as described in Secs.\ \ref{subsec:circeqmisalign} and \ref{sec:leadingorderine}. The spin precession introduces the frequency $\Upsilon_s$ into the motion.  Orbital quantities can then be described using expansions of the form 
\begin{align}
f(\lambda) & =\sum_{j=-1}^{1}\sum_{n=-\infty}^{\infty}f_{jn} e^{-ij \Upsilon_s\lambda} e^{-in (\hat{\Upsilon}_r+\Upsilon_r^S)\lambda}\;. \label{eq:radialexpprecess}
\end{align}

The spin precession induces out-of-plane motion, which we describe by introducing the new variable $\delta{\vartheta}_{S}$, as in Secs.\ \ref{subsec:circeqmisalign} and \ref{sec:leadingorderine}. The orbit can therefore be parameterized by
\begin{align}
r & =\frac{pM}{1+e\cos\left(w_{r}+\delta\hat{\chi}_r+\delta\chi_r^S\right)}\;,\label{eq:parameqprecr} \\ 
\theta &=\frac{\pi}{2}+ \delta\vartheta_S\label{eq:parameqprectheta}\;.
\end{align}
The spin contribution to the radial anomaly angle, $\delta\chi_{r}^{S}$, consists of purely radial oscillations,
\begin{align}
\delta\chi_{r}^{S} & =\sum_{n=-\infty}^{\infty}\delta\chi_{r,n}^{S}e^{-in w_r}\;;\label{eq:deltachireqprec}
\end{align}
the Fourier expansion for $\delta\vartheta_S$ depends in addition on the frequency $\Upsilon_s$,
\begin{align}
\delta\vartheta_S &= \sum_{j=-1}^1\sum_{n=-\infty}^{\infty}\delta\vartheta_{S,jn} e^{-in w_r} e^{-ij w_s}\;.\label{eq:deltachitheta2}
\end{align}
We have introduced $w_{s} = \Upsilon_s\lambda$.

\begin{widetext}

As in Sec.\ \ref{sec:eqplanealign}, we write the axial and temporal components of the 4-velocity in the form Eq.\ (\ref{eq:deltautphis}) and use  Eq.\ (\ref{eq:Reqs}) to find $\delta u_{\phi}^S$ and $\delta u_{t}^S$. We insert Eqs.\ (\ref{eq:parameqprecr}), (\ref{eq:parameqprectheta}) and (\ref{eq:utuphi}) into Eq.\ (\ref{eq:mp1linear}) and linearize in spin. Similarly to Sec.\ \ref{sec:eqplanealign}, the radial component of Eq.\ (\ref{eq:mp1linear}) has the form 
\begin{align}
\mathcal{F}_r \frac{d^2 \delta\chi_r^S}{d \lambda^2}+\mathcal{G}_r \frac{d \delta\chi_r^S}{d \lambda}+\mathcal{G}_{\vartheta}  \frac{d \delta\vartheta_S}{d \lambda}+\mathcal{H}_r \delta\chi_r^S+\mathcal{H}_{\vartheta}\delta\vartheta_S+\mathcal{I}_{1r} \Upsilon_r^S +\mathcal{I}_2 u^S_{t,0} +\mathcal{I}_3 u^S_{\phi, 0}+\mathcal{J} =0\;,\label{eq:radiallinMP2}
\end{align} where $\mathcal{F}_r$, $\mathcal{G}_r$, $\mathcal{G}_{\vartheta}$, $\mathcal{H}_r$,  $\mathcal{H}_{\vartheta}$, $\mathcal{I}_{1r}$, $\mathcal{I}_{2}$, $\mathcal{I}_{3}$ and $\mathcal{J}$ are all functions of known quantities. For nearly equatorial orbits, $\mathcal{G}_{\vartheta}=\mathcal{H}_{\vartheta}=0$.  This is not the case for generic orbit geometry, which we discuss in a companion paper \cite{Paper2}; we include these functions in Eq.\ (\ref{eq:radiallinMP2}) in order to lay out the structure we need for the generic case.

When the small body's spin is misaligned with the orbit, the body's motion takes it out of the equatorial plane.  This requires us to include the $\theta$-component of Eq.\ (\ref{eq:mp1linear}) in our analysis.  We linearize this equation in spin, yielding
\begin{align}
\mathcal{Q}_{\vartheta}\frac{d^2 \delta\vartheta_S}{d \lambda^2}+\mathcal{S}_r \frac{d \delta\chi_r^S}{d \lambda}+\mathcal{S}_{\vartheta}\frac{d \delta\vartheta_S}{d \lambda}+\mathcal{T}_r\delta\chi_r^S+\mathcal{T}_{\vartheta}\delta\vartheta_S+\mathcal{U}_{1r}\Upsilon_r^S +\mathcal{U}_{2}u^S_{t,0} +\mathcal{U}_{3}u^S_{\phi, 0}+\mathcal{V}=0\;.
\label{eq:polarlinMP2}
\end{align}
In (\ref{eq:polarlinMP2}), the functions $\mathcal{Q}_{\vartheta}$, $\mathcal{S}_r$, $\mathcal{S}_{\vartheta}$, $\mathcal{T}_r$,  $\mathcal{T}_{\vartheta}$, $\mathcal{U}_{1r}$ $\mathcal{U}_{2}$, $\mathcal{U}_{3}$ and $\mathcal{V}$ all depend on known quantities.  For nearly equatorial orbits, $\mathcal{S}_r = \mathcal{S}_\vartheta = \mathcal{T}_r = \mathcal{U}_{1r} = \mathcal{U}_{2} = \mathcal{U}_{3} = 0$.  This is not the case for the more generic orbits which we discuss in a companion paper \cite{Paper2}.  As in our discussion of the spin-aligned case, we use $u^{\alpha}u_{\alpha} = -1$ to obtain a linear-in-spin constraint which we write
\begin{align}
\mathcal{K}_r \frac{d \delta\chi_r^S}{d \lambda}+\mathcal{K}_{\vartheta} \frac{d \delta\vartheta_S}{d \lambda}+\mathcal{M}_r \delta\chi_r^S+\mathcal{M}_{\vartheta} \delta\vartheta_S+\mathcal{N}_{1r}\Upsilon_r^S +\mathcal{N}_2 u^S_{t,0} +\mathcal{N}_3 u^S_{\phi, 0}+\mathcal{P}=0\;.
\label{eq:udotu2}
\end{align}
Here, $\mathcal{K}_r$,  $\mathcal{K}_{\vartheta}$, $\mathcal{M}_r$, $\mathcal{M}_{\vartheta}$, $\mathcal{N}_{1r}, \mathcal{N}_2$, $\mathcal{N}_{3}$ and $\mathcal{P}$ are again all functions of known quantities. For nearly equatorial orbits, $L_\theta = M_\theta = 0$.  We list the Schwarzschild limit of all these functions in App.\ \ref{sec:coefficientfunctions}, and include Kerr versions in our supplementary material.

\end{widetext}

We can now write $\mathcal{F}_r$, $\mathcal{G}_r$, $\mathcal{G}_{\vartheta}$, $\mathcal{H}_r$,  $\mathcal{H}_{\vartheta}$, $\mathcal{I}_{1r}$, $\mathcal{I}_{2}$, $\mathcal{I}_{3}$, $\mathcal{J}$, $\mathcal{Q}_{\vartheta}$, $\mathcal{S}_r$, $\mathcal{S}_{\vartheta}$, $\mathcal{T}_r$,  $\mathcal{T}_{\vartheta}$, $\mathcal{U}_{1r}$, $\mathcal{U}_{2}$, $\mathcal{U}_{3}$, $\mathcal{V}$, $\mathcal{K}_r$,  $\mathcal{K}_{\vartheta}$, $\mathcal{M}_r$, $\mathcal{M}_{\vartheta}$, $\mathcal{N}_{1r}$, $\mathcal{N}_2$, $\mathcal{N}_{3}$ and $\mathcal{P}$ as Fourier expansions of the form given in Eq.\ (\ref{eq:radialexpprecess}). We insert these expansions, along with Eqs.\ (\ref{eq:deltachireqprec}) and (\ref{eq:deltachitheta2}), into Eqs. (\ref{eq:radiallinMP2}), (\ref{eq:polarlinMP2}) and (\ref{eq:udotu2}).  This turns these differential equations in linear algebraic ones; as in our discussion of aligned orbits in Sec.\ \ref{sec:eqplanealign}, we gather terms into matrix form, and then solve for the for the unknown variables $\delta\chi_r^S$, $\delta\vartheta_S$, $\Upsilon_r^S$, $u_{t,0}^S$ and $u_{\phi, 0}^S$. Further details about the matrix system corresponding to Eq.\ (\ref{eq:polarlinMP2}) are provided in Appendix \ref{sec:matrixsystem} and the explicit solution given for $n_{\text{max}} = 1$.

As discussed in Secs.\ \ref{subsec:circeqmisalign} and \ref{sec:leadingorderine}, when the small body's spin is misaligned from the orbit, qualitatively distinct behaviour arises due to the spin's precession.  For the nearly equatorial case, non-trivial polar motion $\delta\vartheta_S$ emerges, varying with the spin precession frequency $\Upsilon_s$.  Note, though, that in the expansion (\ref{eq:deltachitheta2}) we do not include harmonics at frequency $\Upsilon_\theta$.  Such harmonics can in principle be present, as we saw in Eqs.\ (\ref{eq:Utheta_linecc}), (\ref{eq:Alambda}), and (\ref{eq:Blambda}).  In the present analysis, we have only considered initial conditions such that the amplitude of the $\Upsilon_\theta$ harmonics are suppressed.  In our companion study \cite{Paper2}, we examine motion with $\delta\vartheta_S$ governed by the completely general form (\ref{eq:flambdaFourier}).  The motion in this case has harmonics of all three frequencies are present.

In the left panel of Fig.\ \ref{fig:exampleorbitseqprec}, we show $r$ and $\theta$ for a small body with misaligned spin; an equatorial geodesic with the same radial turning points is overplotted for comparison.  The form of $\delta\chi_r^S$ and $\delta\vartheta_S$ for this orbit is shown in the right panels of Fig.\ \ref{fig:exampleorbitseqprec}.  As in Sec.\ \ref{sec:eqplanealign}, there are two main ways in which the radial motion of the spinning body differs from that of the geodesic with the same turning points: the radial frequency is shifted, and the shape of the orbit is modified by $\delta\chi_r^S$.  We have actually hidden the first effect by shifting the spinning-body orbit's radial frequency --- the solid curve in Fig.\ \ref{fig:exampleorbitseqprec} is a spinning-body orbit with the radial frequency $\Upsilon_r= \hat\Upsilon_r + \Upsilon^S_r$ replaced with $\hat\Upsilon_r$.  This allows us to more clearly show the impact of the shifted radial anomaly oscillation $\delta\chi_r^S$ --- notice that the shifted geodesic sometimes moves faster, and sometimes slower, than the spinning-body orbit with which it is plotted.  The frequency shift $\Upsilon^S_r$ is exactly the same as for the equivalent aligned case except with $s$ replaced by $s_{\parallel}$.  The harmonic content of $\cos \theta$ is more complicated, exhibiting a beat between $\Upsilon_r$ and $\Upsilon_s$.  We also plot $u_t^S$ and $u_{\phi}^S$ alongside the corrections to the spinning body's energy $\delta E^S$ and orbital angular momentum $\delta L_z^S$ in the right panels of Fig.\ \ref{fig:exampleorbitseqprec}. 

Figure \ref{fig:residualplote} displays the convergence of an orbit with aligned spin, while Figure \ref{fig:residualplotemisaligned} shows the convergence of an orbit with misaligned spin, where both orbits have the same radial turning points.  We call the discrepancy between the exact result and our value for a certain $n_{\text{max}}$ the ``residuals". These residuals are normalized by the exact value of the quantity we are computing, so the values for $\Upsilon_r^S$, $u_{t, 0}^S$ and $u_{\phi, 0}^S$ are directly comparable. As $n_{\text{max}}$ increases, the residuals decrease and approach closer to the true value, as expected. The convergence trend is identical for both the aligned and misaligned cases, except for the highest value of $n_{\text{max}}$ for each of the different eccentricities. At this point, the working precision of the calculation is insufficient and the computation breaks down due to rounding error.

\section{Summary and future work}
\label{sec:summary}

In this work, we have studied equatorial and nearly equatorial orbits of spinning bodies around black holes in detail.  Such orbits reduce to equatorial ones when the orbiting body is non-spinning.  When the spin is aligned with the orbit, the motion is confined to the equatorial plane.  When the spin vector is misaligned, it precesses with Mino-time frequency $\Upsilon_s$, and the motion acquires a polar oscillation $\delta\vartheta_S$ whose magnitude is $\mathcal{O}(S)$.  The solution in this case appears to diverge on ``resonances,'' orbits for which the radial and spin frequencies combine to be commensurate with the polar oscillation frequency: $\hat\Upsilon_r + \Upsilon_s = \Upsilon_\theta$.  In fact, the amplitude of the driving force vanishes at such frequencies, and the system is well behaved, in keeping with past work which demonstrated that nothing ``interesting'' happens during spin-orbit resonances at least when considering the motion to leading order in spin \cite{Witzany2019_2,Zelenka2020}.  Sections \ref{sec:simpleorbits} and \ref{sec:slightlyecc} presented analytic descriptions of nearly equatorial orbits that are circular and slightly eccentric respectively.  In Sec.\ \ref{sec:spinbodyfreqdom}, we introduced a frequency-domain description of nearly equatorial orbits with arbitrary eccentricity. 

In a companion paper, we use this frequency-domain approach to describe completely fully generic orbits --- orbits that are both inclined and eccentric, with the small body's spin arbitrarily oriented \cite{Paper2}.  It is worth remarking that, for the nearly equatorial orbits we consider here, spinning-body orbits share the same radial turning points as some equatorial geodesic orbit.  For the nearly equatorial case, this ``reference geodesic'' which shares the orbit's turning points serves as a particularly convenient point of comparison in analyzing the spinning body's orbit.  This analysis becomes more complicated in the generic case, for which neither the polar nor the radial libration ranges coincide in general with those of a geodesic.  We can nonetheless define a ``reference geodesic'' whose turning points coincide with the spinning body's orbit in an orbit-averaged sense; details are given in Ref.\ \cite{Paper2}.  We use this framework to compute corrections arising from the small body's spin to the orbital frequencies $\Upsilon_r$ and $\Upsilon_\theta$ for generic orbits in Ref.\ \cite{Paper2}.  In addition, we present a detailed comparison between our approach and the methods presented in Ref.\ \cite{Witzany2019_2} for the case of equatorial, spin-aligned orbits in Appendix B of the companion paper \cite{Paper2}.


Results in Ref.\ \cite{Ruangsri2016} suggest that the behavior near resonance of terms which are quadratic in spin plays a critical role in the emergence of chaotic motion via the KAM theorem.  This is supported by Ref.\ \cite{Zelenka2020} which contains a detailed numerical study of the growth of resonances and chaos for spinning-body motion in a Schwarzschild spacetime.  By using the techniques discussed here to provide a very accurate formulation of the linear-in-spin aspect of spinning-body orbits, we plan to extend work in Ref.\ \cite{Ruangsri2016} by investigating the behaviour of the quadratic in spin terms in the frequency domain.  We hope this may clarify the precise manner in which nonlinear terms in the spinning-body equations of motion push such orbits from integrable to chaotic behavior in a Kerr background.

Another avenue for future work is to incorporate secondary spin into gravitational waveform models.  An osculating geodesic integrator \cite{Pound2008,Gair2011} can be used to generate spinning-body worldlines.  Any perturbed system of the form $Dp^{\alpha}/d\tau =\delta f^{\alpha}$ can be described using an osculating geodesic framework, so long as $\delta f^{\alpha}$ is sufficiently small.  In the EMRI limit we are interested in, both the spin-curvature force $f_S^{\alpha}$ and the self-force effects are small, so it should be possible to fold both into a forcing term and build a spinning-body inspiral.  Such a framework has been developed for Schwarzschild orbits, and is presented in Ref.\ \cite{Warburton2017}; we hope to use a similar approach to model completely generic spinning-body Kerr inspirals.  Ultimately, one hopes to build a fully self consistent self-force driven inpiral, and it is encouraging that the first steps have been taken in this direction \cite{mathews2021selfforce}.

\section*{Acknowledgements}

This work has been supported by NASA ATP Grant 80NSSC18K1091, and NSF Grant PHY-1707549 and PHY-2110384.  We are very grateful to Leo Stein and Sashwat Tanay for reading a draft of this paper and providing helpful comments; we are particularly grateful for comments regarding the possible impact of resonances in the low eccentricity limit, which helped us to uncover well-hidden typos in several equations. We are also very grateful to Vojt\v{e}ch Witzany for reading this manuscript and providing very helpful comments, and to Viktor Skoup\'{y} whose feedback and checks of our analysis uncovered a typographical error in one of our equations.

\appendix

\section{Explicit expression for the radial shift of a spinning body's orbit}
\label{sec:secularterms}

In this paper, we seek periodic solutions to the linear-in-spin Mathisson-Pappaptrou equations.  As outlined in Sec.\ \ref{sec:spinbodyfreqdom}, we characterize the radial coordinate of spinning-body orbits using the parameterization
\begin{equation}
r(\lambda)=\frac{pM}{1+e\cos(w_r+\delta \hat{\chi}_r(w_r)+\delta \chi^S_r)}\;,\label{rparamagain}
\end{equation}
where 
\begin{align}
\delta \hat{\chi}_{r}(w_r) &=\sum_{n=-\infty}^{\infty}\delta \hat{\chi}_{r,n} e^{-in w_r}\;, \label{eq:deltahatchi2}\\
\delta \chi^S_{r}& =\sum_{n=-\infty}^{\infty}\delta \chi^S_{r,n} e^{-in w_r}\;, \label{eq:deltachiS2}
\end{align}
and where 
\begin{equation}
w_r=(\hat{\Upsilon}_r+\Upsilon_r^S)\lambda\;.
\label{eq:wr2}
\end{equation}
The quantities written with hat accents, $\hat\Upsilon_r$ and $\delta\hat\chi_r$, are computed using geodesic quantities --- $\hat\Upsilon_r$ is the Mino-time radial frequency for the geodesic with semi-latus rectum $p$ and eccentricity $e$, and $\delta\hat\chi_r$ describes the oscillating contribution to the true anomaly for that geodesic.  The quantities $\delta \chi^S_r$ and $\Upsilon_r^S$ are both $\mathcal{O}(S)$.

Although $\delta\hat\chi_r$ is computed using geodesic quantities, notice that as implemented in this formula we include $\mathcal{O}(S)$ terms in it via the mean anomaly angle $w_r$.  The Fourier coefficients $\delta \hat{\chi}_{r,n}$ are identical to those for a geodesic orbit, but the angle $w_r$ in the exponent of Eq.\ (\ref{eq:deltahatchi2}) includes an $\mathcal{O}(S)$-term associated with the impact of the small body's spin on the orbit, $\Upsilon_r^S$.  This takes into account the fact that the spinning-body orbit's frequencies are shifted by $\Upsilon_r^S$ from those of the geodesic which shares its radial turning points.

Our goal in this Appendix is examine how the spinning body's orbit is shifted from the trajectory of the geodesic which shares the same turning points.  To expedite this comparison, in this Appendix we write the function $\delta\hat{\chi}_r$ with an argument of either $w_r$ or $\hat{w}_r$, where $\hat{w}_r=\hat{\Upsilon}_r\lambda$.  When we use $\delta\hat{\chi}_r(w_r)$, this is the function which parameterizes in part the true anomaly of a spinning body's orbit.  This function's form is given explicitly by Eq.\ (\ref{eq:deltahatchi2}); it oscillates in phase with the radial motion $r(\lambda)$ of the spinning body.  On the other hand, $\delta\hat{\chi}_r(\hat{w}_r)$ is the function that appears in the parameterization (\ref{eq:geodesicorbitagain}) of the geodesic orbit.  It is identical to the form in Eq.\ (\ref{eq:deltahatchi2}) except with $w_r\rightarrow\hat{w}_r$.  It oscillates in phase with the radial motion $\hat{r}(\lambda)$ of the geodesic orbit.

As discussed in Sec.\ \ref{sec:SpinDev}, we can define the difference between the spinning body's orbit and that of the geodesic which shares its turning points as follows:
\begin{equation}
\delta r_S(\lambda) \equiv r(\lambda) - \hat{r}(\lambda)\;, \label{eq:deltarS}
\end{equation}
where $r(\lambda)$ describes the radial motion of a spinning body's orbit, and $\hat{r}(\lambda)$ describes the radial motion of the geodesic which shares its radial turning points.  Note that $\delta r_S(\lambda)=\mathcal{O}(S)$.

We expect $\delta r_S(\lambda)$ to contain secularly growing terms due to the difference in frequencies between the geodesic and the spinning body's motion.  For the parameterization defined in Eq.\ (\ref{rparamagain}), the explicit expressions for $\hat{r}(\lambda)$ and $\delta r_S(\lambda)$ are:
\begin{equation}
    \hat{r}(\lambda)=\frac{pM}{1+e\cos \left(\hat{w}_r+\delta \hat{\chi}_r(\hat{w}_r)\right)}\;
    \label{eq:geodesicorbitagain}
\end{equation}
and 
\begin{align}
    \delta r_S(\lambda)&=pMe\frac{\Upsilon_r^S\lambda\left(1-i\sum_{n}n\delta \hat{\chi}_{r,n} e^{-in\hat{w}_r}\right)+\delta \chi^S_r}{\left(1+e\cos \left(\hat{w}_r+\delta \hat{\chi}_r(\hat{w}_r)\right)\right)^2}\nonumber\\
    &\times \sin\left(\hat{w}_r+\delta\hat{\chi}_r(\hat{w}_r)\right)\;.  \label{eq:secular}
\end{align}
The secular growth of $\delta r_S$ apparent in Eq.\ (\ref{eq:secular}) is a somewhat troublesome mathematical artefact of the fact that we are comparing two integrable systems that have slightly different frequencies.  This is troublesome because we would like to think of the spinning body's orbit as ``close to'' the geodesic which shares its turning points.  Though this describes the behavior of $\delta r_S$ for small $\lambda$, this quantity evolves such that it eventually cannot be regarded as a perturbation.

To address this, we compare the two solutions in such a way that we avoid secularly growing terms, following a Poincare-Lindstedt-type approach.  We begin by shifting the frequency of the geodesic solution so that it matches the frequency of the spinning-body orbit.  Let us define
\begin{equation}
    \hat r_{\rm shift}(\lambda) = \frac{pM}{1+e\cos \left(w_r+\delta \hat{\chi}_r(w_r)\right)}\;.
    \label{eq:geodesicorbitshifted}
\end{equation}
This is just Eq.\ (\ref{eq:geodesicorbitagain}), but with the geodesic mean anomaly $\hat w_r = \hat\Upsilon_r\lambda$ replaced by the mean anomaly $w_r = (\hat\Upsilon_r + \Upsilon^S_r)\lambda$.  We then define
\begin{equation}
    \delta r^{\rm shift}_S(\lambda) = r(\lambda) - \hat r_{\rm shift}(\lambda)\;,
    \label{eq:deltarSshift1}
\end{equation}
where again $r(\lambda)$ describes the radial motion of a spinning body's orbit.  We introduce the superscript label ``shift'' to distinguish this quantity from that introduced in Eq.\ (\ref{eq:deltarS}), noting that its frequency is shifted from the geodesic frequency.  Using Eqs.\ (\ref{rparamagain}) and (\ref{eq:geodesicorbitshifted}), we find
\begin{equation}
\delta r_S^{\rm shift}(\lambda) = pMe\frac{\delta \chi_r^S(w_r)\sin\left[w_r+\delta\hat{\chi}_r(w_r)\right]}{\left(1+e\cos\left[w_r+\delta\hat{\chi}_r(w_r)\right]\right)^2}\;.
\label{eq:deltarSshift}
\end{equation}
The quantity $\delta r_S^{\rm shift}(\lambda)$ does not grow secularly, but is instead periodic at the radial period $\Lambda_r = 2\pi/\left(\hat{\Upsilon}_r + \Upsilon_r^S\right)$.  We can use Fourier expansions quite naturally to describe $\delta r_S^{\rm shift}(\lambda)$ which is advantageous for the frequency-domain approach we use in this paper.

\section{Comparison with Saijo et al., 1998: Aligned spin, equatorial orbits}

\label{sec:Saijocomparison}

Considerable work has been done previously on equatorial orbits with aligned spin.  Almost all such work uses the equations of motion describing a spinning body confined to the equatorial plane that were derived by Saijo et al.\ \cite{Saijo1998}.  Saijo et al.\ use the conserved quantities $E^S$, $L^S$, $S^2=S^{\alpha}S_{\alpha}$ and $-\mu^2=p^{\alpha}p_{\alpha}$ in order to derive these equations; their full derivation is in Ref.\ \cite{Saijo1998} (see also Refs.\ \cite{Hackmann2014} and \cite{1976Tod} for similar related discussion).  We present the equations for Kerr spacetime below in Eqs.\ (\ref{eq:Saijo}) -- (\ref{eq:Lambdas}). 

The radial component of Eq.\ (\ref{eq:mp1}), taking the limit of a body confined to an equatorial orbit with aligned spin, can be written
\begin{align}
\Sigma_s\Lambda_s\frac{dr}{d\tau}=\pm\sqrt{R_s}\;,\label{eq:Saijo}
\end{align}
where 
\begin{align}
R_s&=P_s^2-\Delta \left(\frac{\Sigma_s^2}{r^2}+\left[L_z^S-(a+s_\parallel\mu )E^S\right]^2\right)\;,\\
P_s&=\left[(r^2+a^2)+as_\parallel\mu \left(1+\frac{M}{r}\right)\right]E^S \nonumber\\
&-\left(a+\frac{s_\parallel\mu M}{r}\right)L^S_z\;,\\
\Sigma_s&=r^2\left( 1-\frac{ s_\parallel^2\mu^2M }{r^3} \right)\;, \\
\Lambda_s&=1-\frac{3 s_\parallel^2\mu^2M r\left[L^S_z-(a+s_\parallel\mu )E^S\right]^2}{\Sigma_s^3}\; \label{eq:Lambdas}.
\end{align}
We begin our discussion with the Schwarzschild limit, for which we find particularly compact and convenient expressions. 

\subsection{Schwarzschild spacetime}
\label{sec:schwSaijo}

Linearizing in the small body's spin, Eq.\ (\ref{eq:Saijo}) reduces to a simple form, as presented in Appendix B.3 of Ref.\ \cite{Favata2011}.  We reproduce the result here in our notation, noting that our parameter $s_\parallel$ is dimensionless, and so differs from the correspond spin parameter used in Ref.\ \cite{Favata2011} by a factor of $\mu$:
\begin{align}
\left( \frac{dr}{d\tau}\right)^2 &= (E^S)^2-\left(V^{\rm Schw}_{\rm eff}\right)^2\nonumber\\&+2s_\parallel \mu\frac{\hat E \hat L_z}{r^2}\left(1-\frac{3M}{r}\right)+\mathcal{O}(S^2)\;,\label{eq:drdtauSchw}
\end{align}
where $V^{\rm Schw}_{\rm eff}$ is the usual effective potential for the Schwarzschild metric, but using the angular momentum for a spinning-body orbit:
\begin{align}
\left(V^{\rm Schw}_{\rm eff}\right)^2=\left(1-\frac{2M}{r}\right)\left(1+\frac{(L^S_z)^2}{r^2}\right)\;.
\end{align}
Equation (\ref{eq:drdtauSchw}) is Eq.\ (B14) of Ref.\ \cite{Favata2011}, adapted to our notation and linearizing in spin.

\subsubsection{Circular equatorial orbits}
\label{sec:schwSaijocirc}

To find the energy and angular momentum corresponding for a body in circular orbit with its spin aligned with the orbit, begin by requiring $dr/d\tau = 0$.  This yields a quadratic equation for $E^S$ whose solution to linear order in $s$ is
\begin{align}
(E^S)^2 &= (V^{\rm Schw}_{\rm eff}) - s_\parallel\mu\frac{\hat L_z}{r^2}\left(1 - \frac{3M}{r}\right)
\nonumber\\
&\equiv V^{\rm Schw,spin}_{\rm eff}\;.
\end{align}
Further requiring $\partial V^{\rm Schw,spin}_{\rm eff}/\partial r = 0$ yields the solutions
\begin{align}
E^S & = \frac{r - 2M}{\sqrt{r(r - 3M})} - \frac{s_\parallel\mu}{2r}\left(\frac{M}{r - 3M}\right)^{3/2}\;,\label{eq:EscircSchw}
\\
L_z^S &= \frac{r\sqrt{M}}{\sqrt{r - 3M}} + \frac{s_\parallel\mu}{2}\frac{(r - 2M)(2r - 9M)}{\sqrt{r}(r - 3M)^{3/2}}\;\label{eq:LSscircSchw}.
\end{align}
These expressions match exactly with Eq.\ (\ref{eq:deltaEScirceqaligned}) and (\ref{eq:deltaLzScirceqaligned}) in the limit $a=0$; these expressions can also be found\footnote{Note that Eq.\ (B18) in Ref.\ \cite{Favata2011} contains a typographical error in the denominator; the $r-2m_2$ should be $r-3m_2$.} in Eq.\ (B17) and (B18) of Ref.\ \cite{Favata2011}. Similarly, the expressions in Eqs.\ (54) and (55) of Ref.\ \cite{Hackmann2014} reduce to (\ref{eq:EscircSchw}) in the first order in spin limit.

\subsubsection{Eccentric equatorial orbits}
\label{sec:schwSaijoecc}

Next we consider eccentric equatorial orbits.  We begin again with Eq.\ (\ref{eq:drdtauSchw}), but now multiply by $\Sigma^2 = r^4$, using $d/d\lambda = \Sigma\,d/d\tau$ to change into an expression for $\left(dr/d\lambda\right)^2$:
\begin{align}
\left( \frac{dr}{d\lambda}\right)^2&=r^4(E^S)^2-r\left(r-2M\right)\left(r^2+(L^S_z)^2\right)
\nonumber\\
&+2s_\parallel\mu r {\hat E \hat L_z} \left(r-3M\right)+{\mathcal{O}(S^2)}
\nonumber\\
&\equiv R^{\rm Schw}_s(r)\;.\label{eq:drdlambdaSchw}
\end{align}
With this formulation of Eq.\ (\ref{eq:drdtauSchw}), we can straightforwardly compute $\Upsilon^S_r$, $\delta E^S$ and $\delta L^S$ and compare with results we obtain elsewhere in this work.

We begin by substituting $E^{S}=\hat E + \delta E^S$, $L_{z}^{S}=\hat{L}_z + \delta L_z^S$, with $\delta E^S$ and $\delta L_z^S$ both $\mathcal{O}(S)$, into Eq.\ (\ref{eq:drdlambdaSchw}) and expand to first order in spin, yielding
\begin{align}
R^{\rm Schw}_s(r) = R^{\rm Schw}(r) + \delta R^{\rm Schw}_s(r) + \mathcal{O}(S^2)\;,
\end{align}
where 
\begin{align}
R^{\rm Schw}(r) &= r^4 \hat{E} - r(r - 2M)(r^2 + \hat{L}_z)\;,
\\
\delta R^{\rm Schw}_s(r) &= 2r\left[s_\parallel\mu(r - 3M)\hat{E}\hat{L}_z + r^3\hat{E}\delta E^S\right.
\nonumber\\
&\left. -(r - 2M)\hat{L}_z\delta L_z^S\right]\;.
\end{align}
Using $dr/d\lambda = 0$ at the turning points $r = pM/(1 \pm e)$ yields the well-known results
\begin{align}
    \hat{E} = \sqrt{\frac{(p - 2)^2 - 4e^2}{p(p - 3 - e^2)}}\;, \ \ \hat{L}_z = \frac{pM}{\sqrt{p - 3 - e^2}}\label{eq:hatEhatLSchw}
\end{align}
describing these orbit integrals for Schwarzschild geodesics.  Requiring that $r = pM/(1 \pm e)$ remaining turning points for the spinning bodies orbit, we require $\delta R^{\rm Schw}_s = 0$ at these points as well.  This yields
\begin{align}
\delta E^S &= -\frac{s_\parallel\mu}{M}\frac{(1 - e^2)^2}{2p(p - 3 - e^2)^{3/2}}\;,
\label{eq:ESexactineSchw}\\
\delta L_z^S &= s_\parallel\mu\frac{(2p - 9 - 3e^2)\sqrt{(p - 2)^2 - 4e^2}}{2p^{1/2}(p - 3 - e^2)^{3/2}}\;.
\label{eq:LzSexactineSchw}
\end{align}
These expressions are identical to those in Eqs.\ (46) of Ref.\ \cite{Mukherjee2019} with $r_a = pM/(1 - e)$ and $r_p = pM/(1 + e)$.

Next, we use Eq.\ (3) in Ref.\ \cite{FujitaHikida2009} to calculate $\Lambda_r$, but using $R^{\rm Schw}_s(r)$ as defined in Eq.\ (\ref{eq:drdlambdaSchw}):
\begin{align}
\Lambda_r = 2\int_{r_{min}}^{r_{max}}\frac{dr}{\sqrt{R^{\rm Schw}_s(r)}}\;,\label{eq:lambdar1}
\end{align}
with
\begin{align}
r_{min} = \frac{pM}{1 + e}\;,\ \ r_{max} = \frac{pM}{1 - e}\;.
\end{align}
Using the parameterization of radial motion defined by Eq.\ (\ref{eq:rparam}), we turn equation (\ref{eq:lambdar1}) into an integral over $\chi_r$:
\begin{align}
\Lambda_r = 2\int_{0}^{\pi}\frac{1}{\sqrt{R^{\rm Schw}_s(\chi_r)}}\frac{dr}{d\chi_r}d\chi_r\;,
\label{eq:lambdar2}
\end{align}
where 
\begin{align}
r = \frac{pM}{1 + e\cos{\chi_r}}\;,\ \ \frac{dr}{d\chi_r} = \frac{peM\sin{\chi_r}}{1 + e\cos\chi_r}\;.
\end{align}
Noting that $\Lambda_r = \hat{\Lambda}_r+ \Lambda^S_r$, we break this integral into geodesic and $\mathcal{O}(S)$ pieces:
\begin{align}
\hat{\Lambda}_r &= 2\int_{0}^{\pi}\frac{1}{\sqrt{R^{\rm Schw}(\chi_r)}} \frac{dr}{d\chi_r} d\chi_r\;,
\label{eq:lambdarG}\\
\Lambda^S_r & = -\int_{0}^{\pi}\frac{\delta R^{\rm Schw}_s(\chi_r)} {R^{\rm Schw}(\chi_r)^{3/2}} \frac{dr}{d\chi_r}d\chi_r
\;.\label{eq:lambdarS}
\end{align}
The definitions $\Upsilon_r = 2\pi/\Lambda_r$ and $\Upsilon_r = \hat{\Upsilon}_r+ \Upsilon^S_r$ yielding
\begin{align}
   \hat{\Upsilon}_r=\frac{2\pi}{\hat{\Lambda}_r}\;,\ \ \Upsilon^S_r=-\frac{2\pi \Lambda^S_r}{\hat{\Lambda}_r^2}\;.
\end{align}
This allows us to at last evaluate $\Upsilon^S_r$ as a simple quadrature:
\begin{equation}
\Upsilon^S_r = \frac{2\pi}{\hat{\Lambda}_r^2}\int_{0}^{\pi}\frac{\delta R^{\rm Schw}_s(r)}{R^{\rm Schw}(r)^{3/2}} \frac{dr}{d\chi_r}d\chi_r\;,
\end{equation}
which we write explicitly as
\begin{align}
&\Upsilon^S_r = -\frac{2\pi s_\parallel\mu}{\hat{\Lambda}_r^2M^2}
\nonumber\\
&\quad\int_{0}^{\pi}\frac{(e^2 - 3 - 2e\cos\chi_r)\sqrt{(p - 2)^2 - 4e^2}}{p\sqrt{p - 3 - e^2}(p - 6 - 2e\cos\chi_r)^{3/2}}d\chi_r\;.
\label{eq:UpsilonrsexactineSchw}
\end{align}
Equations (\ref{eq:ESexactineSchw}), (\ref{eq:LzSexactineSchw}) and (\ref{eq:UpsilonrsexactineSchw}) expanded to second-order in eccentricity, yield expressions that match Eqs.\ (\ref{eq:deltaEs2ndine}), (\ref{eq:deltaLzs2ndine}) and (\ref{eq:UpsilonrS2ndine}). 

\subsection{Kerr spacetime}
\label{sec:kerrSaijo}

We now consider Eq.\ (\ref{eq:Saijo}) to leading order in spin, but for general Kerr parameter $a$:
\begin{align}
\left( \frac{dr}{d\lambda}\right)^2 &= [E^S(r^2 + a^2) - aL^S_z]^2
\nonumber\\
&-\Delta[r^2 + (L_z^S - aE^S)^2]
\nonumber\\
&+2as_\parallel\mu M\frac{\left[{\hat L_z}^2 - 2 a {\hat E}{\hat L_z} + a^2{\hat E}^2\right]}{r}
\nonumber\\
&+2s_\parallel\mu r {\hat E}\left[{\hat L_z}(r - 3M) + 3M a {\hat E}\right] + \mathcal{O}(S^2)
\nonumber\\
&\equiv R^{\rm Kerr}_s(r)\;.\label{eq:drdlambdaKerr}
\end{align}

\subsubsection{Circular equatorial orbits}
\label{sec:kerrSaijocirc}

To compute the energy and axial angular momentum of a spinning body in an aligned circular Kerr orbit, we need to find $E^S$ and $L_z^S$ such that $R^{\text{Kerr}}_s(r) = 0$ and $\partial R^{\text{Kerr}}_s(r)/\partial r=0$.  This gives expressions that match Eqs.\ (\ref{eq:deltaEScirceqaligned}) and (\ref{eq:deltaLzScirceqaligned}).  Hackmann et al.\ also have expressions for $E^S$ and $L_z^S$ that are exact in spin for general $a$; compare Eqs.\ (48) and (49) of Ref.\ \cite{Hackmann2014}.  Piovano et al.\ likewise provide $E^S$ and $L_z^S$ in slightly different notation; compare Eqs.\ (59) and (60) of Ref.\ \cite{Piovano2020_2}. 

\subsubsection{Eccentric equatorial orbits}
\label{sec:kerrSaijoecc}

As in our Schwarzschild analysis, we insert $E^{S} = \hat E + \delta E^S$, $L_{z}^{S} = \hat{L}_z + \delta L_z^S$ into Eq.\ (\ref{eq:drdlambdaKerr}) and expand to first order in spin, yielding
\begin{align}
R^{\rm Kerr}_s(r) = R^{\rm Kerr}(r)+\delta R^{\rm Kerr}_s(r) + \mathcal{O}(S^2)\;,
\end{align}
where 
\begin{align}
R^{\rm Kerr}(r) &= [\hat{E}(r^2 + a^2) - a\hat{L}_z]^2 - \Delta[r^2 + (\hat{L}_z - a\hat{E})^2]\;,
\end{align}
and where
\begin{align}
\delta R^{\rm Kerr}_s &= 2\biggl\{as_\parallel\mu M\frac{\left[{\hat L_z}^2 - 2 a {\hat E}{\hat L_z} + a^2{\hat E}^2\right]}{r}
\nonumber\\
&+ s_\parallel\mu r {\hat E}\left[{\hat L_z}(r - 3M) + 3aM {\hat E}\right]
\nonumber\\
& +ar \left[2M\left(\hat{E}\delta L^S + \hat{L}_z\delta E^S\right) - a\hat{E} \delta E^S(r + 2M) \right]
\nonumber\\
&+r \left[\hat{L}_z\delta L^S (r - 2M) - \hat{E}\delta E^S r^3 \right]\biggr\}\;.
\end{align}
Note that $R^{\rm Kerr}(r)$ is given by Eq.\ (\ref{eq:geodr}) with $\hat{Q} \rightarrow 0$.  Expressions for $\hat{E}$ and $\hat{L}_z$ which are exact in eccentricity are given in Eqs.\ (A.1) and (A.2) of Ref.\ \cite{vandeMeent2019}.

As in the Schwarzschild analysis, we solve for $\delta E^S$ and $\delta L_z^S$ by requiring $\delta R^{\rm Kerr}_s = 0$ at $r = pM/(1 \pm e)$.  This yields closed-form expressions for $\delta E^S(p,e)$ and  $\delta L_z^S(p,e)$ analogous to Eqs.\ (\ref{eq:ESexactineSchw}) and (\ref{eq:LzSexactineSchw}) which apply for general $a$, but are quite lengthy and cumbersome.  We refer the reader to Eqs.\ (81) and (83) of Ref.\ \cite{Mukherjee2019} for expressions for $E^S$ and $L_z^S$ to first order in small body spin derived by Mukherjee et al., as well as to Eqs.\ (38) and (39) of Ref.\ \cite{Skoupy2021} for exact-in-$S$ expressions for $E^S$ and $L_z^S$ derived by Skoup\'{y}  et al.  Both Skoup\'{y}  et al.\ and Mukherjee et al.\ write their expressions in terms of $r_a = pM/(1 - e)$ and $r_p = pM/(1 + e)$.  To first order in $e$, the results for $\delta E^S$ and $\delta L_z^S$ reduce to Eqs.\ (\ref{eq:deltaEScirceqaligned}) -- (\ref{eq:deltaLzScirceqaligned}), but with $v = \sqrt{1/p}$.
 
We evaluate $\Upsilon^S_r$ using a formulation analogous to what was done in Sec.\ \ref{sec:schwSaijo}, replacing the Schwarzschild function $R^{\rm Schw}_s(r)$ with $R^{\rm Kerr}_s(r)$:
\begin{equation}
\Upsilon^S_r = \frac{2\pi}{\hat{\Lambda}_r^2}\int_{0}^{\pi}\frac{\delta R^{\rm Kerr}_s(r)}{R^{\rm Kerr}(r)^{3/2}}\frac{dr}{d\chi_r}d\chi_r\;.
\label{eq:UpsilonrSKerrexactine}
\end{equation}
Expanded to first order in eccentricity, this reproduces Eq.\ (\ref{eq:kerrlinecc_UpsilonS_r}).

\begin{widetext}
\section{Explicit frequency-domain expressions}
\subsection{Coefficient functions}
\label{sec:coefficientfunctions}

In Sec.\ \ref{sec:eqplanealign}, we examine spinning-body motion in the equatorial plane using a frequency-domain expansion.  We linearize the radial component of the first Matthisson-Papapetrou equation (\ref{eq:mp1linear}) in small-body spin and re-express it in terms of quantities which are unknown (i.e., $\delta\chi_r^S$, $\Upsilon_r^S$, $u_{t,0}^S$ and $u_{\phi,0}^S$) and Fourier coefficients of functions along geodesics (i.e., $\mathcal{F}_r$, $\mathcal{G}_r$, $\mathcal{H}_r$, $\mathcal{I}_{1r}$, $\mathcal{I}_{2}$, $\mathcal{I}_{3}$ and $\mathcal{J}$).  This yields Eq.\ (\ref{eq:radiallinMP}).

We also linearize the constraint $u^\alpha u_\alpha = -1$ in small-body spin, writing down the corresponding equation (\ref{eq:udotucoeff}) in terms of the same set of unknowns as well as coefficients $\mathcal{K}_r$, $\mathcal{M}_r$, $\mathcal{N}_{1r}$, $\mathcal{N}_2$, $\mathcal{N}_3$ and $\mathcal{P}$ that likewise arise from known geodesics.  We follow a similar procedure in Sec.\ \ref{sec:ecceqprecess} to compute the nearly equatorial motion of a precessing spinning body.  In this case, we also linearize the $\theta$-component of Eq.\ (\ref{eq:mp1linear}) in small-body spin, obtaining Eq.\ (\ref{eq:polarlinMP2}).  In the nearly equatorial limit, the only non-zero coefficients in this equation are $\mathcal{Q}_{\vartheta}$, $T_{\vartheta}$ and $\mathcal{V}$.

In this Appendix, we provide explicit expressions for the Schwarzschild case of the various functions which we then expand in the Fourier domain.  These expressions for the functions appearing in Eq.\ (\ref{eq:radiallinMP}) are given by:
\begin{align}
\mathcal{F}_r(\lambda)&=\frac{e p \sin \hat{\chi}_r}{(1+e \cos \hat{\chi}_r)^2}\;,\ \ \
\mathcal{G}_r(\lambda)=\frac{e p \left(\delta\hat{\chi}_r'(\lambda )+\hat{\Upsilon}_r\right) (e (\cos (2\hat{\chi}_r)+3)-2 (p-2) \cos \hat{\chi}_r)}{(1+e \cos \hat{\chi}_r)^2 (2 e \cos \hat{\chi}_r-p+2)}\;,\\
\mathcal{H}_r(\lambda)&=-\frac{e p}{4 (e \cos \hat{\chi}_r+1)^3 (p-2-2 e \cos \hat{\chi}_r)^2}
\biggl\{-2 \sin \hat{\chi}_r \biggl[ 2 \hat{L}_z^2 \left(e^2 (p-3)-(p-2)^2\right)\nonumber\\
&+\hat{\Upsilon}_r^2 \left(e^2 (15-6 p)-2 (p-2)^2\right)+4 \hat{E}^2 (p-3) p^2\biggr]
+e \biggl[2 \sin (2\hat{\chi}_r) \biggl(\hat{L}_z^2 \left(2 e^2+p^2-8 p+12\right)\nonumber\\&-
\hat{\Upsilon}_r^2 \left(7 e^2+(p-2) p\right)+6 \hat{E}^2 p^2\biggr)+e \biggl(e \left(2 \hat{L}_z^2-\hat{\Upsilon}_r^2\right) \sin (4 \hat{\chi}_r)+\sin (3 \hat{\chi}_r) \left(2 (2 p-3) \hat{\Upsilon}_r^2-4 \hat{L}_z^2 (p-3)\right)\biggr)\biggr]\nonumber\\
&-2 \delta\hat{\chi}_r'(\lambda ) \sin \hat{\chi}_r \left(\delta\hat{\chi}_r'(\lambda )+2 \hat{\Upsilon}_r\right) \biggl[e^2 (e \cos (3 \hat{\chi}_r)+(6-4 p) \cos (2\hat{\chi}_r))+e \left(15 e^2+2 (p-2) p\right) \cos \hat{\chi}_r\nonumber\\&-2 \left(e^2 (4 p-9)+(p-2)^2\right)\biggr]+2 \delta\hat{\chi}_r''(\lambda ) (e (\cos (2\hat{\chi}_r)-3)-2 \cos \hat{\chi}_r) (p-2-2 e \cos \hat{\chi}_r)^2 \biggr\}\;,\\
\mathcal{I}_{1r}(\lambda)&=\frac{e p \left(\Xi_2(\lambda) \sin \hat{\chi}_r\right)}{(e \cos \hat{\chi}_r+1)^2}+\frac{e p \left(\Xi_1(\lambda)+1\right) \left(\delta\hat{\chi}_r'(\lambda )+\hat{\Upsilon}_r\right) (e (\cos (2\hat{\chi}_r)+3)-2 (p-2) \cos \hat{\chi}_r)}{(e \cos \hat{\chi}_r+1)^2 (2 e \cos \hat{\chi}_r-p+2)}\;,\\
\mathcal{I}_2(\lambda)&=-\frac{2 \hat{E} p^3}{(e \cos \hat{\chi}_r+1)^2 (p-2-2 e \cos \hat{\chi}_r)}\;,\ \ \
\mathcal{I}_3(\lambda)=\hat{L}_z \left(4-\frac{2 p}{e \cos \hat{\chi}_r+1}\right)\;,\\
\mathcal{J}(\lambda)&=\frac{3\hat E \hat L_z S_\theta(1+e \cos \hat{\chi}_r)}{p}-\frac{2 \hat{E} p^3 \delta u^S_t(\lambda )}{(1+e \cos \hat{\chi}_r)^2 (p-2-2 e \cos \hat{\chi}_r)}
-\frac{2 \hat{L}_z \delta u^S_\phi(\lambda )(p-2-2 e \cos \hat{\chi}_r)}{(1+e \cos \hat{\chi}_r)}\;,
\end{align}
where we have defined
\begin{align}
    \Xi_1(\lambda)=-i \sum_{n=-n_\text{max}}^{n_\text{max}} n\delta\hat\chi_{r,n} e^{-i n \hat\Upsilon_r \lambda}\;,\ \ \ \text{and}\ \ \  \Xi_2(\lambda)=- 2\hat\Upsilon_r\sum_{n=-n_\text{max}}^{n_\text{max}} n^2\delta\hat\chi_{r,n} e^{-i n \hat\Upsilon_r \lambda}\;.
\end{align}
Here $\Xi_1(\lambda)$ and $\Xi_2(\lambda)$ are functions that depend on the Fourier coefficients of geodesic radial true anomaly $\delta \hat\chi_r$.  We also write down the expressions for the functions which appear in Eq.\ (\ref{eq:udotucoeff}) explicitly, again limiting ourselves here to the Schwarzschild limit:
\begin{align}
\mathcal{K}_r(\lambda)&=\frac{2 e^2 \sin ^2\hat{\chi}_r \left(\delta\hat{\chi}_r'(\lambda )+\hat{\Upsilon}_r\right)}{p (p-2-2 e \cos \hat{\chi}_r)}\;,\\
\mathcal{M}_r(\lambda)&=\frac{e}{2 p^2 (p-2-2 e \cos \hat{\chi}_r)^2} \biggl\{ \sin \hat{\chi}_r \left(4 \hat{L}_z^2 \left(e^2 (p-3)-(p-2)^2\right)-5 e^2 p \hat{\Upsilon}_r^2+4 \hat{E}^2 p^3\right)\nonumber\\
&+e \biggl[e \left(\sin (3 \hat{\chi}_r) \left(4 \hat{L}_z^2 (p-3)-p \hat{\Upsilon}_r^2\right)-2 e \hat{L}_z^2 \sin (4 \hat{\chi}_r)\right)-2 \sin (2\hat{\chi}_r) \left(\hat{L}_z^2 \left(2 e^2+p^2-8 p+12\right)-(p-2) p \hat{\Upsilon}_r^2\right)\biggr]\nonumber\\
&-2 e p \delta\hat{\chi}_r'(\lambda ) \sin \hat{\chi}_r \left(\delta\hat{\chi}_r'(\lambda )+2 \hat{\Upsilon}_r\right) (e (\cos (2\hat{\chi}_r)+3)-2 (p-2) \cos \hat{\chi}_r)\biggr\}\;, \\
\mathcal{N}_{1r}(\lambda)&=\frac{2 e^2 (\Xi_1(\lambda)+1) \left(\delta\hat{\chi}_r'(\lambda )+\hat{\Upsilon}_r\right)\sin ^2\hat{\chi}_r}{p (p-2-2 e \cos \hat{\chi}_r)}\;,\ \ \
\mathcal{N}_2(\lambda)=\frac{2p \hat{E} }{p-2-2 e \cos \hat{\chi}_r}\;, \\
\mathcal{N}_3(\lambda)&=\frac{2 \hat{L}_z (1+e \cos\hat{\chi}_r)^2}{p^2}\;, \ \ \ \text{and} \ \ \
\mathcal{P}(\lambda)=\frac{2p \hat{E}  \delta u^S_t(\lambda )}{p-2-2 e \cos \hat{\chi}_r}+\frac{2 \hat L_z \delta u_\phi^S(\lambda)\left(1+e\cos\hat \chi_r\right)^2}{p^2}\;.
\end{align}
In the nearly equatorial limit, the only non-zero functions which appear in Eq.\ (\ref{eq:polarlinMP2}) are
\begin{align}
\mathcal{Q}_{\vartheta}(\lambda)&=1\;, \ \ \  T_{\vartheta}(\lambda)=\hat L_z^2\;, \label{eq:QT} \\
\mathcal{V}(\lambda)&=\frac{3\hat L_z (1+e\cos\hat\chi_r)\left(S_r\hat E(p-2-2 e \cos \hat{\chi}_r)+e S_t (\hat \Upsilon_r +\delta\hat\chi_r'(\lambda))\sin\hat\chi_r\right)}{p(p-2-2e\cos\hat\chi_r)}\;. \label{eq:ScriptV}
\end{align}
In the Supplemental Material accompanying this paper, we include a \textit{Mathematica} notebook which computes these expressions for general Kerr (i.e., for $a \ne 0)$ \cite{SupplementalMaterial}.

\subsection{Matrix System}
\label{sec:matrixsystem}

As discussed in Sec.\ \ref{sec:eqplanealign}, our procedure to solve for the spinning body's orbit in the frequency domain involves writing the functions $\mathcal{F}_r$, $\mathcal{G}_r$, $\mathcal{H}_r$, $\mathcal{I}_{1r}$, $\mathcal{I}_{2}$, $\mathcal{I}_{3}$ $\mathcal{J}$, $\mathcal{K}_r$, $\mathcal{M}_r$, $\mathcal{N}_{1r}$, $\mathcal{N}_2$, $\mathcal{N}_{3}$ and $\mathcal{P}$ as Fourier expansions of the form
\begin{align}
f(\lambda) & =\sum_{n=-n_\text{max}}^{n_\text{max}}f_n e^{-in \hat\Upsilon_r \lambda} \label{eq:radialexpagain}\;.
\end{align}
We similarly express the unknown function $\delta \chi_r^S(\lambda)$ as a Fourier expansion,
\begin{align}
\delta \chi_r^S(\lambda) & =\sum_{n=-n_\text{max}}^{n_\text{max}}\delta \chi_{r,n}^S e^{-in \hat\Upsilon_r \lambda} \label{eq:deltachirsexpagain}\;,
\end{align}
aiming to solve for its Fourier coefficients $\delta \chi_{r,n}^S$.

To do so, we insert expansions (\ref{eq:radialexpagain}) and (\ref{eq:deltachirsexpagain}) into Eqs.\ (\ref{eq:radiallinMP}) and (\ref{eq:udotucoeff}).  This yields a system of linear equations in the frequency-domain which allows us to solve for the unknown variables $\delta\chi_r^S$, $\Upsilon_r^S$, $u_{\phi,0}^S$ and $u_{t,0}^S$. This system of equations can be written
\begin{equation}
   \mathbf{M}\cdot\mathbf{v}+\mathbf{c}=0\;.\label{eq:matrixeq}
\end{equation}

To get a sense of the character of this system of equations, for the choice $n_\text{max}=1$, the matrix $\mathbf{M}$ and vectors $\mathbf{v}$ and $\mathbf{c}$ are given explicitly by
\begin{align}
 \mathbf{M}&=   \left(
\begin{array}{ccccc}
-\mathcal{F}_{r,0} \hat{\Upsilon}_r^2-i \mathcal{G}_{r,0}\hat{\Upsilon}_r- \mathcal{H}_{r,0} & 0 & \mathcal{I}_{1r,1} & \mathcal{I}_{2,1} & \mathcal{I}_{3,1} \\
 -\mathcal{F}_{r,-1} \hat{\Upsilon}_r^2-i \mathcal{G}_{r,-1} \hat{\Upsilon}_r-\mathcal{H}_{r,-1}& -\mathcal{F}_{r,1} \hat{\Upsilon}_r^2+i \mathcal{G}_{r,1} \hat{\Upsilon}_r+\mathcal{H}_{r,1} & \mathcal{I}_{1r,0} & \mathcal{I}_{2,0} & \mathcal{I}_{3,0} \\
 0 & -\mathcal{F}_{r,0} \hat{\Upsilon}_r^2+i \mathcal{G}_{r,0}\hat{\Upsilon}_r+\mathcal{H}_{r,0} & \mathcal{I}_{1r,-1} & \mathcal{I}_{2,-1} & \mathcal{I}_{3,-1} \\
\mathcal{M}_{r,0}-i \mathcal{K}_{r,0} \hat{\Upsilon}_r & 0 & \mathcal{N}_{1r,1} & \mathcal{N}_{2,1} & \mathcal{N}_{3,1} \\
 \mathcal{M}_{r,-1}-i \mathcal{K}_{r,-1} \hat{\Upsilon}_r & \mathcal{M}_{r,1}+i \mathcal{K}_{r,1} \hat{\Upsilon}_r & \mathcal{N}_{1r,0} & \mathcal{N}_{2,0} & \mathcal{N}_{3,0} \\
 0 &\mathcal{M}_{r,0}+i \mathcal{K}_{r,0} \hat{\Upsilon}_r & \mathcal{N}_{1r,-1} & \mathcal{N}_{2,-1} & \mathcal{N}_{3,-1} \\
\end{array}
\right)\;,\\
\mathbf{v}&=\left(
\begin{array}{c}
 \delta \chi^S_{r,1}\\
 \delta \chi^S_{r,-1} \\
\Upsilon^S_r \\
u^S_{t,0}\\
u^S_{\phi,0} \\
\end{array}
\right)\;,\; \;\text{and}\; \;
\mathbf{c}=\left(
\begin{array}{c}
 \mathcal{J}_1 \\
 \mathcal{J}_0 \\
 \mathcal{J}_{-1} \\
  \mathcal{P}_1 \\
  \mathcal{P}_0 \\
 \mathcal{P}_{-1} \\
\end{array}
\right)\;.
\end{align}
Note that $\mathbf{M}$ is not a square matrix; the system is slightly overconstrained. We use the {\tt PseudoInverse} {\it Mathematica} function to find the values of $\delta\chi_{r,1}^S$, $\delta\chi_{r,-1}^S$, $\Upsilon_r^S$, $u_{t,0}^S$, $u_{\phi,0}^S$ that satisfy the system of the equations to within a certain tolerance.  (We strongly emphasize that $n_{\rm max} = 1$ is too small to accurate describe spinning-body orbits in almost all cases; this is merely used to illustrate the character of this system of linear equations.)

In the case of a nearly equatorial orbit, the polar and radial equations decouple such that we can solve Eq.\ (\ref{eq:matrixeq}) above independently of the equation for the $\theta$-motion. The $\theta$-equation  (\ref{eq:polarlinMP2}) has only three non-zero coefficients in the nearly equatorial limit, Eqs.\ (\ref{eq:QT}) -- (\ref{eq:ScriptV}). We insert the values for $\mathcal{Q}_{\vartheta}$ and $T_{\vartheta}$ and write $\mathcal{V}$ as a Fourier expansion of the form
\begin{align}
f(\lambda) & =\sum_{j=-1}^{1}\sum_{n=-n_\text{max}}^{n_\text{max}}f_{jn} e^{-ij \Upsilon_s \lambda} e^{-in \hat\Upsilon_r \lambda} \label{eq:precessingexpagain}\;.
\end{align}
We also write $\delta\vartheta_S$ as a Fourier expansion,
\begin{align}
\delta\vartheta_S(\lambda) & =\sum_{j=-1}^{1}\sum_{n=-n_\text{max}}^{n_\text{max}}\delta\vartheta_{S,jn} e^{-ij \Upsilon_s \lambda} e^{-in \hat\Upsilon_r \lambda} \label{eq:deltathetaprecessingexp}\;.
\end{align}
We take $n_{\text{max}}=1$ again, obtaining the following solution for Fourier coefficients of $\delta\vartheta_S$:
\begin{align}
\left(
\begin{array}{c}
 \delta\vartheta_{S,-1,-1}\\
  \delta\vartheta_{S,0,-1} \\
\delta\vartheta_{S,1,-1} \\
\delta\vartheta_{S,-1,0} \\
\delta\vartheta_{S,0,0}  \\
\delta\vartheta_{S,1,0}  \\
\delta\vartheta_{S,-1,1}  \\
\delta\vartheta_{S,0,1}  \\
\delta\vartheta_{S,1,1}  \\
\end{array}
\right)&=
-\left(
\begin{array}{c}
 \frac{\mathcal{V}_{-1,-1}}{\hat L_z^2-(\hat\Upsilon_r+\Upsilon_s)^2}\\
   \frac{\mathcal{V}_{0,-1}}{\hat L_z^2-\hat\Upsilon_r^2} \\
\frac{\mathcal{V}_{1,-1}}{\hat L_z^2-(\hat\Upsilon_r-\Upsilon_s)^2} \\
 \frac{\mathcal{V}_{-1,0}}{\hat L_z^2-\Upsilon_s^2} \\
0  \\
 \frac{\mathcal{V}_{1,0}}{\hat L_z^2-\Upsilon_s^2}  \\
\frac{\mathcal{V}_{-1,1}}{\hat L_z^2-(\hat\Upsilon_r-\Upsilon_s)^2} \\
 \frac{\mathcal{V}_{0,1}}{\hat L_z^2-\hat\Upsilon_r^2}  \\
\frac{\mathcal{V}_{1,1}}{\hat L_z^2-(\hat\Upsilon_r+\Upsilon_s)^2}  \\
\end{array}
\right)\;.
\end{align}

\end{widetext}

\bibliographystyle{unsrt}
\bibliography{FreqDomSpinningBody}
\end{document}